\documentclass[
 reprint,
showpacs,
nofootinbib,
 amsmath,amssymb,
 aps
prd,
]{revtex4-1}

\usepackage{graphicx}
\usepackage{dcolumn}
\usepackage{bm}
\usepackage{aas_macros}
\usepackage{xfrac}
\usepackage{color}
\usepackage{subfigure}
\usepackage{enumitem} 
\usepackage{mathrsfs}
\usepackage{gensymb}
\usepackage{hyperref}
\usepackage{mathtools}
\usepackage{amssymb}
\usepackage{multirow}
\usepackage[dvipsnames]{xcolor}

\newcommand\Tstrut{\rule{0pt}{3ex}}         
\newcommand\Bstrut{\rule[-2ex]{0pt}{0pt}}   
\newcommand{\ja}{Paper I}

\begin{document}

\title{Integrated Cosmological Probes: Extended Analysis}

\author{Andrina Nicola}
\email{andrina.nicola@phys.ethz.ch}
\author{Alexandre Refregier}
\author{Adam Amara}
\affiliation{Department of Physics, ETH Zurich, Wolfgang Pauli-Strasse 27, CH-8093 Zurich, Switzerland}

\date{\today}

\begin{abstract}

Recent progress in cosmology has relied on combining different cosmological probes. 
In an earlier work, we implemented an integrated approach to cosmology where the probes are combined into a common framework at the map level. 
This has the advantage of taking full account of the correlations between the different probes, to provide a stringent test of systematics and 
of the validity of the cosmological model. We extend this analysis to include not only CMB temperature, galaxy clustering, weak lensing from SDSS but also CMB lensing, weak lensing from the DES SV survey, Type Ia supernova and $H_{0}$ measurements. This yields 12 auto- and cross-power spectra which include the CMB temperature power spectrum, cosmic shear, galaxy clustering, galaxy-galaxy lensing, CMB lensing cross-correlation along with other cross-correlations as well as background probes. Furthermore, we extend the treatment of systematic uncertainties by studying the impact of intrinsic alignments, baryonic corrections, residual foregrounds in the CMB temperature, and calibration factors for the different power spectra.  
For $\Lambda$CDM, we find results that are consistent with our earlier work. Given our enlarged data set and systematics treatment, this confirms the robustness of our analysis and results. 
Furthermore, we find that our best-fit cosmological model gives a good fit to all the data we consider with no signs of tensions within our analysis.
We also find our constraints to be consistent with those found by the joint analysis of the WMAP9, SPT and ACT CMB experiments and the KiDS weak lensing survey. 
Comparing with the Planck Collaboration results, we see a broad agreement, but there are indications of a tension from the marginalized constraints in most pairs of cosmological parameters.
Since our analysis includes CMB temperature Planck data at $10 < \ell < 610$, the tension appears to arise between the Planck high$-\ell$ modes and the other measurements. 
Furthermore, we find the constraints on the probe calibration parameters to be in agreement with expectations, showing that the data sets are mutually consistent. In particular, this yields a confirmation of the amplitude calibration of the weak lensing measurements from SDSS, DES SV and Planck CMB lensing from our integrated analysis.

\end{abstract}

\pacs{98.80.-k, 98.80.Es}

\maketitle

\section{Introduction}

Recent progress in observations have led to the establishment of the standard model for cosmology. In spite of this
progress, some of the key components of the model such as Dark Energy, Dark Matter, inflation and large-scale gravity
remain either not understood or not fully tested. The constraints on this $\Lambda$CDM model and its extensions rely
on the combination of different cosmological probes such as the Cosmic Microwave Background (CMB), galaxy clustering,
weak gravitational lensing and Type Ia supernovae (SNe Ia). This combination is most often performed at the latest
stage in the analysis consisting of combining the likelihoods to infer a joint posterior constraint on the model parameters.

In an earlier work ~\cite{Nicola:2016} (hereafter \ja{}), we implemented an integrated approach to cosmology 
in which the cosmological probes are combined into a common framework at the map level. This has the advantage of taking full account of the correlations between the different probes which
generally probe common survey volumes, to provide a stringent test of systematics through the test of the consistency between
the probes and to yield a test of the validity of the cosmological model. We applied this framework to a combination of the 
CMB temperature from the Planck mission \cite{Planck-Collaboration:2016ad}, galaxy clustering from the eighth data release of the Sloan Digital Sky Survey (SDSS DR8) \cite{Aihara:2011}
and weak lensing from SDSS Stripe 82 \cite{Annis:2014}, making simplifying approximations but also conservative cuts on the data. 

In the present work, we extend the integrated analysis of \ja{} to also include CMB lensing maps from the Planck mission \cite{Planck-Collaboration:2016aa}, the recent weak lensing measurement with the publicly available Dark Energy Survey (DES) Science Verification (SV) data \cite{Jarvis:2016}, SNe Ia data from the joint light curve analysis (JLA) \cite{Betoule:2014} and constraints on the Hubble parameter from the Hubble Space Telescope (HST) \cite{Riess:2011, Efstathiou:2014}. This yields 12 auto and cross power spectra which include the CMB temperature power spectrum, cosmic shear, galaxy clustering, galaxy-galaxy lensing CMB lensing cross-correlation along with other cross-correlations as well as background probes. Furthermore, we extend the treatment of systematic uncertainties and relax some of the approximations as compared to \ja{}. In particular, we study the impact of intrinsic alignments, baryonic corrections, residual foregrounds in the CMB temperature, and calibration factors for the different power spectra. This extended analysis allows us to derive more robust constraints on the $\Lambda$CDM cosmological model and a more thorough test of the consistency between the different probes. Other joint analyses of different sets of cosmological probes have been performed (see references in \ja{} and Refs.~\cite{Giannantonio:2014aa, Giannantonio:2014ab, Soergel:2015}).

This paper is organized as follows. We review the framework for cosmological probe combination employed in this work in Section \ref{sec:framework}. In Section \ref{sec:data}, we describe the data used in this work and Section \ref{sec:theory} describes the theoretical modelling of the cosmological observables. We detail the computation of spherical harmonic power spectra in Section \ref{sec:cls}, while Section \ref{sec:systematics} summarizes the systematic uncertainties considered in this work. The computation of the covariance matrix is described in Section \ref{sec:covmat}. The method for parameter inference is described in Sec.~\ref{sec:methods} and our results on cosmological constraints are presented in Section \ref{sec:constraints}. We conclude in Section \ref{sec:conclusions}. Robustness tests as well as implementation details are deferred to the Appendices. 

\section{Framework for integrated probe combination}\label{sec:framework}

Following \ja, we create projected two-dimensional maps for the large-scale structure (LSS) and CMB probes. We then compute both the spherical harmonic auto-power spectra of these probes as well as the cross-power spectra for physically overlapping surveys. This yields a set of 12 spherical harmonic power spectra, which does not include the auto-power spectrum of the CMB lensing convergence but only its cross-correlations. We complement the observed power spectra with theoretical predictions and an estimate of their covariance matrix and combine these into a Gaussian likelihood. We combine the power spectrum likelihood with the likelihood of SNe Ia distance moduli and a constraint on the Hubble parameter, assuming these probes to be independent. In a last step we compute cosmological parameter constraints in a joint fit to these data. The implementation details for the CMB lensing convergence, weak lensing data from DES SV, SNe Ia and the Hubble constant measurement are described below. For a description of the remaining data the reader is referred to \ja.

\section{Data}\label{sec:data}

\begin{figure*}
\begin{center}
\includegraphics[scale=0.7]{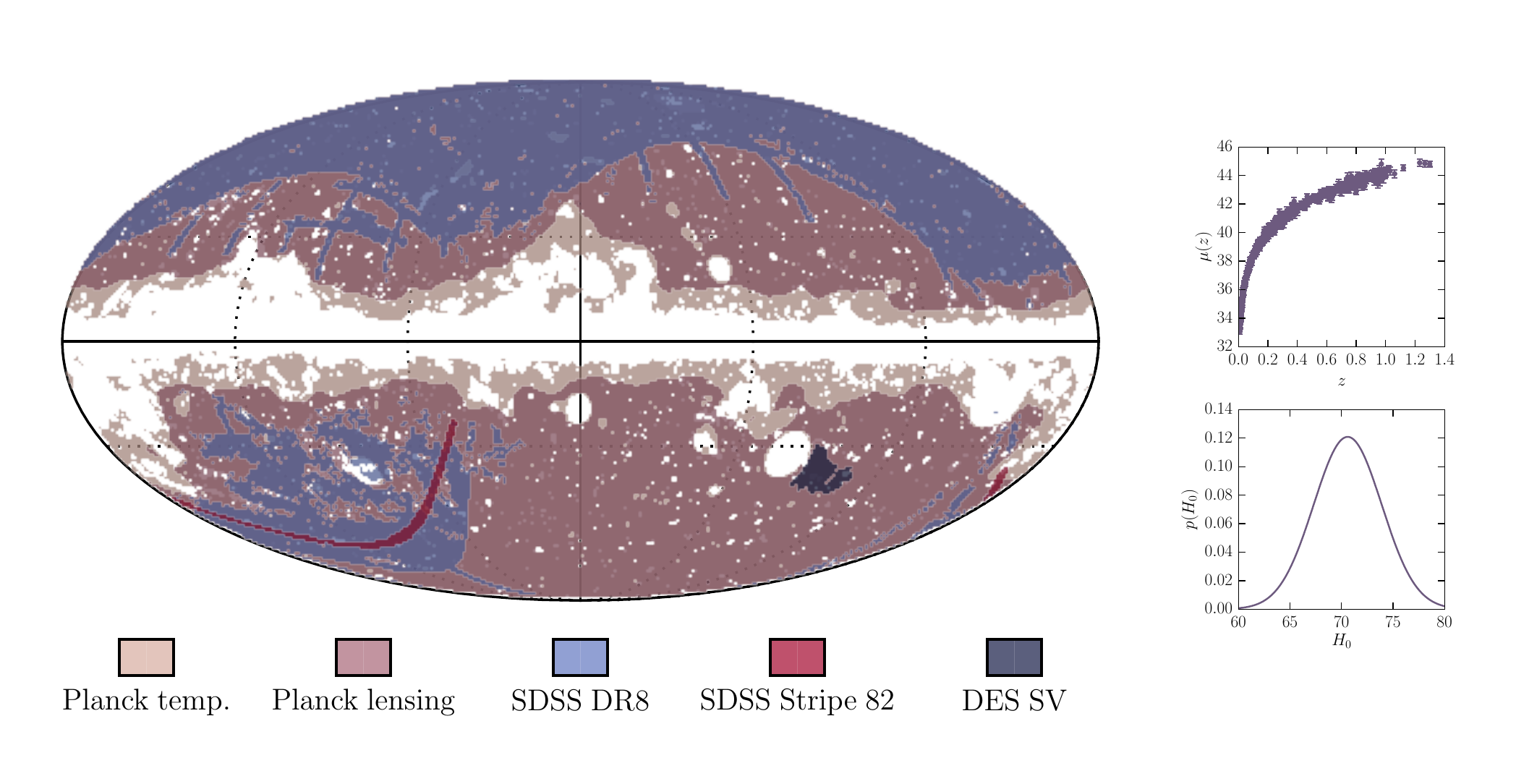}
\caption{Summary of the data used in this work. The left hand side shows an overlay of the footprints of the surveys used in this work: the CMB temperature and CMB lensing convergence from Planck, the galaxy density from SDSS DR8 and weak lensing from SDSS Stripe 82 and DES SV. The right hand side shows the background probes: SNe Ia from JLA and Hubble parameter data from HST. All footprints are shown in Galactic coordinates. The map is shown in Mollweide projection at a HEALPix resolution of $\tt NSIDE$ =  1024. (See Tab.~\ref{tab:data} for references for these different surveys.)} 
\label{fig:footprints}
\end{center}
\end{figure*}

The data used in this analysis is summarized in Table \ref{tab:data} and the footprints of the different surveys are illustrated in Figure \ref{fig:footprints} together with the background probes. We consider the data used in \ja{} namely, the Planck 2015 foreground-reduced CMB temperature anisotropy map derived using the $\tt{Commander}$ algorithm \cite{Planck-Collaboration:2016ab}, a map of the galaxy overdensity field derived using the CMASS1-4 sample from SDSS DR8 \cite{Aihara:2011, Ho:2012, Ross:2011} and a weak lensing shear map derived using SDSS Stripe 82 co-add data \cite{Annis:2014, Lin:2012}. In addition to the cosmological maps we also employ the binary survey masks presented in \ja{}. In this work, we complement these three maps with several data sets as described below.

\begin{table*}
\caption{Summary of the data sets used in our analysis.} \label{tab:data}
\begin{center}
\begin{tabular}{>{\centering}m{1.75cm}|>{\centering}m{1.75cm}|>{\centering}m{8cm}|>{\centering}m{2.5cm}@{}m{0pt}@{}} \hline \hline

\multicolumn{2}{>{\centering}m{3.5cm}|}{CMB temperature} & 
\Tstrut 
Survey: Planck 2015 \cite{Planck-Collaboration:2016ab} \\
Fiducial foreground-reduced map: $\tt{Commander}$ \\
Sky coverage: $f_{\text{sky}} = 0.776$
\Bstrut & \multirow{13}{*}{\centering \ja{}} & 

\tabularnewline \cline{1-3}
\multicolumn{2}{>{\centering}m{3.5cm}|}{Galaxy density} & 
\Tstrut 
Survey: SDSS DR8 \cite{Aihara:2011} \\
Sky coverage: $f_{\mathrm{sky}} = 0.27$ \\
Galaxy sample: CMASS1-4 \\
Number of galaxies: $N_{\mathrm{gal}} = 854\,063$ \\
Photometric redshift range $0.45 \leq z_{\mathrm{phot}} < 0.65$ 
\Bstrut & &

\tabularnewline \cline{1-3}
\multirow{7}{*}{\centering \parbox{1.5cm}{Weak lensing}} & SDSS Stripe 82 & 
\Tstrut 
Survey: SDSS Stripe 82 co-add \cite{Annis:2014}\\
Sky coverage: $f_{\mathrm{sky}} = 0.0069$ \\
Number of galaxies: $N_{\mathrm{gal}} = 3\,322\,915$ \\
Photometric redshift range: $0.1 \lesssim z_{\mathrm{phot}} \lesssim 1.1$ \\
r.m.s. ellipticity per component: $\sigma_{e} \sim 0.43$
\Bstrut &   &

\tabularnewline \cline{2-4}
 & DES & 
\Tstrut 
Survey: DES SV \cite{Jarvis:2016}\\
Sky coverage: $f_{\mathrm{sky}} = 0.0039$ \\
Number of galaxies: $N_{\mathrm{gal}} = 3\,279\,967$ \\
Photometric redshift range: $0.3 < z_{\mathrm{phot}} < 1.3$ \\
r.m.s. weighted ellipticity per component: $\sigma_{e} \sim 0.24$
\Bstrut & Sec.~\ref{subsec:gamma2map} &

\tabularnewline \hline        
\multicolumn{2}{>{\centering}m{3.5cm}|}{CMB lensing}  & 
\Tstrut 
Survey: Planck 2015 \cite{Planck-Collaboration:2016aa} \\
Sky coverage: $f_{\text{sky}} = 0.67$
\Bstrut & Sec.~\ref{subsec:kappamap} &
          
\tabularnewline \hline       
\multicolumn{2}{>{\centering}m{3.5cm}|}{SNe Type Ia} & 
\Tstrut 
Compilation: JLA \cite{Betoule:2014} \\
Number of SNe: $N_{\mathrm{SNe}} = 740$ \\
Redshift range: $0.01 < z < 1.3$
\Bstrut & Sec.~\ref{subsec:sne1a} &

\tabularnewline \hline       
\multicolumn{2}{>{\centering}m{3.5cm}|}{Hubble parameter} & 
\Tstrut 
Distance anchor: NGC 4258 \cite{Humphreys:2013} \\
Number of Cepheids: $N_{\mathrm{Ceph.}} = 600$ \cite{Riess:2011} \\
Number of SNe: $N_{\mathrm{SNe}} = 8$ \cite{Riess:2011} \\
Analysis: \citet{Efstathiou:2014}
\Bstrut & Sec.~\ref{subsec:H0} &

\tabularnewline \hline \hline     

\end{tabular}
\end{center}
\end{table*} 

\subsection{DES Weak lensing}\label{subsec:gamma2map}

We use publicly available data from the DES SV.\footnote{The data is available at: \\$\tt{https://des.ncsa.illinois.edu/releases/sva1}$.} The DES is an ongoing survey, imaging the sky in five photometric bandpasses ($g, r, i, z, Y$) using DECam \cite{Flaugher:2015}. After its five-year duration the DES will have covered approximately 5000 deg$^{2}$ of the southern sky to a limiting magnitude of about 24. The SV data were taken before the start of the main survey and they consist of more than 250 deg$^{2}$ \cite{DESCollaboration:2016aa}. In our analysis we use the largest contiguous area in the DES SV data, which is part of the South Pole Telescope East (SPT-E) field and covers an area of approximately 139 deg$^{2}$. The weak lensing shear for galaxies in the SPT-E region has been measured using two independent shape measurement codes, $\textsc{ngmix}$ \cite{Sheldon:2014} and $\textsc{im3shape}$ \cite{Zuntz:2013}. Both are model-fitting shear measurement codes and are described in Ref.~\cite{Jarvis:2016}. Photometric redshifts (photo-$z$) have been obtained using four different methods as described in Ref.~\cite{Bonnett:2016}. The photometric redshift catalogs both provide the full photo-$z$ probability distribution function (pdf) as well as an estimate of the mean of the pdf for each galaxy. We follow the choice of fiducial catalog of Refs.~\cite{Becker:2016, DES-Collaboration:2015} and perform our analysis using the galaxy shapes measured by $\textsc{ngmix}$ and the photometric redshifts determined using $\tt{SkyNet}$.

Our analysis closely follows the spherical harmonic power spectrum measurement described in Appendix A of Ref.~\cite{Becker:2016}. We select objects passing the SVA1 and the $\textsc{ngmix}$ cuts defined in Ref.~\cite{Jarvis:2016} which fall into any of the three tomographic redshift bins described in Refs.~\cite{Bonnett:2016, Jarvis:2016}. This selection yields $N_{\mathrm{gal}} = 3\,279\,967$ galaxies.

In order to construct the weak lensing shear maps we weight each galaxy's shear by its inverse variance weight described in Ref.~\cite{Jarvis:2016}. The galaxy shapes given in the DES SV shear catalogues are biased estimators of the galaxy shears and the $\textsc{ngmix}$ shape estimates therefore need to be corrected for the sensitivity as described in Ref.~\cite{Jarvis:2016}. Since this correction factor is a noisy estimate of the true correction it cannot be applied on single galaxies. In order to avoid introducing a bias caused by the noisy estimators of the sensitivity we therefore follow Ref.~\cite{Becker:2016} and estimate the weighted average of the galaxy sensitivities in our sample and correct each galaxy shape with this mean correction. 

We then rotate the galaxy shears from equatorial to Galactic coordinates\footnote{The exact rotation applied is given in \ja.} and pixelize them onto HEALPix\footnote{$\tt{http://healpix.sourceforge.net}$.} \cite{Gorski:2005} pixelizations of the sphere choosing a resolution of $\tt{NSIDE}$ = 1024. This resolution corresponds to a pixel area of $11.8$ arcmin$^{2}$. We apply a binary mask constructed from the union of unobserved and empty pixels to both shear maps. The final maps cover a fraction of sky $f_{\mathrm{sky}} = 0.0039$. The mean number of galaxies per pixel is approximately given by $\sfrac{n_{\mathrm{gal}}}{\mathrm{pix}} = 67$, which corresponds to $n_{\mathrm{gal}} = 5.73$ arcmin$^{-2}$. Figure \ref{fig:maps} shows the map of the shear modulus together with a zoom-in region with overlaid whisker plot illustrating the direction of the shear. 

We follow Refs.~\cite{Bonnett:2016, Becker:2016} and estimate the redshift distribution of the galaxies from the sum of the individual galaxy pdfs, weighted by their weak lensing shear weights. The resulting redshift distribution together with the weak lensing window function is shown in the Appendix (Fig.~\ref{fig:dessv-nz}).

\subsection{CMB lensing convergence}\label{subsec:kappamap}

\begin{figure*}
\begin{center}
\includegraphics[scale=0.7]{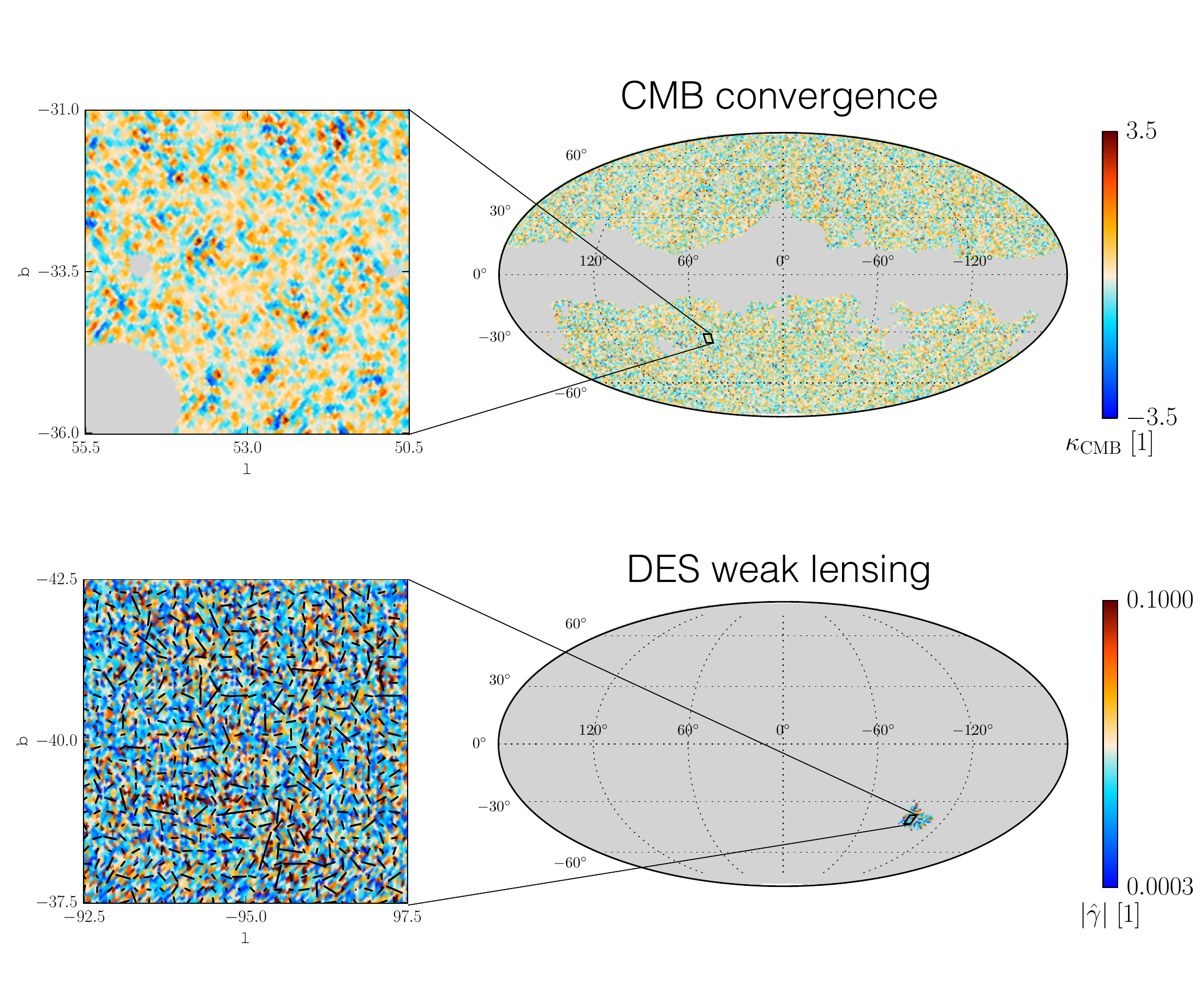}
\caption{New maps used in this analysis in addition to the CMB, galaxy clustering and SDSS weak lensing maps of \ja. The full-sky maps are in Galactic coordinates and are shown in Mollweide projection while the zoom-in versions are in Gnomonic projection. The CMB lensing convergence map derived from the foreground-removed CMB temperature and polarization anisotropy maps from $\tt{SMICA}$ is shown in the top panel. It is masked using the analysis mask provided by the Planck Collaboration. The zoom-in shows an enlarged version of the $5\times5$ deg$^{2}$ region centered on $(\tt{l}, \tt{b})$ $= (53\degree, -33.5\degree)$ shown in the map. The bottom panel shows the map of the weak lensing shear modulus $\vert \hat{\gamma} \vert$ derived from DES SV. Gray regions are masked because they are either unobserved or they do not contain any galaxies at our resolution. The zoom-in shows an enlarged version of the $5\times5$ deg$^{2}$ region centered on $(\tt{l}, \tt{b})$ $= (-95\degree, -40\degree)$ shown in the map. It is overlaid with a whisker plot illustrating the direction of the shear. Both maps are shown at a HEALPix resolution of $\tt NSIDE$ =  1024.} 
\label{fig:maps}
\end{center}
\end{figure*}

CMB lensing causes statistical anisotropies in CMB maps and the lensing potential can be reconstructed from these maps using a quadratic estimator \cite{Okamoto:2003}. We use the CMB lensing potential estimate $\hat{\phi}_{\mathrm{CMB}}$ provided by the Planck Collaboration in their second data release \cite{Planck-Collaboration:2016aa}. This estimator has been derived from the foreground-reduced CMB temperature and polarization maps computed using the $\tt{SMICA}$ algorithm \cite{Planck-Collaboration:2016aa, Planck-Collaboration:2016ab}. The use of both CMB temperature as well as polarization data allows for several CMB lensing potential estimators $(\hat{\phi}_{\mathrm{TT}}, \hat{\phi}_{\mathrm{TE}}, \hat{\phi}_{\mathrm{EE}}, \hat{\phi}_{\mathrm{EB}}, \hat{\phi}_{\mathrm{TB}})$, which can be combined into a minimal-variance estimator. This estimate is given in the form of spherical harmonic coefficients of the CMB lensing convergence $\kappa_{\mathrm{CMB}}$ in the angular multipole range $ 8 \leq \ell \leq 2048$. These are related to the CMB lensing potential $\phi_{\mathrm{CMB}}$ through:
\begin{equation}
\kappa_{\mathrm{CMB}, \ell, m} = \frac{\ell(\ell+1)}{2} \phi_{\mathrm{CMB}, \ell, m}.
\end{equation}
We use these spherical harmonic coefficients to create a HEALPix map of resolution $\tt{NSIDE}$ = 1024 using the HEALPix routine $\tt{alm2map}$. The analysis mask derived by the Planck Collaboration is provided as a HEALPix map of $\tt{NSIDE}$ = 2048. We downgrade this map to a resolution of $\tt{NSIDE}$ = 1024 following the procedure outlined in Ref.~\cite{Planck-Collaboration:2016ab}, which yields a binary analysis mask. We choose the CMB lensing convergence over the CMB lensing potential map since the lensing convergence is more local and should thus be less affected by masking effects arising when computing angular power spectra. The CMB lensing convergence map covers a fraction of sky $f_{\mathrm{sky}} = 0.67$ and is shown in Fig.~\ref{fig:maps}.

\subsection{Type Ia supernovae}\label{subsec:sne1a}

We complement the CMB and LSS data with geometrical constraints on the homogeneous Universe from the distance-redshift relation measured from Type Ia supernovae. We use data from the JLA \cite{Betoule:2014}, which is a compilation of 740 SNe Ia comprising data from SDSS-II \cite{Frieman:2008, Kessler:2009, Sollerman:2009, Lampeitl:2010, Campbell:2013}, the Supernova  Legacy Survey (SNLS) \cite{Astier:2006, Sullivan:2011}, the HST \cite{Riess:2007, Suzuki:2012} and several low-redshift experiments \cite{Betoule:2014}.\footnote{The data can be found at:\\ $\tt{http://supernovae.in2p3.fr/sdss\_snls\_jla/ReadMe.html}$.} The JLA data consist of SNe Ia light curve parameters which can be used to calculate observed distance moduli.  

\subsection{Hubble parameter}\label{subsec:H0}

We also add a local $H_{0}$ measurement from HST \cite{Riess:2011} to our analysis. We use the Hubble parameter estimate by Ref.~\cite{Efstathiou:2014}, which is a revision of the measurement presented in Ref.~\cite{Riess:2011}. Both measurements are derived from Cepheid-calibrated SNe Ia distance moduli but the former uses a revised distance to the anchor NGC 4258 \cite{Humphreys:2013} to calibrate the Cepheid distances. This analysis constrains the Hubble parameter to be given by $H_{0} = 70.6 \pm 3.3$ km s$^{-1}$ Mpc$^{-1}$, where the uncertainties are $1\sigma$ and assumed to be Gaussian.

\section{Model predictions}\label{sec:theory}

The auto- and cross-correlations of the CMB and LSS cosmological probes can be computed theoretically from the primordial power spectrum. In order to compute the power spectra of the cosmological fields $\delta_{g}$, $\gamma$ and $\kappa_{\mathrm{CMB}}$ we employ the Limber approximation \cite{Limber:1953, Kaiser:1992, Kaiser:1998} as in \ja. We further assume a flat cosmological model, i.e. $\Omega_{\mathrm{k}} = 0$. With these approximations the spherical harmonic power spectrum $C_{\ell}^{ij}$ between cosmological probes $i, j \in [ \delta_{g}, \gamma, \kappa_{\mathrm{CMB}}]$ at angular multipole $\ell$ can be expressed as 
\begin{multline}
C_{\ell}^{ij}=\int \mathrm{d} z \; \frac{c}{H(z)} \; \frac{W^{i}\boldsymbol{\left(}\chi(z)\boldsymbol{\right)}W^{j}\boldsymbol{\left(}\chi(z)\boldsymbol{\right)}}{\chi^{2}(z)} \\
\times P^{\mathrm{nl}}_{\delta \delta}\left(k=\frac{\ell+\sfrac{1}{2}}{\chi(z)}, z\right),
\end{multline}
where $H(z)$ is the Hubble parameter, $\chi(z)$ the comoving distance and $c$ denotes the speed of light. Furthermore, $P^{\mathrm{nl}}_{\delta \delta}\left(k, z\right)$ denotes the nonlinear matter power spectrum at redshift $z$ and wave vector $k$ and $W^{i'}\boldsymbol{\left(}\chi(z)\boldsymbol{\right)}$ is the window function for probe $i'$. 

The window functions for $\delta_{g}$ and $\gamma$ are given in \ja. Since the CMB lensing convergence is approximately sourced by a single-lens plane located at the last scattering surface with redshift $z_{*}$ its window function can be expressed as the single-plane limit of the weak lensing shear window function. We therefore have
\begin{equation}
W^{\kappa_{\mathrm{CMB}}}\boldsymbol{\left(}\chi(z)\boldsymbol{\right)} = \frac{3}{2} \frac{\Omega_{\mathrm{m}} H^{2}_{0}}{c^{2}} \frac{\chi(z)}{a} \frac{\chi(z_{*})-\chi(z)}{\chi(z_{*})},
\label{eq:kappawindow}
\end{equation} 
where $\Omega_{\mathrm{m}}$ is the fractional matter density today and $a$ is the scale factor. In our calculations we set $z_{*} = 1090$.

The power spectra involving CMB temperature anisotropies can also be related to the primordial density fluctuations. The expression for the CMB temperature power spectrum is given in \ja. The observed CMB temperature anisotropies are further correlated to tracers of the LSS. For the galaxy overdensity and weak lensing shear this cross-correlation is mainly due to the integrated Sachs-Wolfe (ISW) \cite{Sachs:1967} effect and the resulting cross-power spectra are given in \ja. The cross-correlation between the CMB temperature anisotropies and the CMB lensing convergence is dominated by the ISW but receives further contributions from Doppler effects arising from bulk velocities of electrons scattering the CMB photons and from the Sunyaev-Zel'dovich (SZ) \cite{Sunyaev:1980} effect (for a description of these effects see e.g. Refs.~\cite{Goldberg:1999, Cooray:2000}). The cross-correlation due to the SZ effect is not observable using the foreground-reduced CMB temperature anisotropy maps from Ref.~\cite{Planck-Collaboration:2016ab} but the remaining effects are observable. The cross-power spectrum between CMB temperature anisotropies and CMB lensing convergence can be computed from 
\begin{equation}
\langle a_{\mathrm{T}, \ell m} \; a_{\kappa_{\mathrm{CMB}}, \ell' m'} \rangle = C_{\ell}^{\mathrm{T}\kappa_{\mathrm{CMB}}} \delta_{\ell \ell'}\delta_{m m'},
\end{equation}
where $a_{\mathrm{T}, \ell m}$ denotes the spherical harmonic coefficients of the CMB temperature anisotropies $\Delta \mathrm{T}(\boldsymbol{\theta})$ and $a_{\kappa_{\mathrm{CMB}}, \ell' m'}$ denotes the spherical harmonic coefficients of the CMB lensing convergence defined as
\begin{equation}
\kappa_{\mathrm{CMB}}(\boldsymbol{\theta}) = \int \mathrm{d}z \; \frac{c}{H(z)} \; W^{\kappa_{\mathrm{CMB}}}\boldsymbol{\left(}\chi(z)\boldsymbol{\right)} \; \delta(\chi(z)\boldsymbol{\theta}, z).
\end{equation}

While the observables discussed so far probe cosmic structure formation, SNe Ia mainly probe the background evolution through their distance moduli. The distance modulus $\mu$ of a Type Ia supernova at redshift $z_{\mathrm{SNe}}$ is given by
\begin{equation}
\mu(z_{\mathrm{SNe}}) = 5 \log_{10}\left(\frac{d_{\mathrm{L}}(z_{\mathrm{SNe}})}{10 [\mathrm{pc}]}\right),
\label{eq:distmod}
\end{equation} 
where $d_{\mathrm{L}}(z_{\mathrm{SNe}})$ is the luminosity distance to redshift $z_{\mathrm{SNe}}$.

To compute theoretical predictions for all cosmological observables we follow \ja. We use the publicly available Boltzmann code $\textsc{class}$\footnote{$\tt{http://class\text{-}code.net}$.} \cite{Lesgourgues:2011} to compute the CMB temperature anisotropy power spectra and the cross-correlation between the CMB temperature anisotropies and the CMB lensing convergence. For the other observables we use $\textsc{PyCosmo}$ \cite{Refregier:2016}. As in \ja, we calculate the linear matter power spectrum from the transfer function derived by Ref.~\cite{Eisenstein:1998}. In order to compute nonlinear matter power spectra we use the $\textsc{Halofit}$ fitting function \cite{Smith:2003} with the revisions of Ref.~\cite{Takahashi:2012}. 

\section{Spherical harmonic power spectra}\label{sec:cls}

\begin{table}
\caption{Summary of spherical harmonic power spectrum parameters and angular multipole ranges used in this analysis. The first six power spectra are described in \ja.} \label{tab:clparams}
\begin{center}
\begin{ruledtabular}
\begin{tabular}{ccccc}
Power spectrum & $\theta_{\mathrm{max}}$ [deg] & $\theta_{\mathrm{FWHM}}$ [deg] & $\ell$-range & $\Delta \ell$ \\ \hline \Tstrut           
$C^{\mathrm{TT}}_{\ell}$ & 40 & 20 & $[10, \,610]$ & 30  \\
$C^{\delta_{g} \delta_{g}}_{\ell}$ & 80 & 40 & $[30, \,210]$ & 30  \\
$C^{\gamma_{1} \gamma_{1}}_{\ell}$ & 10 & 5 & $[70, \,610]$ & 60  \\
$C^{\delta_{g}\mathrm{T}}_{\ell}$ & 40 & 20 & $[30, \,210]$ & 30  \\
$C^{\gamma_{1} \mathrm{T}}_{\ell}$ & 10 & 5 & $[70, \,610]$ & 60  \\
$C^{\gamma_{1} \delta_{g}}_{\ell}$ & 10 & 5 & $[30, \,210]$ & 60  \\ \\                  
$C^{\kappa\mathrm{T}}_{\ell}$ & 40 & 20 & $[40, \,400]$ & 60  \\
$C^{\delta_{g}\kappa}_{\ell}$ & 80 & 40 & $[40, \,190]$ & 30  \\
$C^{\gamma_{1} \kappa}_{\ell}$ & 10 & 5 & $[70, \,370]$ & 60  \\
$C^{\gamma_{2} \mathrm{T}}_{\ell}$ & 15 & 7.5 & $[70, \,610]$ & 60  \\
$C^{\gamma_{2} \kappa}_{\ell}$ & 15 & 7.5 & $[70, \,370]$ & 60  \\
$C^{\gamma_{2} \gamma_{2}}_{\ell}$ & 15 & 7.5 & $[70, \,610]$ & 60 
\end{tabular}
\end{ruledtabular}
\end{center}
\end{table} 

Following \ja{} we use $\tt{PolSpice}$\footnote{\tt{http://www2.iap.fr/users/hivon/software/PolSpice/}.} \cite{Szapudi:2001, Szapudi:2001ab, Chon:2004} to measure the demasked spherical harmonic power spectra from the maps. We calculate the auto-power spectrum of the DES SV weak lensing shear map as well as the cross-correlations between the maps discussed in Section \ref{sec:data} and in \ja{} which have overlaps.

We do not include the auto-power spectrum of the CMB lensing convergence in this analysis. This is due to the fact that the auto-power spectrum of the CMB lensing convergence estimator is a biased estimate of the CMB lensing convergence auto-power spectrum because it probes both the connected and the disconnected part of the 4-point function of the CMB temperature anisotropies \cite{Planck-Collaboration:2014aa}. In order to obtain the power spectrum of the CMB lensing convergence, the auto-power spectrum of the estimator thus needs to be corrected for this disconnected bias \cite{Planck-Collaboration:2014aa}, which is beyond the scope of this paper.

In order to compute the power spectra, we follow the method outlined in \ja{} to estimate the values of the maximal angular scale used by $\tt{PolSpice}$ $\theta_{\mathrm{max}}$ and the apodization parameter $\theta_{\mathrm{FWHM}}$. We validate these settings using the Gaussian simulations described in Appendix \ref{sec:ap-mocks}. The demasking procedure used by $\tt{PolSpice}$ leads to biases in the recovered power spectra, as discussed in \ja{}. The kernels that relate average $\tt{PolSpice}$ estimates to the true power spectra can be computed analytically for each choice of $\theta_{\mathrm{max}},\theta_{\mathrm{FWHM}}$ and we take them into account by convolving all theoretical predictions with these kernels. The choice of angular multipole ranges follows that described in \ja{} for all power spectra already included in that analysis. The angular multipole range for power spectra involving the CMB lensing convergence follows the conservative choice described in Ref.~\cite{Planck-Collaboration:2016aa}. The low-$\ell$ limit for the power spectra is chosen to minimize the impact of mean field corrections; the high-$\ell$ limit is chosen because of mild evidence for systematic errors at higher multipoles \cite{Planck-Collaboration:2016aa}. The chosen bin widths largely follow the conservative binning outlined in Ref.~\cite{Planck-Collaboration:2016aa} ($\Delta \ell = 45$) and the binning scheme in \ja, which is chosen to roughly correspond to the width of the $\tt{PolSpice}$ kernels. Where we choose angular multipole bins broader than necessary this is mainly done to reduce the size of the spherical harmonic power spectrum vector. The binning schemes and $\tt{PolSpice}$ parameters used for all power spectra are summarized in Table \ref{tab:clparams}.

All the spherical harmonic power spectra are computed from the maps of resolution $\tt{NSIDE}$ = 1024. They are further corrected for the effect of the HEALPix pixel window function and the power spectra involving the CMB temperature anisotropy map are further corrected for the Planck effective beam window function. The uncertainties are derived from the Gaussian simulations described in Appendix \ref{sec:ap-mocks}. 
 
The power spectra computed in this work are described in more detail below; for a description of the remaining power spectra the reader is referred to \ja.

\subsection{DES SV cosmic shear}

We compute the cosmic shear power spectrum for DES SV using the map and mask described in Section \ref{subsec:gamma2map}. In order to estimate the contribution of shape noise, we follow \ja{} and resort to simulations. We generate 100 noise maps by rotating the galaxy shears by a random angle. We then calculate the power spectra of these maps and our estimator of the shape noise power spectrum is given by the mean power spectrum of the noise maps. 

The weak lensing shear E-mode power spectrum for DES SV is shown in the $4, 4$-panel of Fig.~\ref{fig:cls} and the B-mode power spectrum is shown in the Appendix (Fig.~\ref{fig:cls_b_dessv}). The noise level of DES SV is lower than the one from SDSS Stripe 82 as can be seen by comparing panels $4, 4$ and $3, 3$. This is to be expected from the higher galaxy number density and smaller measurement noise of DES SV data.

In Appendix \ref{sec:ap-robustness-tests} we compare the DES SV cosmic shear power spectra computed from the maps in Galactic and equatorial coordinates. We find discrepancies similar to those found for SDSS Stripe 82 \cite{Nicola:2016}, especially at small angular scales. Since the differences detected are within the uncertainties of the measurement, we use the cosmic shear power spectrum calculated from the maps in Galactic coordinates in our integrated analysis.

\subsection{CMB temperature and DES SV weak lensing shear cross-correlation}

We compute the cross-power spectrum between the DES SV weak lensing shear and the CMB temperature anisotropies using the maps and masks presented in Section \ref{subsec:gamma2map} and \ja. As discussed in \ja{}, we choose to compute cross-correlations using the combined masks of the respective probes rather than the single-probe masks. This is due to the fact that the former approach results in a better recovery of the input cross-power spectra in the Gaussian simulations described in Appendix \ref{sec:ap-mocks}. We therefore mask both maps with the combination of the single-probe masks, which covers a fraction of sky $f_{\mathrm{sky}} \sim  0.0035$. 

The resulting spherical harmonic power spectrum is shown in the $4, 0$-panel of Fig.~\ref{fig:cls}. As can be seen, the noise level is too high to allow for a detection of the ISW from DES SV weak lensing shear. This is in agreement with the results found for SDSS Stripe 82 in \ja. We nevertheless include the power spectrum in our analysis since it provides an upper limit to the ISW signal from weak lensing. 

In Appendix \ref{sec:ap-robustness-tests} we investigate the impact of our choice of fiducial foreground-reduced CMB temperature map by comparing the power spectra obtained using the four different foreground-reduction algorithms employed by Ref.~\cite{Planck-Collaboration:2016ab}. As can be seen from Fig.~\ref{fig:cls_foreground_rem} we find the measured power spectra to be virtually the same.

\subsection{CMB lensing convergence and galaxy overdensity cross-correlation}

To compute the cross-power spectrum between the CMB lensing convergence and the SDSS DR8 galaxy overdensity we use the maps and masks described in Section \ref{subsec:kappamap} and \ja. We mask both maps with their combined mask, which covers a fraction of sky $f_{\mathrm{sky}} \sim  0.26$.

The spherical harmonic cross-power spectrum between the CMB lensing convergence and the galaxy overdensity is shown in the $2, 1$-panel in Fig.~\ref{fig:cls}. We see that we clearly detect a nonzero correlation between the CMB lensing convergence and the galaxy overdensity.

\subsection{CMB lensing convergence and CMB temperature cross-correlation}

To compute the cross-power spectrum between the CMB lensing convergence and the CMB temperature anisotropies we use the maps and masks presented in Section \ref{subsec:kappamap} and \ja. The combined mask of both probes covers a fraction of sky $f_{\mathrm{sky}} \sim  0.65$ and we apply this mask to both maps.

The resulting spherical harmonic power spectrum is shown in the $2, 0$-panel of Fig.~\ref{fig:cls}. Comparing to the results derived in Ref.~\cite{Planck-Collaboration:2016aa}, we find good overall agreement. 

In Appendix \ref{sec:ap-robustness-tests}, we again compare the power spectra obtained from the different foreground-reduced CMB temperature anisotropy maps and find them to agree rather well.

\subsection{CMB lensing convergence and SDSS Stripe 82 weak lensing shear cross-correlation}

We estimate the cross-power spectrum between the CMB lensing convergence and the SDSS Stripe 82 weak lensing shear map using the maps and masks described in Section \ref{subsec:kappamap} and \ja. We mask both maps with their combined mask, which covers a fraction of sky $f_{\mathrm{sky}} \sim  0.0064$.

The spherical harmonic power spectrum is illustrated in the $3, 2$-panel in Fig.~\ref{fig:cls}. We see that the obtained cross-power spectrum is rather noisy and it does not allow for a detection of the correlation between CMB lensing convergence and SDSS Stripe 82 weak lensing shear. This is probably due to the combined effect of a low fractional sky coverage of SDSS Stripe 82 data and significant noise in both the CMB lensing convergence and SDSS Stripe 82 weak lensing shear. We nevertheless include this cross-correlation into our analysis to serve as an upper limit.

\subsection{CMB lensing convergence and DES SV weak lensing shear cross-correlation}

To compute the cross-power spectrum between the CMB lensing convergence and the DES SV weak lensing shear map we use the maps and masks presented in Sections \ref{subsec:gamma2map} and \ref{subsec:kappamap}. The combined mask of both maps covers a fraction of sky $f_{\mathrm{sky}} \sim  0.0037$ and we apply it to both maps.

The $4, 2$-panel in Fig.~\ref{fig:cls} shows the resulting spherical harmonic power spectrum. We see that the signal-to-noise of the cross-correlation is low for the angular scales considered, which we attribute to both a small sky coverage of DES SV data and the noise level in both maps. Nevertheless we include the power spectrum in our analysis since it provides an upper limit to the cross-correlation of the CMB lensing convergence and the DES SV weak lensing shear field.

\begin{figure*}
\begin{center}
\includegraphics[scale=0.5]{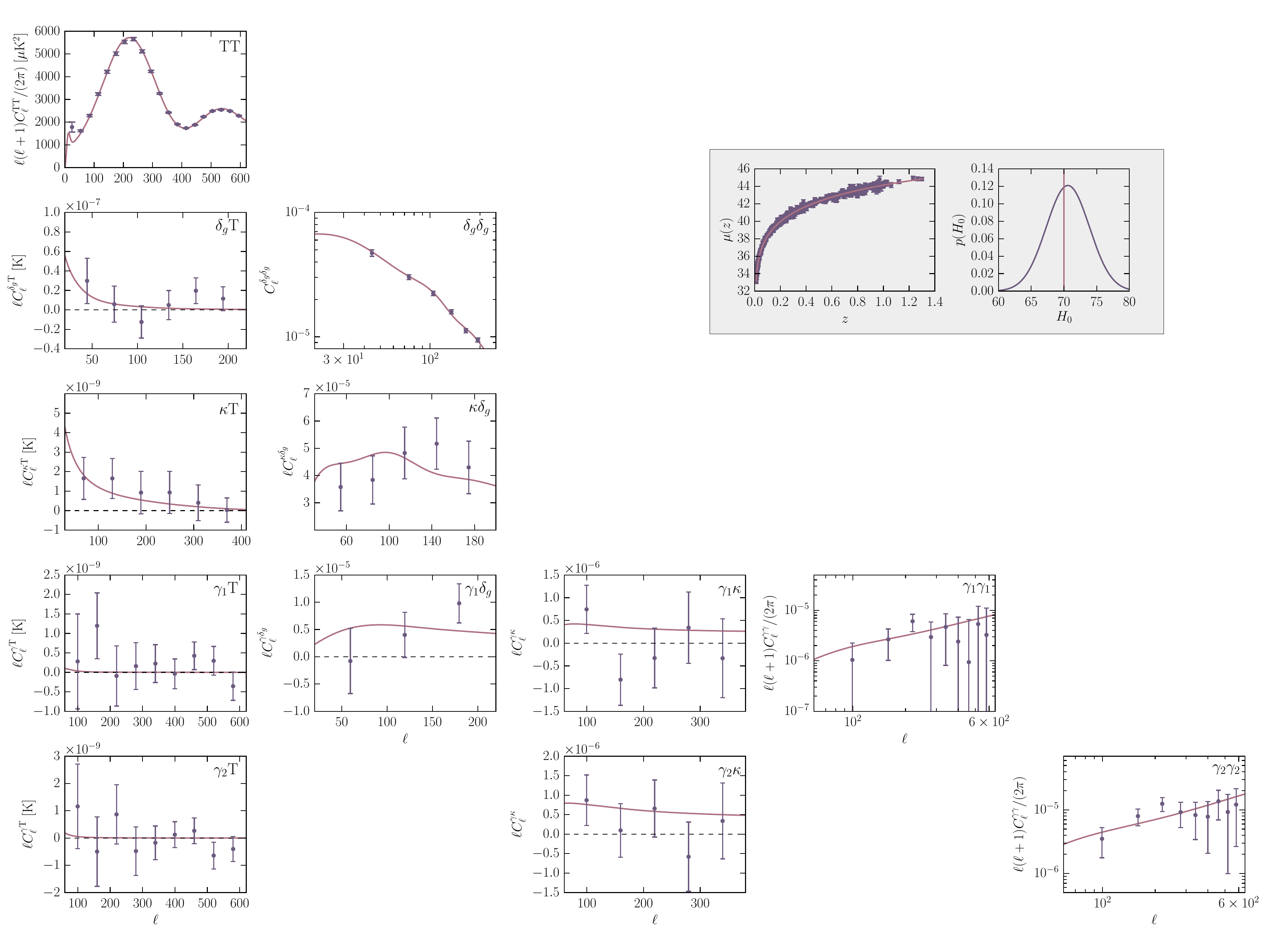}
\caption{Measured auto and cross spherical harmonic power spectra along with background probes for this analysis. Starting from the top-left corner, the $0, 0$-panel shows the CMB temperature anisotropy power spectrum from Planck 2015. The $1, 0$-panel shows the cross-power spectrum between the galaxy overdensity from the SDSS DR8 CMASS sample and the CMB temperature anisotropies from Planck 2015. The $1,1$-panel shows the galaxy clustering power spectrum from the SDSS DR8 CMASS sample. The $2, 0$-panel shows the cross-power spectrum between the CMB lensing convergence and the CMB temperature anisotropies from Planck 2015. The $2, 1$-panel shows the cross-power spectrum between the CMB lensing convergence from Planck 2015 and the galaxy overdensity from the SDSS DR8 CMASS sample. The $3, 0$-panel shows the cross-power spectrum between the weak lensing shear from SDSS Stripe 82 ($\gamma_{1}$) and the CMB temperature anisotropies from Planck 2015. The $3, 1$-panel shows the cross-power spectrum between the weak lensing shear from SDSS Stripe 82 and the galaxy overdensity from the SDSS DR8 CMASS sample. The $3, 2$-panel shows the cross-power spectrum between the weak lensing shear from SDSS Stripe 82 and the CMB lensing convergence from Planck 2015. The $3, 3$-panel shows the cosmic shear power spectrum from SDSS Stripe 82. The $4, 0$-panel shows the cross-power spectrum between the weak lensing shear from DES SV ($\gamma_{2}$) and the CMB temperature anisotropies from Planck 2015. The $4, 2$-panel shows the cross-power spectrum between the weak lensing shear from DES SV and the CMB lensing convergence from Planck 2015. The $4, 4$-panel shows the cosmic shear power spectrum from DES SV. The gray panel in the upper right corner shows the SNe Ia distance moduli and error bars from the JLA and the Hubble constant measurement from Ref.~\cite{Efstathiou:2014}. All the power spectra have been computed using the maps in Galactic coordinates. The solid lines show the theoretical predictions for the best-fit cosmological model determined from the integrated analysis which is summarized in Tab.~\ref{tab:params}. The theoretical predictions for the power spectra have been convolved with the $\tt{PolSpice}$ kernels described in \ja{} and Section \ref{sec:cls}. The error bars for the power spectra are derived from the Gaussian simulations described in Section \ref{sec:covmat} and Appendix \ref{sec:ap-mocks}. The angular multipole ranges and binning schemes used for all the power spectra are summarized in Tab.~\ref{tab:clparams}.
\label{fig:cls}} 
\label{fig:cls}
\end{center}
\end{figure*}

\section{Systematics}\label{sec:systematics}

Cosmological measurements are generally affected by systematics. We parametrize these using eight different nuisance parameters, which we simultaneously fit with the cosmological parameters. A summary of these parameters can be found in Tab.~\ref{tab:params} and they are described separately for each cosmological probe below.

\subsection{CMB temperature anisotropies} 
The foreground-reduced CMB temperature anisotropy maps contain significant contamination from unresolved extragalactic sources, mainly dusty and radio galaxies \cite{Planck-Collaboration:2014ac}. Following Ref.~\cite{Planck-Collaboration:2014af}, these can be modelled as additional, residual power spectra, which become significant at high angular multipoles. Ref.~\cite{Planck-Collaboration:2014af} include two different contributions: a contribution of an unclustered Poisson component $C_{\ell}^{\mathrm{ps}}$ and the contribution of a clustered component $C_{\ell}^{\mathrm{cl}}$. We study the impact of including these power spectra and find that they do not have a significant impact on our results; see Appendix \ref{sec:ap-cmb-foregrounds} for further details. We therefore do not include them in our fiducial analysis.

\subsection{Galaxy overdensity}
The galaxy overdensity field is a biased tracer of the underlying dark matter. We account for this uncertainty with a linear galaxy bias parameter $b$, which relates the galaxy overdensity $\delta_{g}$ to the dark matter overdensity $\delta$ i.e. we set
\begin{equation}
\delta_{g}(k, z) = b \, \delta(k, z).
\end{equation}
In addition, the observed galaxy overdensity is potentially affected by observational and sky systematics. As detailed in \ja{} we correct the galaxy overdensity map for these systematics. We do not find any significant cross-correlation between the foreground-reduced galaxy overdensity map and systematics that could be common to other probes, such as the extinction map from Ref.~\cite{Schlegel:1998} as well as a map of stars detected by SDSS with $\tt{i}$-band magnitudes $18.0 \leq i < 18.5$. We therefore do not include nuisance parameters accounting for foreground contamination of the galaxy overdensity field into our analysis.

\subsection{Weak lensing}

The estimated weak lensing shear of galaxies $\hat{\boldsymbol{\gamma}}$ is prone to multiplicative biases. We parametrize these potential unaccounted calibration uncertainties using a scalar multiplicative bias parameter defined as
\begin{equation}
\hat{\boldsymbol{\gamma}} = (1 + m^{i}_{\mathrm{calib}})\boldsymbol{\gamma},
\end{equation}
where $i \in$ [SDSS, DES].

A further potential contaminant to the observed weak lensing shear signal are intrinsic correlations between the unlensed shapes of galaxies (for reviews see e.g. \cite{Joachimi:2015, Troxel:2015}, for observational detections see e.g. \cite{Heymans:2006, Mandelbaum:2006, Hirata:2007, Faltenbacher:2009, Okumura:2009, Joachimi:2011, Singh:2015}). The measured weak lensing shear in the presence of intrinsic alignments is given by
\begin{equation}
\boldsymbol{\gamma}^{\mathrm{obs}} = \boldsymbol{\gamma}^{G} + \boldsymbol{\gamma}^{I},
\end{equation}
where $\boldsymbol{\gamma}^{G}$ denotes the gravitational and $\boldsymbol{\gamma}^{I}$ the intrinsic part of the shear. To first order, the shapes of galaxies are linearly related to the tidal field in which they form \cite{Catelan:2001}. This gives rise to the so-called linear alignment model \cite{Catelan:2001, Hirata:2004aa, Bridle:2007}. The expected linear alignment signal in this model follows very closely the scale dependence of the weak lensing shear power spectrum as discussed in Appendix \ref{sec:ap-ias}. We therefore choose to model intrinsic alignments as an additional contribution that modifies the amplitude of the weak lensing shear i.e.
 \begin{equation}
\hat{\boldsymbol{\gamma}} = (1 + m^{i}_{\mathrm{calib}} + m^{i}_{\mathrm{IA}})\boldsymbol{\gamma} \triangleq (1 + m^{i}_{\star})\boldsymbol{\gamma},
\end{equation}
where $i \in$ [SDSS, DES]. In order to reduce the number of free parameters we combine these two amplitudes into an effective multiplicative calibration $m^{i}_{\star}$. We include a separate effective multiplicative bias parameter for each of the two weak lensing surveys considered in this work. 

The dark matter power spectrum at small scales is affected by baryonic processes such as feedback from supernovae and active galactic nuclei (AGN) or gas cooling. We investigate the effects of baryonic feedback on the power spectra using the effective halo model prescription of \citet{Mead:2015} (for further details see Appendix \ref{sec:ap-baryons}). We find the effect of baryon feedback to be smaller than the uncertainties on our measurement on all angular scales considered, which means that our data currently is not sensitive to baryonic feedback. Therefore we choose to model the nonlinear matter power spectrum using $\textsc{Halofit}$ in this work and leave the investigation of baryonic feedback to future work.

\subsection{CMB lensing convergence}

The CMB lensing potential estimator described in Sec.~\ref{subsec:kappamap} has a non-trivial, cosmology-dependent response to an input CMB lensing potential \cite{Planck-Collaboration:2014aa, Planck-Collaboration:2016aa}. Ref.~\cite{Planck-Collaboration:2016aa} correct for this bias assuming a fiducial cosmological model \cite{Planck-Collaboration:2016aa}. Therefore, the normalization of the CMB lensing convergence estimator is cosmology-dependent and should be varied alongside the cosmological parameters in a sampling process. To reduce computation time we choose an alternative approach and do not take into account the cosmology-dependence of the CMB lensing convergence amplitude as e.g. Ref.~\cite{Giannantonio:2014ab}. We account for this normalization uncertainty by including a multiplicative bias parameter $m_{\kappa_{\mathrm{CMB}}}$ such that
\begin{equation}
\hat{\kappa}_{\mathrm{CMB}} = (1 + m_{\kappa_{\mathrm{CMB}}}) \kappa_{\mathrm{CMB}},
\end{equation}
and we allow it to vary independently from the cosmological parameters. We leave the introduction of a cosmology-dependent CMB lensing convergence amplitude to future work.
 
\subsection{SNe Ia} 
Type Ia supernovae are not perfect standard candles since their absolute peak magnitude depends on the duration of the SNe explosion, as measured using the stretch parameter $X_{1}$, the color $C$ of the SNe and the properties of the host galaxy. In order to take these effects into account we parametrize the observed distance modulus following Ref.~\cite{Betoule:2014} as:
\begin{equation}
\mu = m_{\mathrm{B}}^{*} - (M_{\mathrm{B}} - \alpha X_{1} + \beta C),
\end{equation}
where $m_{\mathrm{B}}^{*}$ denotes the observed peak magnitude in rest frame $\tt{B}$-band. The parameters $\alpha, \beta$ and $M_{\mathrm{B}}$ are nuisance parameters accounting for the uncertainties in the absolute peak SNe Ia magnitude. Both the absolute magnitude parameter $M_{\mathrm{B}}$ and the parameter $\beta$ were found to depend on the properties of the supernova's host galaxy \cite{Sullivan:2010, Johansson:2013}. In order to take these effects into account, we follow Ref.~\cite{Betoule:2014} and set 
\begin{equation}
M_{\mathrm{B}} = 
    \begin{cases}
      M^{1}_{\mathrm{B}} & \text{if } M_{\text{stellar}} < 10^{10} M_{\odot}, \\
      M^{1}_{\mathrm{B}} + \Delta M       & \text{otherwise}.
    \end{cases}
\end{equation}
This parametrization thus finally gives rise to four different nuisance parameters $\alpha, \beta, M^{1}_{\mathrm{B}}$ and $\Delta M$.

\section{Covariance matrix}\label{sec:covmat}

In order to compute constraints on cosmological parameters from the 12 power spectra described above, we need to estimate the joint covariance matrix of these probes. We follow \ja{} and assume the covariance matrix to be Gaussian. This assumption is justified for the CMB lensing convergence field, as shown by e.g. Ref.~\cite{van-Engelen:2012}. As discussed in \ja{} this is also appropriate for the CMB temperature anisotropy and galaxy density fields at the scales considered but it is only an approximation for the weak lensing shear field. We expect this to be a reasonable approximation since we do not include small angular scales in our analysis and our uncertainties on the cosmic shear power spectrum are dominated by shape noise. We therefore leave the issue of non-Gaussian covariance matrices to future work.

Following \ja{} we compute the covariance matrix employing two different methods: the first is based on a theoretical prediction of the covariance matrix while the second is an empirical method based on Gaussian simulations of the cosmological probes considered in this work. As in \ja{}, we use the empirical covariance matrix for our fiducial analysis and we compute it using 1000 Gaussian simulations, which are described in Appendix \ref{sec:ap-mocks}. We validate the empirical covariance matrix using the theoretical prediction. A more detailed description of both methods can be found in \ja.

\begin{figure}
\begin{center}
\includegraphics[scale=0.3]{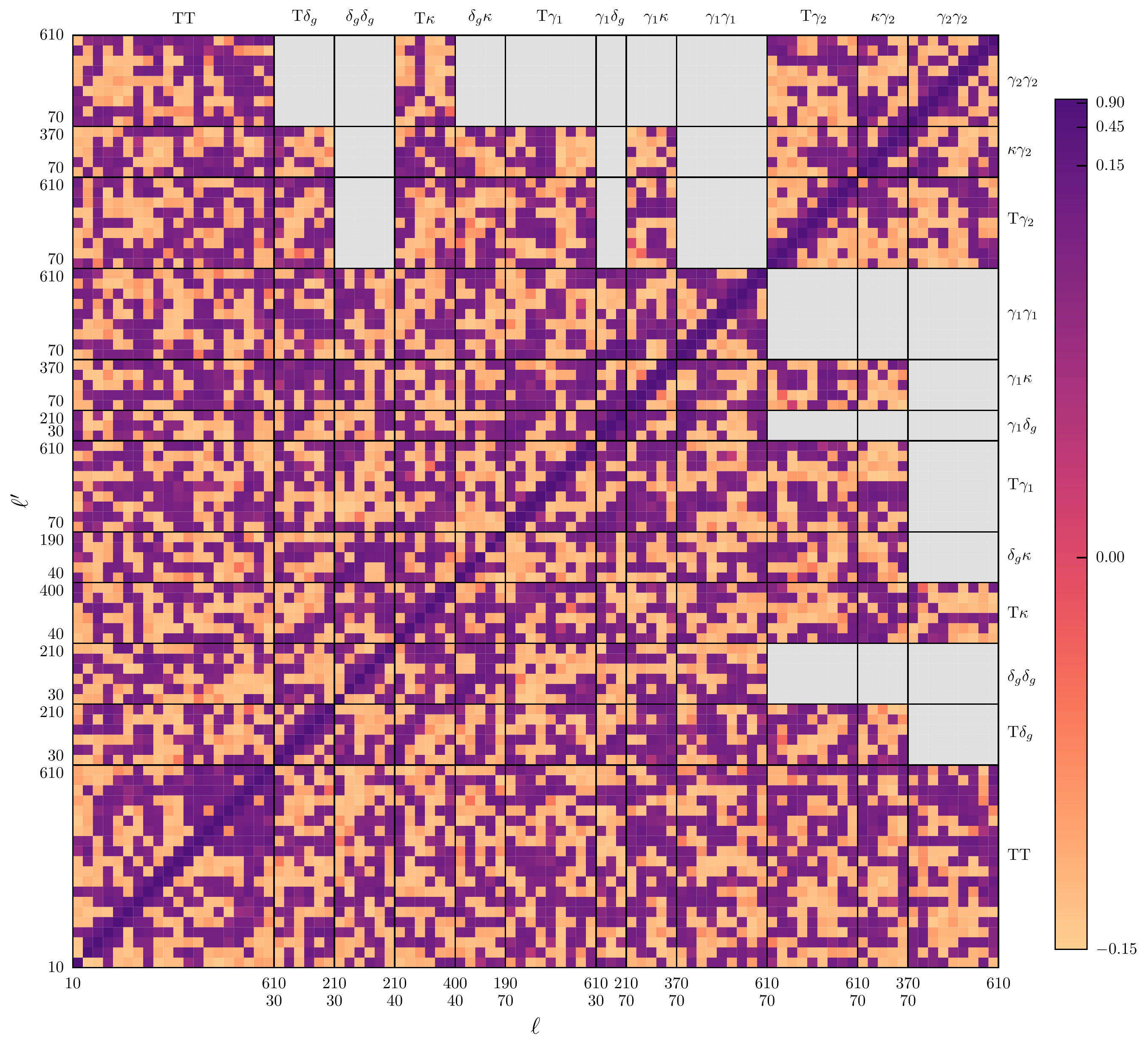}
\caption{Correlation matrix of the spherical harmonic power spectra derived using the Gaussian simulations described in Sec.~\ref{sec:covmat}. Gray regions are set to zero because they correspond to the covariance between non-overlapping surveys. The binning scheme and angular multipole ranges for each probe follow those given in Tab.~\ref{tab:clparams}.} 
\label{fig:covmat}
\end{center}
\end{figure}

Figure \ref{fig:covmat} illustrates the correlation matrix derived from the sample variance of the 1000 Gaussian simulations. We explicitly set sub-covariance matrices of non-overlapping surveys to zero. These regions are marked in gray in the figure.

\section{Parameter inference}\label{sec:methods}

To compute cosmological parameter constraints from the data presented in Sections \ref{sec:data} and \ref{sec:cls} we follow \ja{} and assume the joint likelihood of the 12 spherical harmonic power spectra to be Gaussian i.e.
\begin{multline}
\mathscr{L}(D \vert \theta) = \frac{1}{[(2\pi)^{d}\det{C_{G}}]^{\sfrac{1}{2}}} \\
\times e^{-\frac{1}{2}(\mathbf{C}^{\mathrm{obs}}_{\ell}-\mathbf{C}^{\mathrm{theor}}_{\ell})^{\mathrm{T}}C^{-1}_{G}(\mathbf{C}^{\mathrm{obs}}_{\ell}-\mathbf{C}^{\mathrm{theor}}_{\ell})},
\label{eq:pslikelihood}
\end{multline}
where $C_{G}$ denotes the Gaussian covariance matrix described in Sec.~\ref{sec:covmat}. $\mathbf{C}^{\mathrm{theor}}_{\ell}$ denotes the theoretical prediction for the spherical harmonic power spectrum vector of dimension $d=92$ and $\mathbf{C}^{\mathrm{obs}}_{\ell}$ is the observed power spectrum vector, defined as 
\begin{multline}
\mathbf{C}^{\mathrm{obs}}_{\ell} = (C^{\mathrm{TT}}_{\ell} \; C_{\ell}^{\delta_{g}\mathrm{T}} \; C_{\ell}^{\delta_{g} \delta_{g}} \; C_{\ell}^{\kappa_{\mathrm{CMB}}\mathrm{T}} \; C_{\ell}^{\kappa_{\mathrm{CMB}}\delta_{g}} \; C^{\gamma_{1} \mathrm{T}}_{\ell} \; C^{\gamma_{1}\delta_{g}}_{\ell}  \\
C_{\ell}^{\gamma_{1} \kappa_{\mathrm{CMB}}} \; C^{\gamma_{1} \gamma_{1}}_{\ell} \; C^{\gamma_{2} \mathrm{T}}_{\ell} \; C_{\ell}^{\gamma_{2} \kappa_{\mathrm{CMB}}} \; C^{\gamma_{2} \gamma_{2}}_{\ell})_{\mathrm{obs}}.
\label{eq:psvector}
\end{multline} 
As in \ja{} we neglect the potential non-Gaussian nature of the weak lensing likelihood. We estimate the joint covariance matrix $C_{G}$ both from simulations as well as analytically as described in Sec.~\ref{sec:covmat}. We compute it for a fiducial $\Lambda$CDM cosmological model with parameter values $\{h,\, \Omega_{\mathrm{m}}, \,\Omega_{\mathrm{b}}, \, n_{\mathrm{s}}, \,\sigma_{8}, \,\tau_{\mathrm{reion}}, \,T_{\mathrm{CMB}}\} = \{0.7, \,0.3, \,0.049, \,1.0, \,0.88, \,0.078, \,2.275 \,\mathrm{K}\}$, where $h$ is the dimensionless Hubble parameter, $\Omega_{\mathrm{m}}$ is the fractional matter density today, $\Omega_{\mathrm{b}}$ is the fractional baryon density today, $n_{\mathrm{s}}$ denotes the scalar spectral index, $\sigma_{8}$ is the r.m.s. of linear matter fluctuations in spheres of comoving radius $8 \,h^{-1}$ Mpc, $\tau_{\mathrm{reion}}$ denotes the optical depth to reionization and $T_{\mathrm{CMB}}$ is the mean temperature of the CMB today. We assume no systematic uncertainties in our fiducial model except a linear galaxy bias i.e. $\{b,\, m^{\mathrm{SDSS}}_{*}, \, m^{\mathrm{DES}}_{*}, \, m_{\kappa_{\mathrm{CMB}}}, \, A_{\mathrm{ps}}, \, A_{\mathrm{cl}}\} = \{2., \,0., \,0., \,0., \,0., \,0.\}$. As described in \ja{} we employ the corrections described in Refs.~\cite{Kaufman:1967, Anderson:2003, Hartlap:2007} to debias the inverse of the empirical covariance matrix and we neglect the cosmology dependence of the covariance in our sampling process. 

We further assume a Gaussian likelihood for the SNe Ia distance moduli $\boldsymbol{\mu}$ i.e.
\begin{multline}
\mathscr{L}(D \vert \theta) = \frac{1}{[(2\pi)^{d}\det{C}]^{\sfrac{1}{2}}} \\
\times e^{-\frac{1}{2}(\boldsymbol{\mu}^{\mathrm{obs}}-\boldsymbol{\mu}^{\mathrm{theor}})^{\mathrm{T}}C^{-1}(\boldsymbol{\mu}^{\mathrm{obs}}-\boldsymbol{\mu}^{\mathrm{theor}})},
\label{eq:snelikelihood}
\end{multline}
where $d = N_{\mathrm{SNe}}$, $\boldsymbol{\mu}^{\mathrm{obs}}$ is the vector of observed SNe Ia distance moduli and $\boldsymbol{\mu}^{\mathrm{theor}}$ denotes the theoretical prediction. The covariance matrix $C$ contains both statistical as well as systematic errors and is constructed following Ref.~\cite{Betoule:2014}\footnote{The covariance matrix can be found at:\\ $\tt{http://supernovae.in2p3.fr/sdss\_snls\_jla/ReadMe.html}$.}. We note that $C$ depends on the values of the nuisance parameters $\alpha$ and $\beta$ and thus needs to be reevaluated in each step when sampling parameters.

We combine the power spectra together with the SNe Ia and the Hubble constant measurement assuming them to be independent i.e. we multiply the likelihoods. We note that this is an approximation since the residuals of the SNe Ia distance moduli are affected by weak gravitational lensing and redshift space distortions. It will be interesting to investigate these correlations for future surveys (see e.g. \cite{Scovacricchi:2016}).

From the combined likelihood we compute cosmological parameter constraints in the framework of a flat $\Lambda$CDM cosmological model. All our constraints are derived assuming a vanishing neutrino mass i.e. $\sum m_{\nu} = 0.00$ eV. We sample the likelihood in a Monte Carlo Markov Chain (MCMC) with $\tt{CosmoHammer}$ \cite{Akeret:2012} varying 14 different parameters. We vary the six cosmological parameters $\{h,\, \Omega_{\mathrm{m}}, \,\Omega_{\mathrm{b}}, \, n_{\mathrm{s}}, \,\sigma_{8}, \,\tau_{\mathrm{reion}}\}$ as well as the eight nuisance parameters $\{b, \, m^{\mathrm{SDSS}}_{*}, \, m^{\mathrm{DES}}_{*}, \, m_{\kappa_{\mathrm{CMB}}}, \, \alpha, \, \beta, \, M^{1}_{\mathrm{B}}, \, \Delta M\}$, which are described in Sec.~\ref{sec:systematics}. We note that our fiducial constraints do not include the nuisance parameters $A_{\mathrm{ps}}$, $A_{\mathrm{cl}}$ and we recall that we further neglect several sources of systematic uncertainties such as photometric redshift uncertainties, stochastic and scale-dependent galaxy bias \cite{Pen:1998, Tegmark:1998, Dekel:1999} or baryonic effects on the matter power spectrum. Also, we do not model intrinsic alignments but include them as part of our effective multiplicative bias. The data combination considered weakly constrains the optical depth to reionization and we therefore follow \ja{} and apply a Gaussian prior of $\tau_{\mathrm{reion}} = 0.089 \pm 0.02$. As discussed in \ja{} this corresponds to an enlarged WMAP9 prior \cite{Hinshaw:2013}. We further assume Gaussian priors on the effective multiplicative bias parameters $m_{*}^{i}$, $i \in [\mathrm{SDSS}, \mathrm{DES}]$, as $m_{*}^{i} = 0. \pm 0.22$. The width of the prior is given $\sigma^{2}(m_{*}^{i}) = \sigma^{2}(m_{\mathrm{calib}}^{i}) + \sigma^{2}(m_{\mathrm{IA}}^{i})$. Assuming a maximal calibration uncertainty of $\sigma(m_{\mathrm{calib}}^{i}) = 0.1$ and $\sigma(m_{\mathrm{IA}}^{i}) = 0.2$ gives $\sigma(m_{*}^{i}) \approx 0.22$. The choice of $\sigma(m_{\mathrm{calib}}^{i})$ is motivated by Ref.~\cite{Hirata:2003}, who found the multiplicative bias for the shape measurement method of SDSS Stripe 82 data to lie in the range $m_{\mathrm{calib}}^{\mathrm{SDSS}} \in [-0.08, 0.13]$. We choose a conservative approach and apply the same uncertainty to the DES SV galaxy shears, even though the reported calibration uncertainty is $\sigma(m_{\mathrm{calib}}^{\mathrm{DES}}) = 0.05$ \cite{Jarvis:2016}. The choice for $\sigma(m_{\mathrm{IA}}^{i})$ follows from the discussion in Appendix \ref{sec:ap-ias}, from which we can see that the contribution of intrinsic alignments amounts to maximally $20 \%$ of the measured weak lensing signal in our fiducial model. We nevertheless choose to center the prior on  $m^{\mathrm{SDSS}}_{*}, \, m^{\mathrm{DES}}_{*}$ on 0 rather than on the value expected from our fiducial model for intrinsic alignments in order to confirm that the data moves the posterior. We allow for intrinsic alignments by broadening the prior as compared to the expectation from only multiplicative calibration uncertainties. For all other parameters we assume flat priors, which are summarized in Tab.~\ref{tab:params}.

We compute all parameter constraints using the covariance matrix derived from the Gaussian simulations. The results using the theoretical covariance matrix or a covariance matrix derived using a cosmological model with parameter values similar to the ones derived for our best-fit\footnote{$\{h,\, \Omega_{\mathrm{m}}, \,\Omega_{\mathrm{b}}, \, n_{\mathrm{s}}, \,\sigma_{8}, \,\tau_{\mathrm{reion}}, \,T_{\mathrm{CMB}}\} = \{0.699, \,0.278, \,0.0455, \,0.975, \,0.799, \,0.0792, \,2.275 \,\mathrm{K}\}$ and $\{b,\, m^{\mathrm{SDSS}}_{*}, \, m^{\mathrm{DES}}_{*}, \, m_{\kappa_{\mathrm{CMB}}}, \, A_{\mathrm{ps}}, \, A_{\mathrm{cl}}\} = \{2.13, \,-0.142, \,0., \,0., \,0., \,0.\}$.} are shown in the Appendix (Fig.~\ref{fig:constraints-ja-diff-covs}) and we find them to be consistent with our fiducial results. This confirms that the computation of the empirical covariance matrix has converged with 1000 simulations. It further verifies that our results are not sensitive to the slightly high value of $\sigma_{8}$ chosen for our fiducial simulations.

In order to investigate the impact of residual foregrounds on our analysis, we compute constraints on cosmological parameters from CMB temperature data alone both including the parameters $A_{\mathrm{ps}}$, $A_{\mathrm{cl}}$ and neglecting them. As can be seen from the Appendix (Fig.~\ref{fig:constraints-ja-cmb}) we find no significant difference between the constraints derived from the extended parameter set and the ones derived ignoring these additional degrees of freedom. This suggests that the low-$\ell$ CMB temperature anisotropy power spectrum is not affected by these foregrounds. We therefore choose to not include contamination from residual extragalactic point sources into our fiducial analysis.

We further validate our analysis of the CMB temperature anisotropies by comparing our fiducial CMB-only parameter constraints to those obtained from running the official Planck likelihood \cite{Planck-Collaboration:2016ae} with nuisance parameters fixed to the best-fit values derived by Ref.~\cite{Planck-Collaboration:2016ae} (TT+lowP). Figure \ref{fig:constraints-ja-cmb} shows the comparison between the CMB temperature constraints derived in this work with the constraints derived from the Planck likelihood for $\ell_{\mathrm{max}} \simeq 610$. We show the results obtained from the Planck likelihood both with $\ell_{\mathrm{min}} = 10$ and $\ell_{\mathrm{min}} = 30$. This is due to the fact that in the Planck likelihood the angular multipoles for $\ell < 30$ are unbinned while for $\ell \geq 30$ the power spectrum is binned into bins of $\Delta \ell = 21$, which is more comparable to our binning scheme of $\Delta \ell = 30$. As can be seen we find reasonable agreement between the derived parameter constraints. This is not the case when we extend the angular multipole range and we therefore use $\ell_{\mathrm{max}} = 610$ as in \ja{} for our fiducial analysis.

\section{Cosmological constraint results}\label{sec:constraints}

\begin{figure*}
\begin{center}
\includegraphics[scale=0.5]{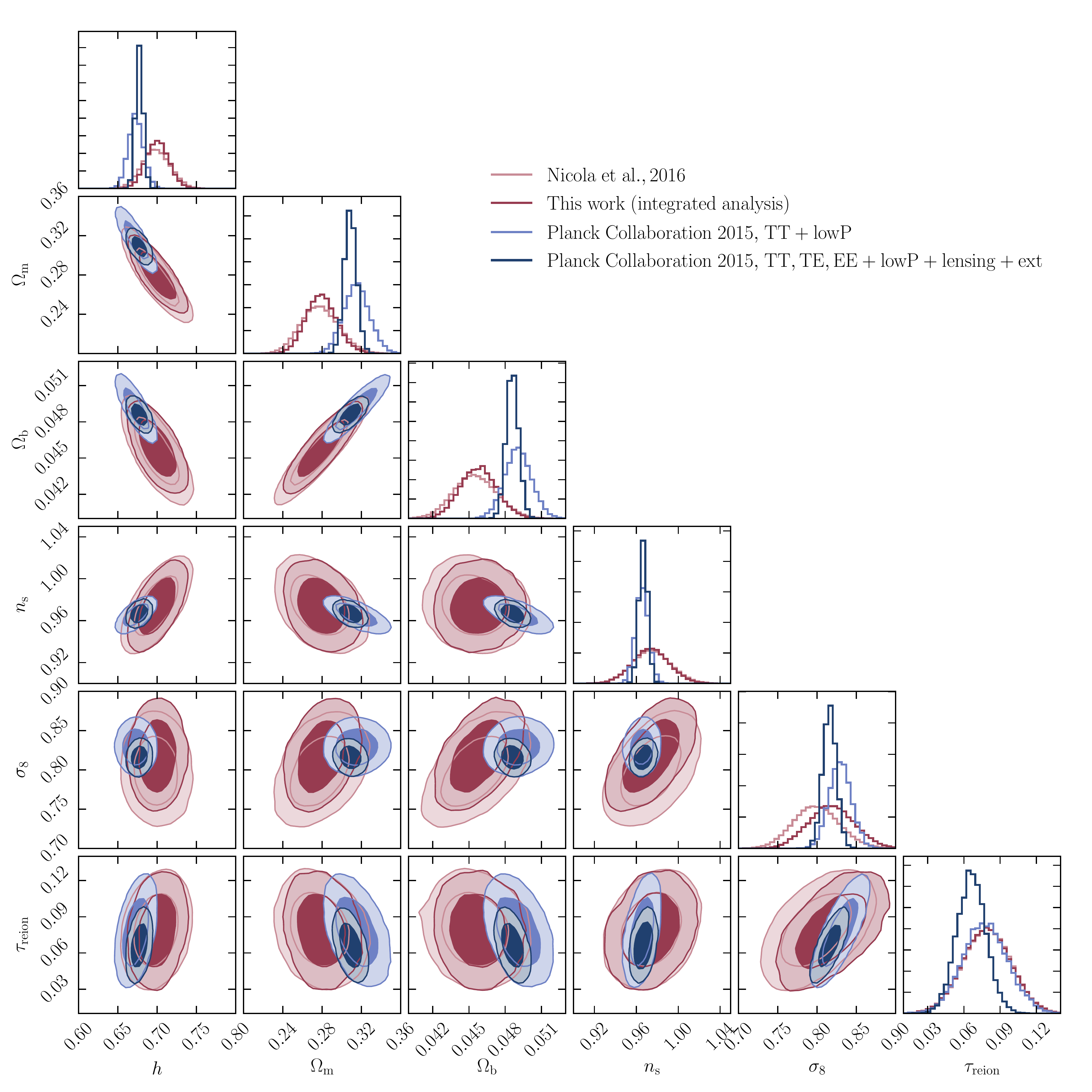}
\caption{Comparison of the constraints obtained from the integrated analysis from \ja{} to the constraints obtained in this work and the constraints obtained by the Planck Collaboration \cite{Planck-Collaboration:2016ae} using only CMB data (TT+lowP) or adding external data (TT,TE,EE+lowP+lensing+ext). The constraints from \ja{} are marginalized over $b, m^{\mathrm{SDSS}}_{\mathrm{calib}}$ while the constraints from this work are marginalized over all nuisance parameters given in Tab.~\ref{tab:params}. The Planck constraints are marginalized over all nuisance parameters. In each case the inner (outer) contour shows the $68 \%$ c.l. ($95 \%$ c.l.).}
\label{fig:constraints-old-ja-vs-ja-vs-Planck}
\end{center}
\end{figure*}

Figure \ref{fig:constraints-old-ja-vs-ja-vs-Planck} shows our integrated cosmological parameter constraints along with those from \ja. The associated means of the posterior distributions for all the parameters along with their $68 \%$ confidence limits (c.l.) are given in Tab.~\ref{tab:params}. As can be seen from the figure, our constraints with the new data and greater flexibility for systematics agree very well with our previous findings\footnote{In \ja{} we included massive neutrinos in our fiducial model in contrast to this work, but we find the constraints without massive neutrinos to be consistent. The constraints of \ja{} with and without massive neutrinos differ at the percent-level in the best-fit value for $\sigma_{8}$, with the constraints with massive neutrinos being lower in $\sigma_{8}$.}, demonstrating the robustness of our results. In fact, we also find that the results are robust to removing random probes from our analysis. We note that the slight tension between CMB temperature anisotropies and weak lensing discussed in \ja{} is resolved when accounting for the expected effects of intrinsic alignments. This can be seen from the fact that broadening the prior on the multiplicative calibration bias parameter in \ja{} recovers a value consistent with the expectation from intrinsic alignments and results in a better fit to the power spectra involving weak gravitational lensing, as discussed in \ja. Furthermore we see from Fig.~\ref{fig:cls} that there is no noticeable tension between CMB temperature anisotropies and weak lensing in the extended analysis.

In Fig.~\ref{fig:constraints-old-ja-vs-ja-vs-Planck}, we also compare our constraints to those derived by the Planck Collaboration \cite{Planck-Collaboration:2016ae}. For the latter, we show the constraints derived from the combination of CMB temperature anisotropies with the Planck low-$\ell$ polarization likelihood (TT+lowP) and the constraints when also including the Planck polarization power spectra, CMB lensing and external data sets (TT,TE,EE+lowP+lensing+BAO+JLA+$H_{0}$). While we see a general broad agreement, a global tension is nevertheless apparent in most of the panels. Our analysis indeed continues to prefer a lower value of $\Omega_{\mathrm{m}}, \Omega_{\mathrm{b}}$ and $\sigma_{8}$ as well as a higher value of $h$. This tension is at similar levels to what has been found by other groups (e.g. \cite{Hildebrandt:2016} and references therein). Since our analysis includes a subset of the Planck data, the tension appears to be with the Planck data that we have not included, namely low-$\ell$ polarization data, and CMB temperature anisotropy data for $\ell \in [2, \; 9]$ and $\ell \in [611, \; 2508]$, where the latter has the greatest impact. This parameter shift induced by the Planck high-$\ell$ measurements has also been reported and studied by others, including \cite{Planck-Collaboration:2016ac} and \cite{Addison:2016}.

\begin{figure*}
\begin{center}
\includegraphics[scale=0.5]{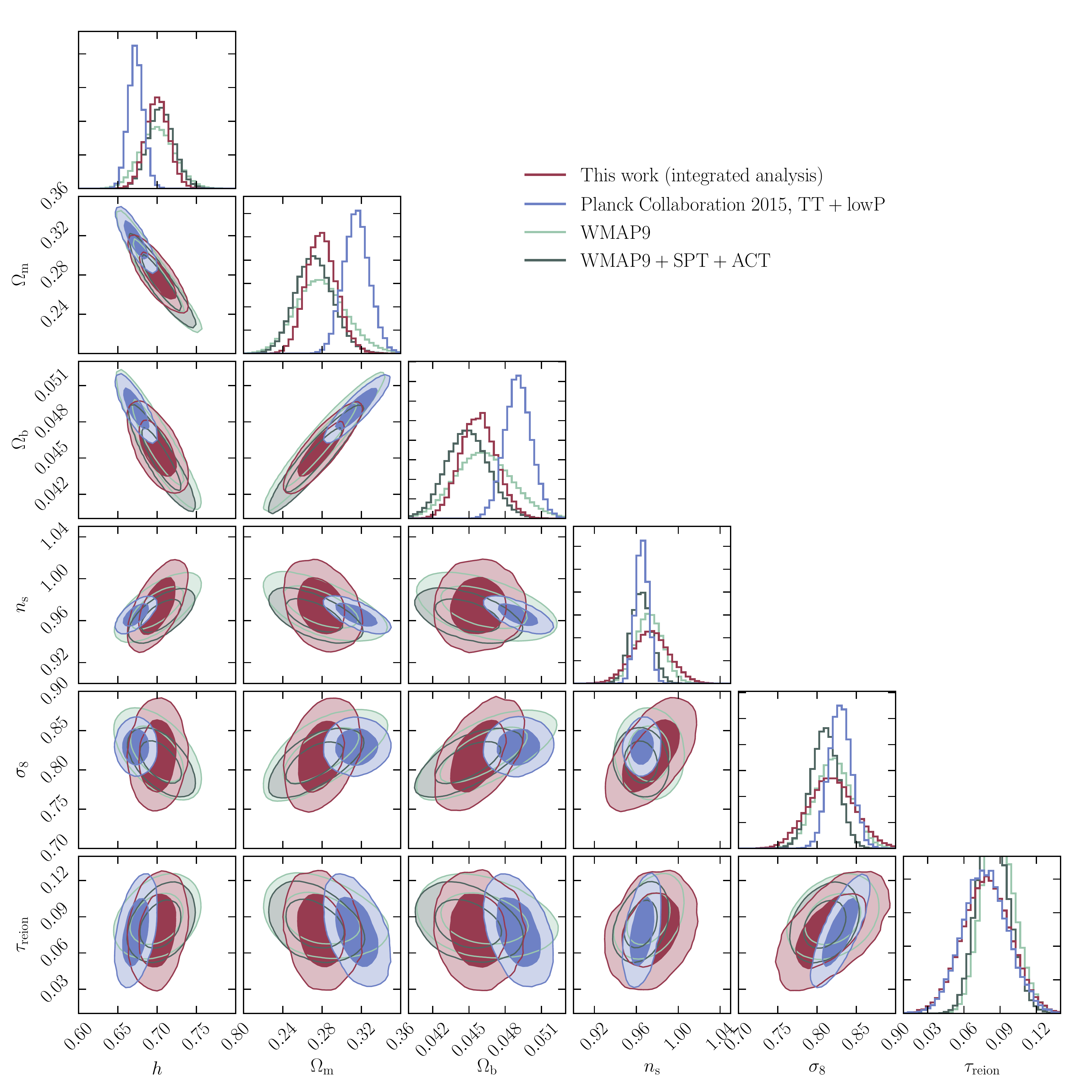}
\caption{Comparison of the constraints obtained in this work to the constraints obtained by the Planck Collaboration \cite{Planck-Collaboration:2016ae} using only CMB data (TT+lowP) and the constraints obtained by WMAP9 \cite{Hinshaw:2013} both using multipoles $\ell <  1200$ (WMAP9) and combined with high-$\ell$ data from SPT and ACT (WMAP9+SPT+ACT). The constraints from this work are marginalized over all nuisance parameters given in Tab.~\ref{tab:params}. The Planck and the WMAP9 constraints are marginalized over all nuisance parameters. In each case the inner (outer) contour shows the $68 \%$ c.l. ($95 \%$ c.l.).}
\label{fig:constraints-ja-vs-Planck-vs-wmap}
\end{center}
\end{figure*}

Exploring this further, Fig.~\ref{fig:constraints-ja-vs-Planck-vs-wmap} shows the comparison of our integrated analysis with the constraints from WMAP9 \cite{Hinshaw:2013} and WMAP9 combined with high-$\ell$ data from the Atacama Cosmology Telescope (ACT) \cite{Fowler:2010, Das:2011} and the South Pole Telescope (SPT) \cite{Keisler:2011, Reichardt:2012}. We also show the constraints from the Planck Collaboration \cite{Planck-Collaboration:2016ae}. We see that our results are in good agreement with WMAP9 and the combination of WMAP9, ACT and SPT and are consistent with the already highlighted tension with the Planck high-$\ell$ measurement \cite{Addison:2016, Planck-Collaboration:2016ac}.

\begin{figure}
\begin{center}
\includegraphics[scale=0.5]{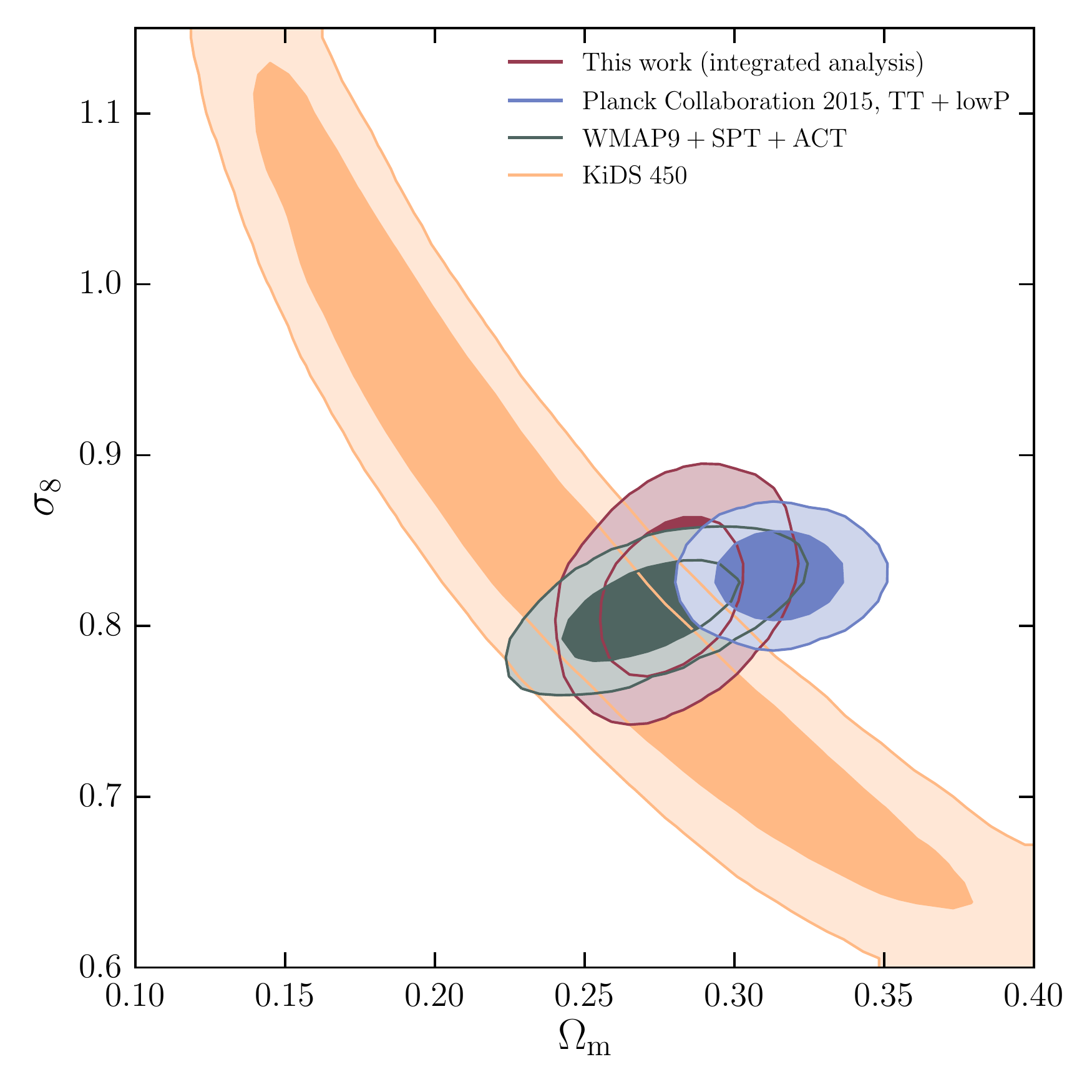}
\caption{Comparison of the constraints on $\Omega_{\mathrm{m}}$ and $\sigma_{8}$ obtained in this work and the constraints obtained by the Planck Collaboration \cite{Planck-Collaboration:2016ae} using only CMB data (TT+lowP), the constraints obtained by WMAP9 \cite{Hinshaw:2013} combined with high-$\ell$ data from SPT and ACT (WMAP9+SPT+ACT) and the constraints obtained from cosmic shear measurements from KiDS-450 \cite{Hildebrandt:2016}. The constraints from this work are marginalized over all nuisance parameters given in Tab.~\ref{tab:params}. The Planck, WMAP9+SPT+ACT and KiDS constraints are marginalized over all nuisance parameters. In each case the inner (outer) contour shows the $68 \%$ c.l. ($95 \%$ c.l.).}
\label{fig:constraints-s8-omegam}
\end{center}
\end{figure}

In Figure \ref{fig:constraints-s8-omegam} we focus on one of the panels, showing the $\Omega_{\mathrm{m}}-\sigma_{8}$ plane, and now include the results from KiDS \cite{Hildebrandt:2016}. We also see that our results are in good agreement with KiDS.

\begin{figure*}
\begin{center}
\includegraphics[scale=0.3]{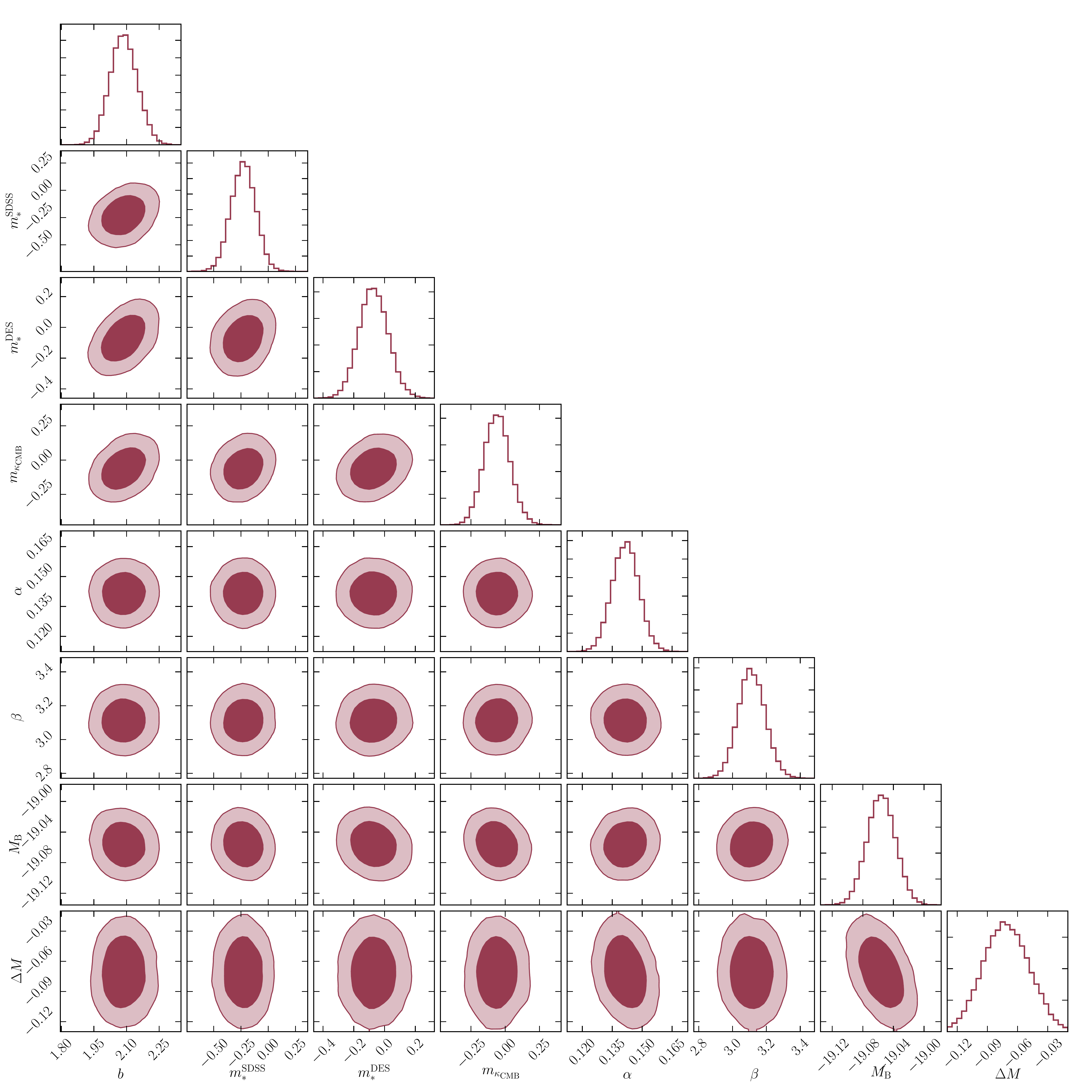}
\caption{Constraints on the nuisance parameters described in Tab.~\ref{tab:params} obtained in this work. These have been marginalized over all cosmological parameters. The inner (outer) contour shows the $68 \%$ c.l. ($95 \%$ c.l.).}
\label{fig:constraints-ja-nuisance-params}
\end{center}
\end{figure*}

\begin{figure*}
\begin{center}
\includegraphics[scale=0.5]{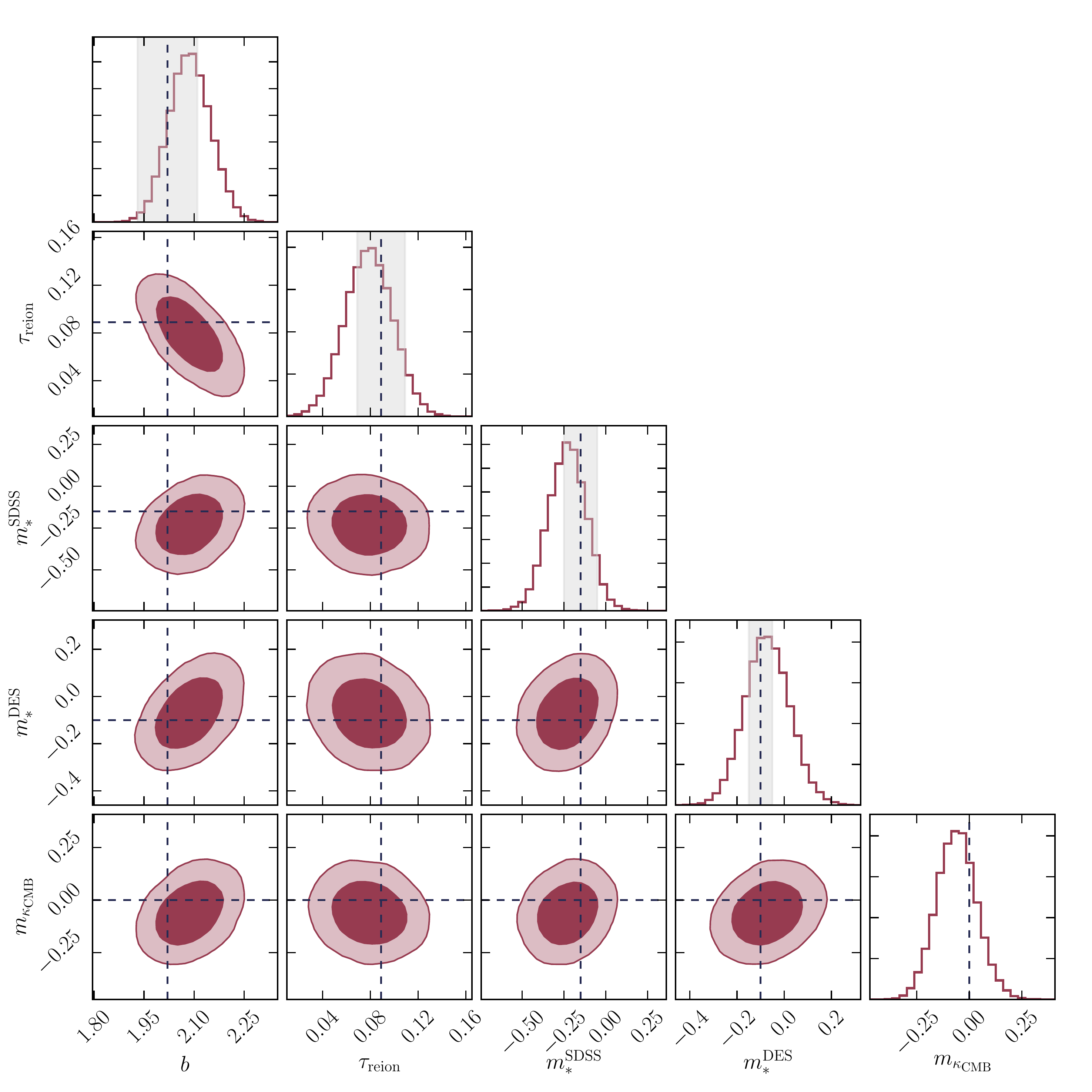}
\caption{Constraints on the nuisance parameters described in Tab.~\ref{tab:params} affecting the amplitude of the power spectra as well as $\tau_{\mathrm{reion}}$ obtained in this work. These have been marginalized over all remaining parameters. For the linear galaxy bias parameter $b$, the dashed line and the gray band show the mean and $1\sigma$ uncertainty of the tomographic galaxy bias derived in Ref.~\cite{Ho:2012} for the CMASS1-4 sample. The dashed line and the gray band for $\tau_{\mathrm{reion}}$ show the mean and $1\sigma$ uncertainty of the adopted prior. While for $m^{\mathrm{SDSS}}_{*}$ and $m^{\mathrm{DES}}_{*}$, the dashed lines and gray bands show the effective multiplicative bias expected from intrinsic alignments together with the $1\sigma$ calibration uncertainties for both SDSS Stripe 82 and DES SV. For  $m_{\kappa_{\mathrm{CMB}}}$ finally, the dashed line is centered on $m_{\kappa_{\mathrm{CMB}}} = 0$. The inner (outer) contour shows the $68 \%$ c.l. ($95 \%$ c.l.).}
\label{fig:constraints-ja-nuisance-params-amplitude}
\end{center}
\end{figure*}

The constraints on the nuisance parameters varied in our analysis are shown in Figures \ref{fig:constraints-ja-nuisance-params} and \ref{fig:constraints-ja-nuisance-params-amplitude}. From Tab.~\ref{tab:params} we see that we find values of the effective weak lensing shear calibration parameter of $m^{\mathrm{SDSS}}_{*} = -0.229 \pm 0.113$ and $m^{\mathrm{DES}}_{*} = -0.0708\substack{+0.0953 \\ -0.0946}$. The obtained value for the effective multiplicative bias of SDSS Stripe 82 is broadly consistent with an overall intrinsic alignment contribution of $-15 \%$ and a calibration uncertainty of around $5 \%$. The effective multiplicative bias for DES SV is slightly lower than would be expected from our fiducial model, which would suggest a lower limit on $m^{\mathrm{DES}}_{*}$ of around $-10 \%$ coming solely from intrinsic alignments. This discrepancy could be due to cancellations between the two components of the effective bias. We find a value of the multiplicative bias of the CMB lensing convergence of $m_{\kappa_{\mathrm{CMB}}} = -0.0598\substack{+0.0941 \\ -0.0946}$, which is again consistent with the expectation of $m_{\kappa_{\mathrm{CMB}}} = 0$. Finally, our constraint on the galaxy bias $b$ for the SDSS CMASS1-4 sample is consistent with the mean of the tomographic bias parameters reported in Ref.~\cite{Ho:2012}. We further find the constraints on different nuisance parameters to be only slightly correlated between one another. This shows that the combination of auto and cross correlations can be used for cross-calibration of different surveys and cosmological probes.

The theoretical predictions for the best-fitting cosmological model together with the measured data are shown in Fig.~\ref{fig:cls}. As can be seen, the theoretical predictions fit the data rather well and there is no sign of a tension between different data sets. 

\begin{table*}
\caption{Parameters varied in the MCMC with their respective priors and posterior means. The uncertainties denote the $68 \%$ c.l..} \label{tab:params}
\begin{center}
\begin{ruledtabular}
\begin{tabular}{ccc}
Parameter & Prior & Posterior mean\\ \hline \Tstrut                             
$h$ & flat $\in [0.2, \,1.2]$ & $0.700 \pm 0.014$ \\ 
$\Omega_{\mathrm{m}}$ & flat $\in [0.1, \,0.7]$ & $0.279 \pm 0.015$ \\
$\Omega_{\mathrm{b}}$ & flat $\in [0.01, \,0.09]$ & $0.0458 \pm 0.0015$ \\
$n_{\mathrm{s}}$ & flat $\in [0.1, \,1.8]$ & $0.974\substack{+0.018 \\ -0.017}$ \\
$\sigma_{8}$ & flat $\in [0.4, \,1.5]$ & $0.819 \pm 0.029$ \\ \\
$\tau_{\mathrm{reion}}$ & Gaussian with $\mu = 0.089$, $\sigma = 0.02$\footnote{This corresponds to a WMAP9 \cite{Hinshaw:2013} prior with increased variance to accommodate the Planck results.} & $0.0787\substack{+0.0200 \\ -0.0199}$ \\
$b$ & flat $\in [1., \,3.]$ & $2.09 \pm 0.06$ \\
$m^{\mathrm{SDSS}}_{*}$ & Gaussian with $\mu = 0.0$, $\sigma = 0.22$ & $-0.229 \pm 0.113$ \\
$m^{\mathrm{DES}}_{*}$ & Gaussian with $\mu = 0.0$, $\sigma = 0.22$ & $-0.0708\substack{+0.0953 \\ -0.0946}$ \\
$m_{\kappa_{\mathrm{CMB}}}$ & flat $\in [-0.5, \,0.5]$ & $-0.0598\substack{+0.0941 \\ -0.0946}$ \\
$\alpha$ & flat $\in [0.1, \,0.2]$ & $0.142 \pm 0.007$ \\
$\beta$ & flat $\in [2., \,4.]$ & $3.11 \pm 0.08$ \\
$M^{1}_{\mathrm{B}}$ & flat $\in [-25., \,-10.]$ & $-19.06 \pm 0.02$ \\
$\Delta M$ & flat $\in [-0.13, \,-0.01]$ & $-0.0711\substack{+0.0230 \\ -0.0227}$ 
\end{tabular}
\end{ruledtabular}
\end{center}
\end{table*} 

\section{Conclusions}\label{sec:conclusions}

In this work, we have extended our framework for integrated cosmological probe analysis presented in \ja. We have combined data from CMB temperature anisotropies, galaxy clustering, weak lensing and CMB lensing focusing on two-point statistics and taking into account the cross-correlations between the different probes. We used CMB temperatures from Planck 2015 \cite{Planck-Collaboration:2016ad}, galaxy clustering data from SDSS DR8 \cite{Aihara:2011}, weak lensing data from both SDSS Stripe 82 \cite{Annis:2014} and DES SV \cite{Jarvis:2016} and CMB lensing data from
Planck 2015 \cite{Planck-Collaboration:2016aa}. We have further included SNe Ia distance moduli from the JLA \cite{Betoule:2014} and Hubble constant measurements from HST \cite{Riess:2011, Efstathiou:2014}. 
For all the probes of the inhomogeneous Universe we have computed 12 spherical harmonic power spectra and we have combined them into a joint likelihood assuming both a Gaussian covariance matrix as well as a Gaussian likelihood. We have then combined this likelihood with the Gaussian SNe Ia likelihood assuming these two data sets to be independent.
We have extended our treatment of systematic uncertainties and relaxed some of the approximations used in \ja{}. Furthermore, we have also studied the impact of intrinsic alignments, baryonic corrections, residual foregrounds in the CMB temperature, and calibration factors for the different power spectra. This extended analysis allows us to derive more robust constraints on the cosmological model and a more thorough test of the consistency between the different probes.

From this analysis, we have computed cosmological parameter constraints for a flat $\Lambda$CDM cosmological model marginalizing over eight nuisance parameters. We have made several simplifying approximations throughout this analysis: We have assumed the joint covariance matrix of the considered cosmological probes to be Gaussian. Further, we have only included systematic uncertainties from linear galaxy bias, multiplicative biases due to either calibration or intrinsic alignments in the weak lensing shear, uncertainties in the SNe Ia intrinsic luminosities and multiplicative calibration uncertainties in the CMB lensing convergence. We have not taken into account uncertainties due to photometric redshift errors or additive weak lensing calibration bias and we have not included intrinsic alignment modelling in our analysis. Since we have found our data to be insensitive to baryonic effects on the matter power spectrum and residual foregrounds in the CMB temperature anisotropy maps, we have not included these effects into our analysis. Finally our theoretical predictions are all computed using the Limber approximation. Due to the conservative cuts we have applied on the data we do not believe these effects to significantly affect our conclusions. We find results that are consistent with those presented in \ja, which, given our enlarged data set and systematics treatment, confirms the robustness of our analysis and results. Furthermore, we find our data to be well fit by our best-fit cosmological model and we do not see any sign of tension between the data sets considered in our analysis. 

We also find that our constraints are consistent with those derived by other analyses such as the joint analysis of the WMAP9, SPT and ACT CMB experiments and the KiDS weak lensing survey \cite{Hildebrandt:2016}. Comparing the obtained constraints to those from the Planck Collaboration \cite{Planck-Collaboration:2016ae}, we find a broad agreement but also tensions in the marginalized constraints in most pairs of cosmological parameters. In particular, we find a lower value of $\Omega_{\mathrm{m}}, \Omega_{\mathrm{b}}$ and $\sigma_{8}$ as well as a higher value of $h$.  Since our analysis includes low-$\ell$ Planck CMB temperature data ($\ell \in [10, 610]$), the tension appears to arise between the Planck high-$\ell$ modes and the other measurements. 

We further find the constraints on the probe calibration parameters to be largely independent and in agreement with expectation. This shows that these data sets are mutually consistent. This also yields a confirmation of the amplitude calibration of the weak lensing measurements from SDSS, DES SV and Planck CMB lensing from our integrated analysis.

Future cosmological surveys will provide data with unprecedented precision. It will thus be interesting to extend the framework presented here to models beyond $\Lambda$CDM as well as extend this framework to include 3-dimensional tomographic information, further data or higher-order statistics. 

\begin{acknowledgments}

We would like to thank Duncan Hanson for his help with the Planck CMB lensing maps and Donnacha Kirk for result comparison. We would further like to thank the referee for very carefully reviewing our manuscript and for providing useful comments, which helped us improve the quality and clarity of the paper. This work was supported in part by SNF grants 200021\_16930 and 200021\_143906. 

Some of the results in this paper have been derived using the HEALPix (\citet{Gorski:2005}) package. The color palettes employed in this work are taken from $\tt{http://colorpalettes.net}$. We further acknowledge the use of the color map provided by \cite{Planck-Collaboration:2016ad}. The contour plots have been created using $\tt{corner.py}$ \cite{ForemanMackey:2016}. Fig.~\ref{fig:footprints} has been created using the $\tt{footprint}$ package\footnote{See \tt{https://github.com/nasa-lambda/cmb\_footprint}.}.

This project used public archival data from the Dark Energy Survey (DES). Funding for the DES Projects has been provided by the U.S. Department of Energy, the U.S. National Science Foundation, the Ministry of Science and Education of Spain, the Science and Technology Facilities Council of the United Kingdom, the Higher Education Funding Council for England, the National Center for Supercomputing Applications at the University of Illinois at Urbana-Champaign, the Kavli Institute of Cosmological Physics at the University of Chicago, the Center for Cosmology and Astro-Particle Physics at the Ohio State University, the Mitchell Institute for Fundamental Physics and Astronomy at Texas A$\&$M University, Financiadora de Estudos e Projetos, Funda\c{c}\~{a}o Carlos Chagas Filho de Amparo \`{a} Pesquisa do Estado do Rio de Janeiro, Conselho Nacional de Desenvolvimento Cient\'{i}fico e Tecnol\'{o}gico and the Minist\'{e}rio da Ci\^{e}ncia, Tecnologia e Inova\c{c}\~{a}o, the Deutsche Forschungsgemeinschaft and the Collaborating Institutions in the Dark Energy Survey. The Collaborating Institutions are Argonne National Laboratory, the University of California at Santa Cruz, the University of Cambridge, Centro de Investigaciones En\'{e}rgeticas, Medioambientales y Tecnol\'{o}gicas-Madrid, the University of Chicago, University College London, the DES-Brazil Consortium, the University of Edinburgh, the Eidgen\"{o}ssische Technische Hochschule (ETH) Z\"{u}rich, Fermi National Accelerator Laboratory, the University of Illinois at Urbana-Champaign, the Institut de Ci\`{e}ncies de l'Espai (IEEC/CSIC), the Institut de F\'{i}sica d'Altes Energies, Lawrence Berkeley National Laboratory, the Ludwig-Maximilians Universit\"{a}t M\"{u}nchen and the associated Excellence Cluster Universe, the University of Michigan, the National Optical Astronomy Observatory, the University of Nottingham, the Ohio State University, the University of Pennsylvania, the University of Portsmouth, SLAC National Accelerator Laboratory, Stanford University, the University of Sussex, and Texas A$\&$M University. 

Funding for the SDSS and SDSS-II has been provided by the Alfred P. Sloan Foundation, the Participating Institutions, the National Science Foundation, the U.S. Department of Energy, the National Aeronautics and Space Administration, the Japanese Monbukagakusho, the Max Planck Society, and the Higher Education Funding Council for England. The SDSS Web Site is http://www.sdss.org/.

The SDSS is managed by the Astrophysical Research Consortium for the Participating Institutions. The Participating Institutions are the American Museum of Natural History, Astrophysical Institute Potsdam, University of Basel, University of Cambridge, Case Western Reserve University, University of Chicago, Drexel University, Fermilab, the Institute for Advanced Study, the Japan Participation Group, Johns Hopkins University, the Joint Institute for Nuclear Astrophysics, the Kavli Institute for Particle Astrophysics and Cosmology, the Korean Scientist Group, the Chinese Academy of Sciences (LAMOST), Los Alamos National Laboratory, the Max-Planck-Institute for Astronomy (MPIA), the Max-Planck-Institute for Astrophysics (MPA), New Mexico State University, Ohio State University, University of Pittsburgh, University of Portsmouth, Princeton University, the United States Naval Observatory, and the University of Washington.

Funding for SDSS-III has been provided by the Alfred P. Sloan Foundation, the Participating Institutions, the National Science Foundation, and the U.S. Department of Energy Office of Science. The SDSS-III web site is http://www.sdss3.org/.

SDSS-III is managed by the Astrophysical Research Consortium for the Participating Institutions of the SDSS-III Collaboration including the University of Arizona, the Brazilian Participation Group, Brookhaven National Laboratory, Carnegie Mellon University, University of Florida, the French Participation Group, the German Participation Group, Harvard University, the Instituto de Astrofisica de Canarias, the Michigan State/Notre Dame/JINA Participation Group, Johns Hopkins University, Lawrence Berkeley National Laboratory, Max Planck Institute for Astrophysics, Max Planck Institute for Extraterrestrial Physics, New Mexico State University, New York University, Ohio State University, Pennsylvania State University, University of Portsmouth, Princeton University, the Spanish Participation Group, University of Tokyo, University of Utah, Vanderbilt University, University of Virginia, University of Washington, and Yale University.

Based on observations obtained with Planck (http://www.esa.int/Planck), an ESA science mission with instruments and contributions directly funded by ESA Member States, NASA, and Canada.

\end{acknowledgments}

\appendix

\section{Impact of intrinsic alignments}\label{sec:ap-ias}

The shapes of unlensed galaxies have been found to exhibit significant correlations, which are called intrinsic alignments \cite{Heymans:2006, Mandelbaum:2006, Hirata:2007, Faltenbacher:2009, Okumura:2009, Joachimi:2011, Singh:2015}. Since weak lensing shear assumes all correlations between galaxy shapes to be due to gravitational lensing, any measurement of weak lensing will be biased by the presence of intrinsic alignments. These affect both the cosmic shear power spectrum as well as any cross-correlation between LSS probes and the weak lensing shear. 

The observed cosmic shear power spectrum in the presence of intrinsic alignments can be written as
\begin{equation}
C_{\ell}^{\gamma \gamma ,\mathrm{obs}} = C_{\ell}^{\gamma \gamma} + 2C_{\ell}^{\mathrm{GI}} + C_{\ell}^{\mathrm{II}}, 
\end{equation}
where $C_{\ell}^{\gamma \gamma}$ is the cosmic shear power spectrum. $C_{\ell}^{\mathrm{GI}}$ denotes correlations between the intrinsic alignments of foreground galaxies and the weak lensing shear of background galaxies \cite{Hirata:2004aa}. These arise because the gravitational field causing the intrinsic alignments is the same as that giving rise to the weak lensing shear and they are called gravitational-intrinsic (GI) correlations. Finally $C_{\ell}^{\mathrm{II}}$ denotes correlations between shapes of neighbouring galaxies which arise because these form under the influence of similar tidal gravitational fields \cite{Catelan:2001}. These correlations are termed intrinsic-intrinsic (II) galaxy alignments.

Since intrinsic galaxy alignments are due to the large-scale gravitational field in which galaxies form, these will be correlated to any tracer of the LSS. Therefore we should expect intrinsic alignment contributions to the cross-correlation between the weak lensing shear and both the galaxy overdensity and the CMB lensing convergence. The observed cross-correlation between the galaxy overdensity and the weak lensing shear is given by
\begin{equation}
C_{\ell}^{\gamma \delta_{g},\mathrm{obs}} = C_{\ell}^{\gamma \delta_{g}} + C_{\ell}^{\mathrm{I} \delta_{g}},
\end{equation}
where $C_{\ell}^{\gamma \delta_{g}}$ denotes the contribution due to weak lensing and $C_{\ell}^{\mathrm{I} \delta_{g}}$ is due to intrinsic galaxy alignments.

The cross-correlation between the CMB lensing convergence and the weak lensing shear can be written as
\begin{equation}
C_{\ell}^{\kappa_{\mathrm{CMB}} \gamma ,\mathrm{obs}} = C_{\ell}^{\kappa_{\mathrm{CMB}} \gamma} + C_{\ell}^{\kappa_{\mathrm{CMB}} \mathrm{I}},
\end{equation}
where again $C_{\ell}^{\kappa_{\mathrm{CMB}} \gamma}$ is the signal coming from weak gravitational lensing and $C_{\ell}^{\kappa_{\mathrm{CMB}} \mathrm{I}}$ denotes the intrinsic alignment contribution.

In order to model these effects and investigate the impact of intrinsic alignments on the spherical harmonic power spectra used in our analysis we follow Refs.~\cite{DES-Collaboration:2015, Hildebrandt:2016} and adopt the nonlinear alignment model \cite{Hirata:2004aa, Bridle:2007}. In this model the nonlinear intrinsic alignment power spectra are parametrized as
\begin{equation}
\begin{aligned}
P_{\mathrm{II}} &= F^{2}(z) P^{\mathrm{nl}}_{\delta\delta}(k, z), \\
P_{\delta\mathrm{I}} &= F(z) P^{\mathrm{nl}}_{\delta\delta}(k, z).
\end{aligned}
\end{equation}
The function $F(z)$ parametrizes the response of a galaxy shape to an external gravitational tidal field and is given by
\begin{equation}
F(z) = -A_{\mathrm{IA}} C_{1} \rho_{\mathrm{crit}} \frac{\Omega_{\mathrm{m}}}{D(z)},
\end{equation}
where we have neglected any redshift- or luminosity dependence of intrinsic alignments. $\rho_{\mathrm{crit}}$ is the critical density of the Universe at $z=0$, $D(z)$ denotes the linear growth factor normalised to unity today and $C_{1} = 5 \times 10^{-14} \, h^{-2} M_{\odot}^{-1}$ Mpc$^{3}$ is a normalization constant. The amplitude of the intrinsic alignments is determined by the free parameter $A_{\mathrm{IA}}$.

In this model the intrinsic alignment spherical harmonic power spectra for the cosmic shear become (see e.g. \cite{Joachimi:2010})
\begin{equation}
\begin{aligned}
C_{\ell}^{\mathrm{II}} &= \int \mathrm{d}z \; \frac{H(z)}{c} \; \frac{n^{\gamma}(z) \; n^{\gamma}(z)}{\chi^{2}(z)} \;P_{\mathrm{II}}\left(k=\frac{\ell+\sfrac{1}{2}}{\chi(z)}, z\right), \\
C_{\ell}^{\mathrm{GI}} &= \int \mathrm{d}z \; \frac{W^{\gamma}\boldsymbol{\left(}\chi(z)\boldsymbol{\right)} \;n^{\gamma}(z)}{\chi^{2}(z)} \;P_{\delta\mathrm{I}}\left(k=\frac{\ell+\sfrac{1}{2}}{\chi(z)}, z\right),
\end{aligned}
\end{equation}
where $n^{\gamma}(z)$ denotes the normalized redshift selection function of the weak lensing survey under consideration.

The intrinsic alignment contribution to the cross-correlation of the galaxy overdensity and the weak lensing shear can be written as \cite{Joachimi:2010}
\begin{multline}
C_{\ell}^{\mathrm{I}\delta_{g}} = \int \mathrm{d}z \; \frac{H(z)}{c} \; \frac{n^{\delta_{g}}(z) \; n^{\gamma}(z)}{\chi^{2}(z)} \\
\times b \; P_{\delta\mathrm{I}}\left(k=\frac{\ell+\sfrac{1}{2}}{\chi(z)}, z\right).
\end{multline}

Analogously the contribution of intrinsic alignments to the cross-correlation between the CMB lensing convergence and weak lensing shear is given by \cite{Hall:2014, Troxel:2014}
\begin{multline}
C_{\ell}^{\kappa_{\mathrm{CMB}}\mathrm{I}} = \int \mathrm{d}z \; \frac{W^{\kappa_{\mathrm{CMB}}}\boldsymbol{\left(}\chi(z)\boldsymbol{\right)} \; n^{\gamma}(z)}{\chi^{2}(z)} \\
\times P_{\delta\mathrm{I}}\left(k=\frac{\ell+\sfrac{1}{2}}{\chi(z)}, z\right).
\end{multline} 

\begin{figure*}
\begin{center}
\includegraphics[scale=0.55]{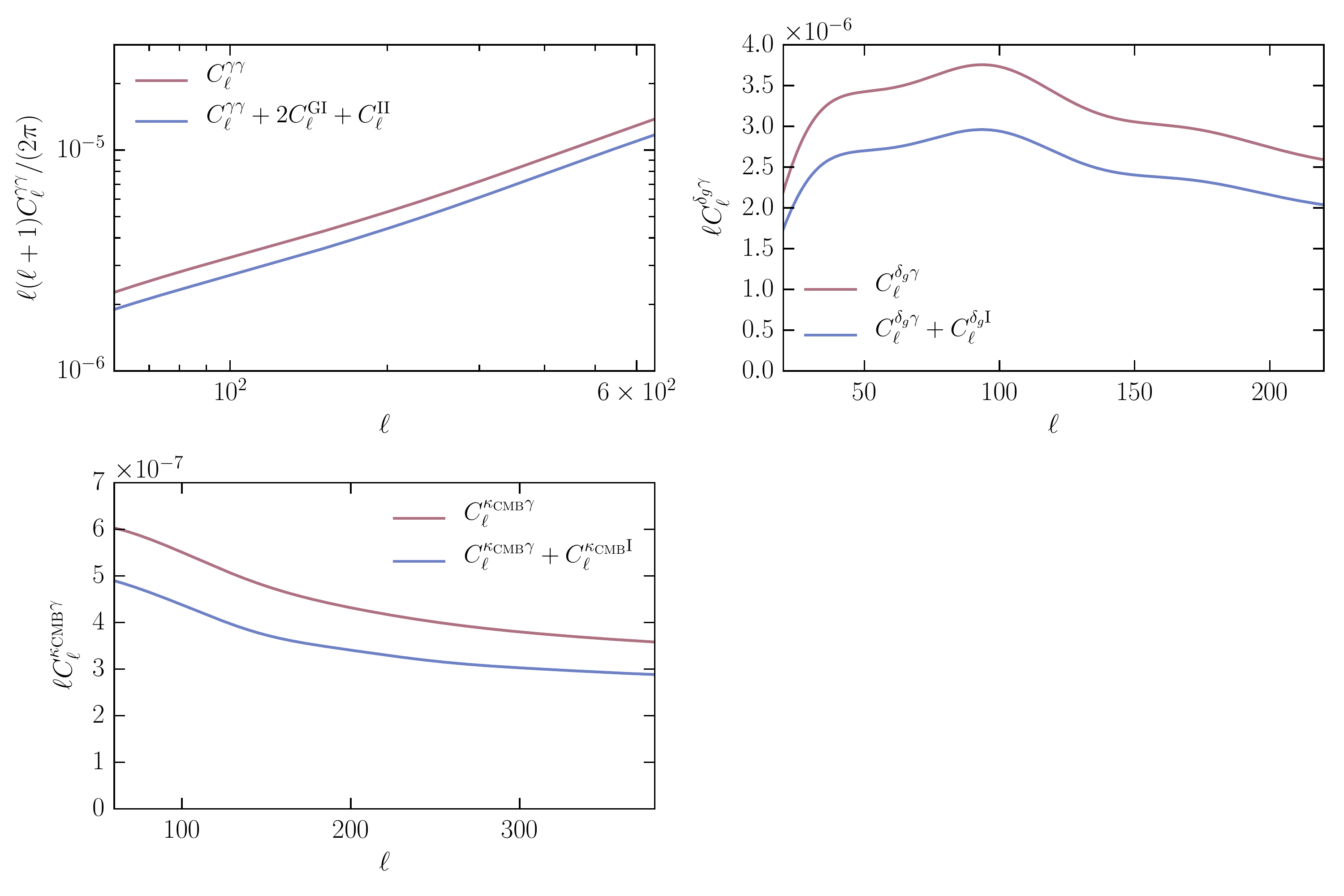}
\caption{Impact of intrinsic galaxy alignments on the theoretical spherical harmonic power spectra involving weak lensing data from SDSS Stripe 82 as estimated using the nonlinear alignment model.}
\label{fig:ias}
\end{center}
\end{figure*}

Figure \ref{fig:ias} shows the linear alignment contribution to $C_{\ell}^{\gamma \gamma ,\mathrm{obs}}, C_{\ell}^{\gamma \delta_{g},\mathrm{obs}}$ and $C_{\ell}^{\kappa_{\mathrm{CMB}} \gamma ,\mathrm{obs}}$ for SDSS Stripe 82 weak lensing shear evaluated for our best-fit cosmological model given in Tab.~\ref{tab:params} and assuming an intrinsic alignment amplitude $A_{\mathrm{IA}} = 1$ and using $\textsc{Halofit}$. Since the magnitude of intrinsic alignments does not depend on calibration uncertainties or galaxy bias we further assume no systematic uncertainties. The choice of intrinsic alignment amplitude is consistent with the recent results of Ref.~\cite{Hildebrandt:2016}. As can be seen from the figure, the observed spherical harmonic power spectra are significantly reduced in the presence of intrinsic alignments. The effect is most pronounced for $C_{\ell}^{\gamma \delta_{g}}$ for which the clustering amplitude is reduced by approximately $20 \%$. We can also see that the scale-dependence of the intrinsic alignment power spectra closely follows the weak lensing shear power spectra. In order to reduce computation time for our parameter estimation we therefore choose to model the intrinsic alignments as an additional multiplicative degree of freedom as described in Section \ref{sec:systematics}.

We note that the results for DES SV are similar but we choose to show the results for SDSS Stripe 82 because the latter shows a more significant impact of intrinsic alignments.

\section{Impact of baryonic processes on the dark matter power spectrum}\label{sec:ap-baryons}

Numerical simulations have shown that baryonic processes such as gas cooling and feedback from both AGN and supernovae have the potential to significantly alter the dark matter power spectrum on small scales (e.g. \cite{van-Daalen:2011}). These processes are still poorly understood which leads to systematic uncertainties on the matter power spectrum at small scales \cite{van-Daalen:2011}. Depending on the chosen model for baryonic physics, the matter power spectrum can be reduced by around $10 \%$ at scales of $k = 1$ Mpc$^{-1}$, which could significantly alter the weak lensing power spectrum due to the broadness of the redshift kernel $W^{\gamma}\boldsymbol{\left(}\chi(z)\boldsymbol{\right)}$. 

\begin{figure*}
\begin{center}
\includegraphics[scale=0.5]{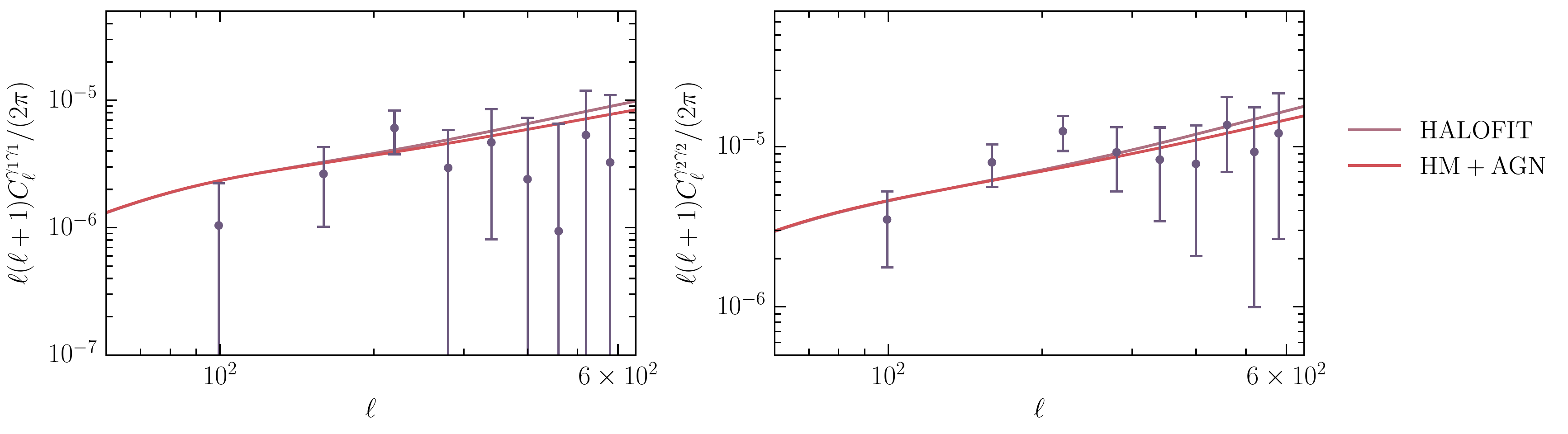}
\caption{Comparison of the theoretical predictions for the cosmic shear power spectra obtained taking into account baryonic corrections to the matter power spectrum as well as without taking them into account. The left hand panel shows the results for SDSS Stripe 82 while the right hand panel shows the results for DES SV. The theoretical predictions for the power spectra have been convolved with the $\tt{PolSpice}$ kernels described in \ja{} and Section \ref{sec:cls}. The error bars are derived from the Gaussian simulations described in Section \ref{sec:covmat} and Appendix \ref{sec:ap-mocks}. The angular multipole range and binning scheme is summarized in Tab.~\ref{tab:clparams}.}
\label{fig:cls_baryons}
\end{center}
\end{figure*}

In order to investigate the impact of baryonic corrections on the power spectra involving weak lensing shear considered in this work we adopt the effective halo model described in Refs.~\cite{Mead:2015, Mead:2016}. This optimized halo model has been shown to accurately reproduce the results from the $\textsc{Cosmic Emu}$ dark matter-only simulations \cite{Heitmann:2010, Heitmann:2014}. Baryonic effects can be incorporated by adjusting the internal structure of the halos. The two free parameters of the model are the amplitude $A$ of the mass-concentration relation and a halo bloating parameter $\eta_{0}$ which controls the density profile of the halo. In order to investigate the impact of baryonic corrections on our analysis we compute the matter power spectrum using the best-fit halo model parameters described in Refs.~\cite{Mead:2015, Mead:2016}. We incorporate baryonic effects using the parameters of the AGN model of Ref.~\cite{Mead:2015}, which is the model that shows the largest deviations from the dark matter-only results. We then compute the cosmic shear power spectra for the surveys considered in this work and compare them to those obtained when neglecting baryonic effects. The results are shown in Fig.~\ref{fig:cls_baryons} for a cosmological model defined by $\{h,\, \Omega_{\mathrm{m}}, \,\Omega_{\mathrm{b}}, \, n_{\mathrm{s}}, \,\sigma_{8}\} = \{0.71, \,0.27, \,0.045, \,0.97, \,0.8\}$ and $\{m^{\mathrm{SDSS}}_{*}, \, m^{\mathrm{DES}}_{*}\} = \{-0.11, \,-0.02\}$. As can be seen from the figure, the amplitude of the cosmic shear power spectra is significantly reduced for small angular scales. On the smallest angular scales considered, the suppression reaches approximately $15 \%$. Comparing the magnitude of this effect to the measured data we see that it is significantly smaller than the size of our error bars and our data is therefore currently not sensitive to baryonic effects. We therefore do not include baryonic corrections into our fiducial model and leave an investigation of those to future work.

\section{Correlated spin-0 and spin-2 fields}\label{sec:ap-mocks}

To validate our analysis as well as to obtain an estimate of the covariance matrix, we need to generate correlated Gaussian realizations of CMB temperature anisotropies $\mathrm{T}$, galaxy overdensity $\delta_{g}$, CMB lensing convergence $\kappa_{\mathrm{CMB}}$ and two weak lensing shear fields $\gamma_{1}, \gamma_{2}$ with auto- and cross-power spectra $\{C^{\mathrm{TT}}_{\ell},\allowbreak C_{\ell}^{\delta_{g}\mathrm{T}},\allowbreak C_{\ell}^{\delta_{g} \delta_{g}},\allowbreak C_{\ell}^{\kappa_{\mathrm{CMB}}\mathrm{T}},\allowbreak C_{\ell}^{\kappa_{\mathrm{CMB}}\delta_{g}},\allowbreak C^{\gamma_{1} \mathrm{T}}_{\ell},\allowbreak C^{\gamma_{1}\delta_{g}}_{\ell},\allowbreak C_{\ell}^{\gamma_{1} \kappa_{\mathrm{CMB}}},\allowbreak C^{\gamma_{1} \gamma_{1}}_{\ell},\allowbreak C^{\gamma_{2} \mathrm{T}}_{\ell},\allowbreak C_{\ell}^{\gamma_{2} \kappa_{\mathrm{CMB}}},\allowbreak C^{\gamma_{2} \gamma_{2}}_{\ell}\}$. To this end we extend the algorithm presented in \ja, which is based on Refs.~\cite{Giannantonio:2008, Cabre:2007}. We further improve the algorithm by including several terms we neglected due to their low amplitude in \ja. 

Due to the sky coverage of the surveys considered in this work, the DES SV weak lensing shear field is not correlated to both the SDSS galaxy overdensity and the SDSS Stripe 82 weak lensing shear field. To take this into account we simulate two separate sets of maps for each weak lensing survey. Both these sets consist of the weak lensing shear maps as well as all the spin-0 maps correlated to those. We then add the spin-0 maps obtained from these two sets of maps to create the final Gaussian realizations. 

The first set of maps consists of a correlated realization of CMB temperature anisotropies, galaxy overdensity, CMB lensing convergence and the SDSS weak lensing shear field and is constructed as follows:
\begin{enumerate}[label=(\roman*)]
\item We first create three correlated HEALPix maps using $\tt synfast$ in polarization mode with the power spectra 
\begin{align*}
C^{00}_{\ell} &= \sfrac{C^{\mathrm{TT}}_{\ell}}{2},\\
C^{\mathrm{EE}}_{\ell} &= \sfrac{C^{\gamma_{1} \gamma_{1}}_{\ell}}{3},\\
C^{\mathrm{BB}}_{\ell} &= 0,\\
C^{0\mathrm{E}}_{\ell} &= C^{\gamma_{1}\mathrm{T}}_{\ell}.
\end{align*}
These maps are denoted $m^{'1}_{i}$, where $i \in \{\mathrm{T}, \gamma_{1}^{1}, \gamma_{1}^{2}\}$.
\item Following Refs.~\cite{Giannantonio:2008, Cabre:2007, Nicola:2016} we then create three maps with a new random seed and the power spectra
\begin{align*}
C^{00}_{\ell} &= C_{\ell}^{\delta_{g} \delta_{g}} - \sfrac{2(C_{\ell}^{\delta_{g}\mathrm{T}})^{2}}{C_{\ell}^{\mathrm{TT}}},\\
C^{\mathrm{EE}}_{\ell} &= \sfrac{C^{\gamma_{1} \gamma_{1}}_{\ell}}{3},\\
C^{\mathrm{BB}}_{\ell} &= 0,\\
C^{0\mathrm{E}}_{\ell} &= C^{\gamma_{1}\delta_{g}}_{\ell} - \sfrac{2 C_{\ell}^{\delta_{g}\mathrm{T}}C_{\ell}^{\gamma_{1}\mathrm{T}}}{C_{\ell}^{\mathrm{TT}}}.
\end{align*}
The second term in the last equation removes unwanted cross-correlations between the spin-0 and spin-2 fields, which would otherwise arise in this process.
These maps are denoted $m^{'2}_{i}$, where $i \in \{\delta_{g}, \gamma_{1}^{1}, \gamma_{1}^{2}\}$.
\item In a next step we create three additional maps with another random seed and power spectra
\begin{align*}
C^{00}_{\ell} &= \sfrac{C_{\ell}^{\kappa_{\mathrm{CMB}}\kappa_{\mathrm{CMB}}}}{2} - \sfrac{(C_{\ell}^{\kappa_{\mathrm{CMB}}\mathrm{T}})^{2}}{(2 C_{\ell}^{\mathrm{TT}})} - \\
& \sfrac{(C_{\ell}^{\kappa_{\mathrm{CMB}}\delta_{g}} - \sfrac{C_{\ell}^{\delta_{g}\mathrm{T}}C_{\ell}^{\kappa_{\mathrm{CMB}}\mathrm{T}}}{C_{\ell}^{\mathrm{TT}}})^{2}}{(C_{\ell}^{\delta_{g} \delta_{g}}-\sfrac{2(C_{\ell}^{\delta_{g}\mathrm{T}})^{2}}{C_{\ell}^{\mathrm{TT}}})},\\
C^{\mathrm{EE}}_{\ell} &= \sfrac{C^{\gamma_{1} \gamma_{1}}_{\ell}}{3},\\
C^{\mathrm{BB}}_{\ell} &= 0,\\
C^{0\mathrm{E}}_{\ell} &= C^{\gamma_{1}\kappa_{\mathrm{CMB}}}_{\ell} - \sfrac{C_{\ell}^{\gamma_{1}\mathrm{T}}C_{\ell}^{\kappa_{\mathrm{CMB}}\mathrm{T}}}{C_{\ell}^{\mathrm{TT}}} - \\
& \sfrac{C_{\ell}^{\gamma_{1}\delta_{g}}C_{\ell}^{\kappa_{\mathrm{CMB}}\delta_{g}}}{[C_{\ell}^{\delta_{g} \delta_{g}} - \sfrac{2(C_{\ell}^{\delta_{g}\mathrm{T}})^{2}}{C_{\ell}^{\mathrm{TT}}}]}.
\end{align*}
As before, unwanted spin-0/spin-2 correlations are removed by adding the second and third term in the last equation. 
These maps are denoted $m^{'3}_{i}$, where $i \in \{\kappa_{\mathrm{CMB}}, \gamma_{1}^{1}, \gamma_{1}^{2}\}$.
\item We create two spin-0 maps generated with the same seed as used for $m^{'1}$ with the power spectra
\begin{align*}
C^{00}_{\ell} &= \sfrac{2(C_{\ell}^{\delta_{g}\mathrm{T}})^{2}}{C_{\ell}^{\mathrm{TT}}}, \\
C^{00}_{\ell} &= \sfrac{(C_{\ell}^{\kappa_{\mathrm{CMB}}\mathrm{T}})^{2}}{(2 C_{\ell}^{\mathrm{TT}})}.
\end{align*}
The map corresponding to the first power spectrum is denoted $m^{'4}_{\delta_{g}}$, while the one corresponding to the second is called $m^{'4}_{\kappa_{\mathrm{CMB}}}$.
\item We create a spin-0 map generated with the same seed as used for $m^{'2}$ with the power spectrum
\begin{equation*}
C^{00}_{\ell} = \sfrac{(C_{\ell}^{\kappa_{\mathrm{CMB}}\delta_{g}} - \sfrac{C_{\ell}^{\delta_{g}\mathrm{T}}C_{\ell}^{\kappa_{\mathrm{CMB}}\mathrm{T}}}{C_{\ell}^{\mathrm{TT}}})^{2}}{(C_{\ell}^{\delta_{g} \delta_{g}}-\sfrac{2(C_{\ell}^{\delta_{g}\mathrm{T}})^{2}}{C_{\ell}^{\mathrm{TT}}})}, 
\end{equation*}
This map is called $m^{'5}_{\kappa_{\mathrm{CMB}}}$.
\item Finally we combine the maps i.e.
\begin{align*}
m^{'}_{\mathrm{T}} &= m^{'1}_{\mathrm{T}}, \\
m^{'}_{\delta_{g}} &= m^{'2}_{\delta_{g}} + m^{'4}_{\delta_{g}}, \\
m^{'}_{\kappa_{\mathrm{CMB}}} &= m^{'3}_{\kappa_{\mathrm{CMB}}} + m^{'4}_{\kappa_{\mathrm{CMB}}} + m^{'5}_{\kappa_{\mathrm{CMB}}}, \\
m^{'}_{\gamma_{1}^{1}} &= m^{'1}_{\gamma_{1}^{1}} + m^{'2}_{\gamma_{1}^{1}} + m^{'3}_{\gamma_{1}^{1}}, \\
m^{'}_{\gamma_{1}^{2}} &= m^{'1}_{\gamma_{1}^{2}} + m^{'2}_{\gamma_{1}^{2}} + m^{'3}_{\gamma_{1}^{2}}.
\end{align*}
\end{enumerate}

We construct a correlated realization of CMB temperature anisotropies, CMB lensing convergence and the DES SV weak lensing shear field analogously:
\begin{enumerate}
\item We first create three correlated HEALPix maps using $\tt synfast$ in polarization mode with the power spectra 
\begin{align*}
C^{00}_{\ell} &= \sfrac{C^{\mathrm{TT}}_{\ell}}{2},\\
C^{\mathrm{EE}}_{\ell} &= \sfrac{C^{\gamma_{2} \gamma_{2}}_{\ell}}{2},\\
C^{\mathrm{BB}}_{\ell} &= 0,\\
C^{0\mathrm{E}}_{\ell} &= C^{\gamma_{2}\mathrm{T}}_{\ell}.
\end{align*}
These maps are denoted $\tilde{m}^{1}_{i}$, where $i \in \{\mathrm{T}, \gamma_{2}^{1}, \gamma_{2}^{2}\}$.
\item Following Refs.~\cite{Giannantonio:2008, Cabre:2007, Nicola:2016} we then create three maps with a new random seed and the power spectra
\begin{align*}
C^{00}_{\ell} &= \sfrac{C_{\ell}^{\kappa_{\mathrm{CMB}}\kappa_{\mathrm{CMB}}}}{2} - \sfrac{(C_{\ell}^{\kappa_{\mathrm{CMB}}\mathrm{T}})^{2}}{(2 C_{\ell}^{\mathrm{TT}})},\\
C^{\mathrm{EE}}_{\ell} &= \sfrac{C^{\gamma_{2} \gamma_{2}}_{\ell}}{2},\\
C^{\mathrm{BB}}_{\ell} &= 0,\\
C^{0\mathrm{E}}_{\ell} &= C^{\gamma_{2}\kappa_{\mathrm{CMB}}}_{\ell} - \sfrac{C_{\ell}^{\kappa_{\mathrm{CMB}}\mathrm{T}}C_{\ell}^{\gamma_{2}\mathrm{T}}}{C_{\ell}^{\mathrm{TT}}}.
\end{align*}
These maps are denoted $\tilde{m}^{2}_{i}$, where $i \in \{\kappa_{\mathrm{CMB}}, \gamma_{2}^{1}, \gamma_{2}^{2}\}$.
\item We create a spin-0 map generated with the same seed as used for $\tilde{m}^{1}$ with the power spectrum
\begin{equation*}
C^{00}_{\ell} = \sfrac{(C_{\ell}^{\kappa_{\mathrm{CMB}}\mathrm{T}})^{2}}{(2 C_{\ell}^{\mathrm{TT}})}.
\end{equation*}
The map is denoted $\tilde{m}^{3}_{\kappa_{\mathrm{CMB}}}$.
\item Finally we combine the maps i.e.
\begin{align*}
\tilde{m}_{\mathrm{T}} &= m^{1}_{\mathrm{T}}, \\
\tilde{m}_{\kappa_{\mathrm{CMB}}} &= \tilde{m}^{2}_{\kappa_{\mathrm{CMB}}} + \tilde{m}^{3}_{\kappa_{\mathrm{CMB}}}, \\
\tilde{m}_{\gamma_{2}^{1}} &= \tilde{m}^{1}_{\gamma_{2}^{1}} + \tilde{m}^{2}_{\gamma_{2}^{1}}, \\
\tilde{m}_{\gamma_{2}^{2}} &= \tilde{m}^{1}_{\gamma_{2}^{2}} + \tilde{m}^{2}_{\gamma_{2}^{2}}.
\end{align*}
\end{enumerate}
 
In a last step we combine both sets of maps i.e.
\begin{align*}
m_{\mathrm{T}} &= m^{'}_{\mathrm{T}} + \tilde{m}_{\mathrm{T}}, \\
m_{\delta_{g}} &= m^{'}_{\delta_{g}}, \\
m_{\kappa_{\mathrm{CMB}}} &= m^{'}_{\kappa_{\mathrm{CMB}}} + \tilde{m}_{\kappa_{\mathrm{CMB}}}, \\
m_{\gamma_{1}^{1}} &= m^{'}_{\gamma_{1}^{1}}, \\
m_{\gamma_{1}^{2}} &= m^{'}_{\gamma_{1}^{2}}, \\
m_{\gamma_{2}^{1}} &= \tilde{m}_{\gamma_{2}^{1}}, \\
m_{\gamma_{2}^{2}} &= \tilde{m}_{\gamma_{2}^{2}}.
\end{align*}

This algorithm yields a set of seven correlated HEALPix maps $\{m_{\mathrm{T}} , m_{\delta_{g}}, m_{\kappa_{\mathrm{CMB}}}, m_{\gamma_{1}^{1}}, m_{\gamma_{1}^{2}}, m_{\gamma_{2}^{1}}, m_{\gamma_{2}^{2}}\}$ with auto- and cross-power spectra given by $\{C^{\mathrm{TT}}_{\ell},\allowbreak C_{\ell}^{\delta_{g}\mathrm{T}},\allowbreak C_{\ell}^{\delta_{g} \delta_{g}},\allowbreak C_{\ell}^{\kappa_{\mathrm{CMB}}\mathrm{T}},\allowbreak C_{\ell}^{\kappa_{\mathrm{CMB}}\delta_{g}},\allowbreak C^{\gamma_{1} \mathrm{T}}_{\ell},\allowbreak C^{\gamma_{1}\delta_{g}}_{\ell},\allowbreak C_{\ell}^{\gamma_{1} \kappa_{\mathrm{CMB}}},\allowbreak C^{\gamma_{1} \gamma_{1}}_{\ell},\allowbreak C^{\gamma_{2} \mathrm{T}}_{\ell},\allowbreak C_{\ell}^{\gamma_{2} \kappa_{\mathrm{CMB}}},\allowbreak C^{\gamma_{2} \gamma_{2}}_{\ell}\}$. As in \ja, we include the HEALPix window function and CMB temperature beam window function into the spherical harmonic power spectra prior to creating the maps.

Note that it is not possible to create maps with negative power spectra using HEALPix. This means that we always need to make sure that all cross-terms remain positive. This can usually be achieved by applying the counter-terms to the power spectra with larger amplitude.

In order to obtain simulations of observed maps, we need to add realistic noise to the signal-only maps. To this end we follow the implementation in \ja{} for the CMB temperature anisotropy, the galaxy overdensity and the weak lensing shear maps. As described in more detail in \ja{}, we add noise to the simulations on the map level. For the CMB temperature anisotropies we add the $\tt{Commander}$ half-mission half-difference (HMHD) map to the signal. The noise maps for the galaxy overdensity consist of randomized maps of the galaxy positions in the data while the weak lensing noise maps are created by rotating the galaxy shears in the data by a random angle. We create noise-only maps for the CMB lensing convergence using the simulated data provided by the Planck Collaboration in Ref.~\cite{Planck-Collaboration:2016aa}. These data consist of a set of 100 simulations of observed spherical harmonic coefficients of the CMB lensing convergence as well as the spherical harmonic coefficients of the input CMB lensing convergence to each simulation. To compute noise-only maps we first compute the difference between observed and true spherical harmonic coefficients. Using those we create HEALPix maps of the CMB lensing convergence reconstruction noise at resolution $\tt{NSIDE}$ = 1024. This approach is an approximation for two reasons. Firstly, it is only exactly valid in the limit of linearity in signal and noise. This is not true in this case and we therefore expect to see differences between the estimated noise power spectra and the true one. Comparing these two we find differences of at most $3.5 \%$, which is an acceptable accuracy for covariance matrix computations. Secondly, the CMB lensing noise power spectrum is cosmology dependent since the main source of noise is the disconnected part of the CMB temperature 4-point function, which depends on the cosmological model assumed in the simulations. Therefore we should strictly only use these maps for the fiducial cosmological model adopted in Ref.~\cite{Planck-Collaboration:2016aa}. Since we only include cross-power spectra involving the CMB lensing convergence we believe the errors due to assuming a different cosmological model for the theory and the noise to be subdominant. We therefore leave the refinement of the CMB lensing convergence noise estimate to future work.

\section{Validation of spherical harmonic power spectrum measurement}\label{sec:ap-mocks-reconst}

\begin{figure*}
\begin{center}
\includegraphics[scale=0.3]{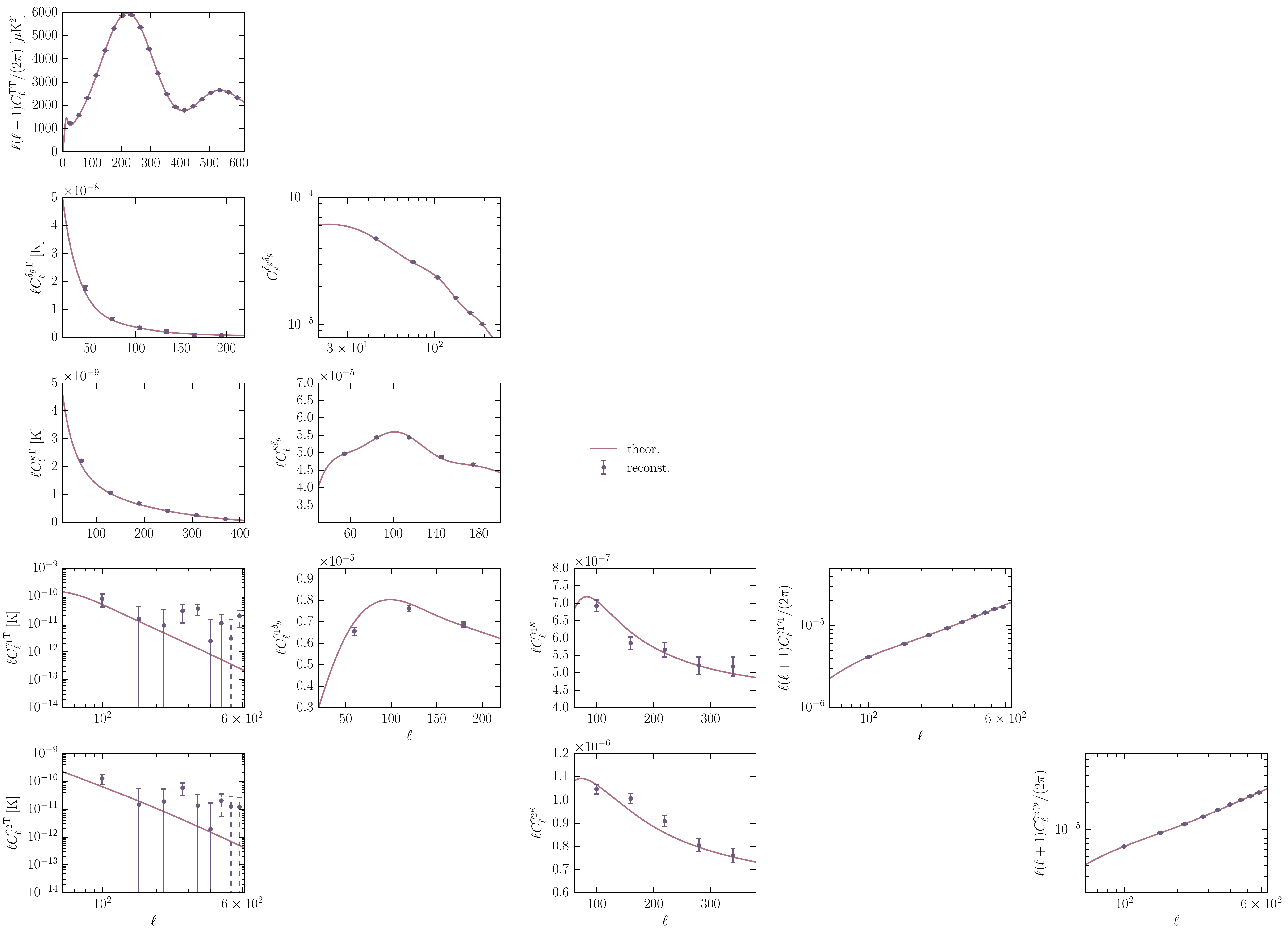}
\caption{Comparison between input theoretical power spectra and the mean reconstruction derived from $N_{\mathrm{sim}} = 1000$ Gaussian simulations as described in Sec.~\ref{sec:ap-mocks-reconst}. The angular sky coverage and the noise level of the simulations closely match the data. The $\tt{PolSpice}$ settings used to compute these power spectra are identical to those applied on the data and the angular multipole range as well as binning schemes match those described in Tab.~\ref{tab:clparams}. Dashed lines denote negative spherical harmonic power spectrum values. The input power spectra have been convolved with the $\tt{PolSpice}$ kernels described in \ja{} and Section \ref{sec:cls}. The error bars are derived from the Gaussian simulations described in Section \ref{sec:covmat} and Appendix \ref{sec:ap-mocks}.}
\label{fig:cls-mock-reconstruction}
\end{center}
\end{figure*}

\begin{figure*}
\begin{center}
\includegraphics[scale=0.5]{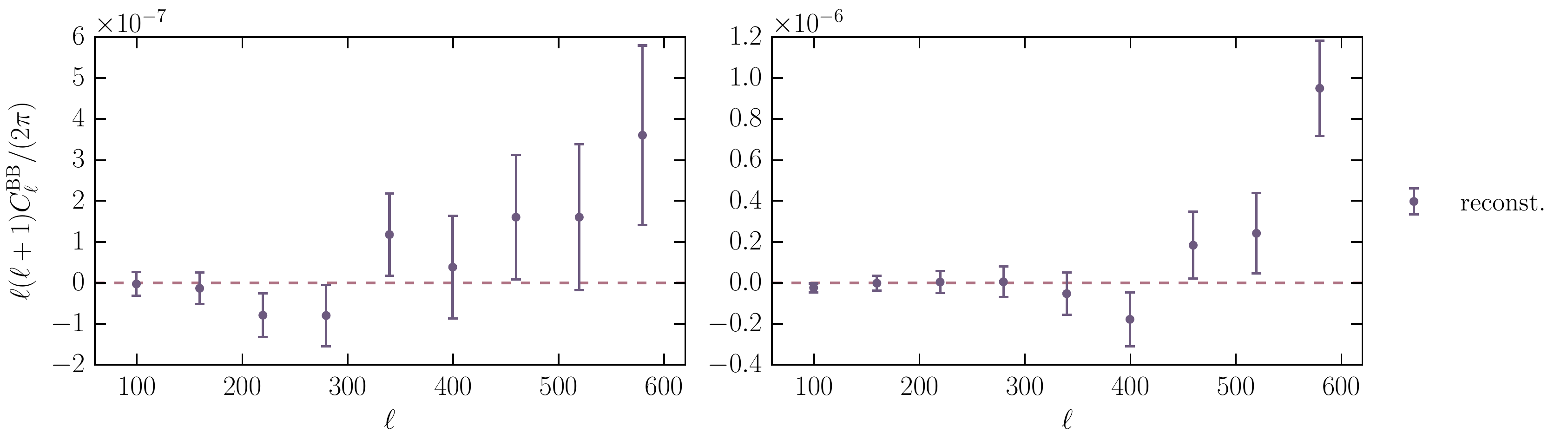}
\caption{Mean of the reconstructed cosmic shear B-modes derived from $N_{\mathrm{sim}} = 1000$ Gaussian simulations as described in Sec.~\ref{sec:ap-mocks-reconst}. The angular sky coverage and the noise level of the simulations closely match the data. The $\tt{PolSpice}$ settings used are identical to those applied on the data and the angular multipole range as well as binning schemes match those described in Tab.~\ref{tab:clparams}. The left hand panel shows the results for the SDSS Stripe 82 simulations, while the right hand panel shows the results for the DES SV simulations. The error bars are derived from the Gaussian simulations described in Section \ref{sec:covmat} and Appendix \ref{sec:ap-mocks}.}
\label{fig:cls-B-mode-mock-reconstruction}
\end{center}
\end{figure*}

As in \ja{} we validate our power spectrum measurement using the Gaussian simulations described in Sec.~\ref{sec:ap-mocks}. We compute $N_{\mathrm{sim}} = 1000$ realizations of signal and noise for our fiducial cosmological model defined by $\{h,\, \Omega_{\mathrm{m}}, \,\Omega_{\mathrm{b}}, \, n_{\mathrm{s}}, \,\sigma_{8}, \,\tau_{\mathrm{reion}}, \,T_{\mathrm{CMB}}\} = \{0.7, \,0.3, \,0.049, \,1.0, \,0.88, \,0.078, \,2.275 \,\mathrm{K}\}$ and $\{b,\, m^{\mathrm{SDSS}}_{*}, \, m^{\mathrm{DES}}_{*}, \, m_{\kappa_{\mathrm{CMB}}}, \, A_{\mathrm{ps}}, \, A_{\mathrm{cl}}\} = \{2., \,0., \,0., \,0., \,0., \,0.\}$. The noise level in the simulations is consistent with the data and we apply the survey masks determined from the data to each simulated map. The spherical harmonic power spectra are computed with the same $\tt{PolSpice}$ settings as used on the data and the noise bias correction is performed as described in Sec.~\ref{sec:cls}. Figures \ref{fig:cls-mock-reconstruction} and \ref{fig:cls-B-mode-mock-reconstruction} show the comparison between the mean of the recovered power spectra and the input theoretical power spectra. The error bars denote the error on the mean of the power spectra and are derived using the Gaussian simulations. We find that the input power spectra are generally recovered rather well, which validates our power spectrum measurement. 

\begin{figure}
\begin{center}
\includegraphics[scale=0.5]{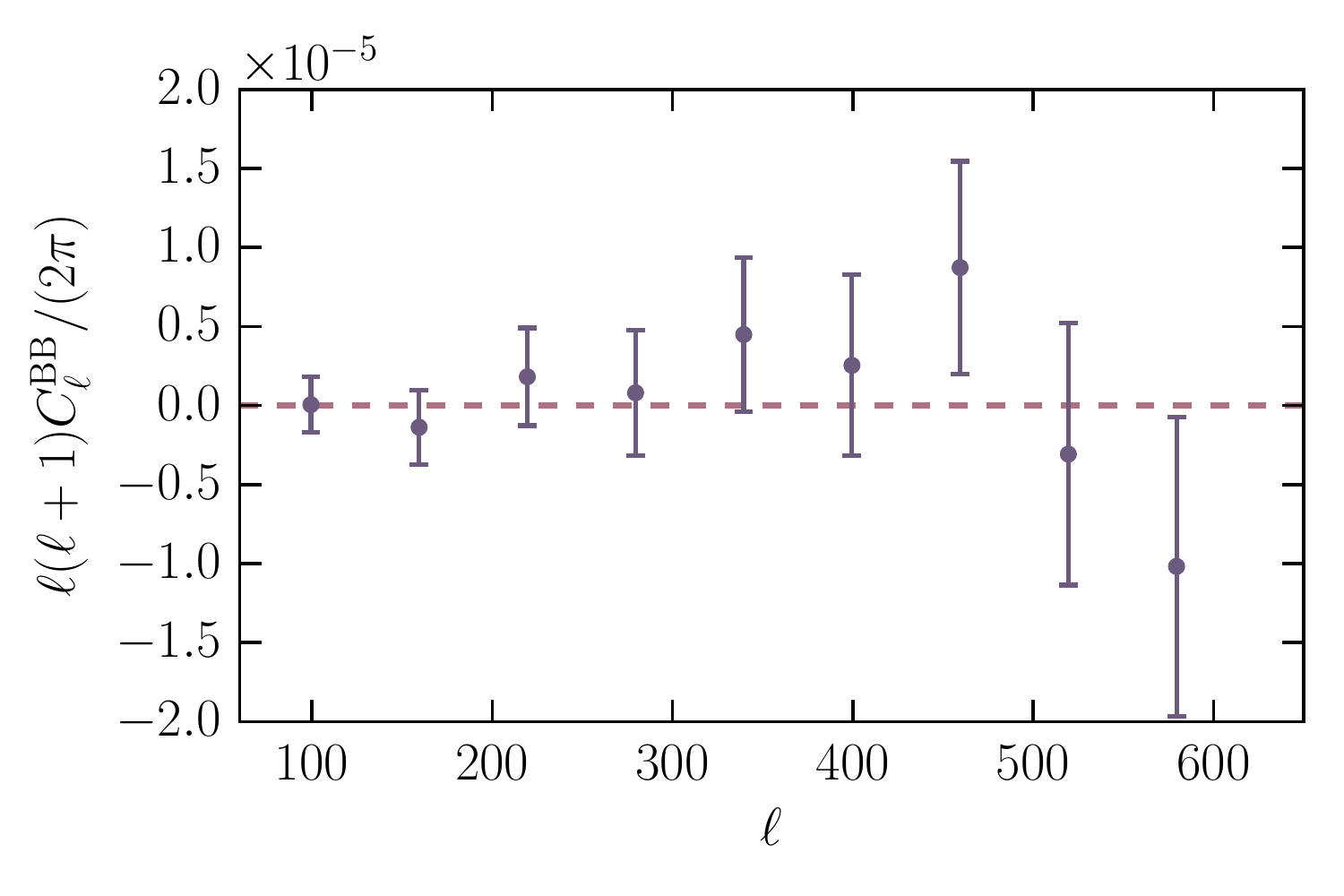}
\caption{Spherical harmonic B-mode power spectrum of DES SV. The error bars are derived from the Gaussian simulations described in Section \ref{sec:covmat} and Appendix \ref{sec:ap-mocks}. The angular multipole range and binning scheme used is summarized in Tab.~\ref{tab:clparams}.}
\label{fig:cls_b_dessv}
\end{center}
\end{figure}

\section{Spherical harmonic power spectrum robustness tests}\label{sec:ap-robustness-tests}

In this section we summarize the robustness tests performed for the measured spherical harmonic power spectra.

\subsection{Comparison between spherical harmonic power spectra in equatorial and Galactic coordinates}

\begin{figure}
\begin{center}
\includegraphics[scale=0.5]{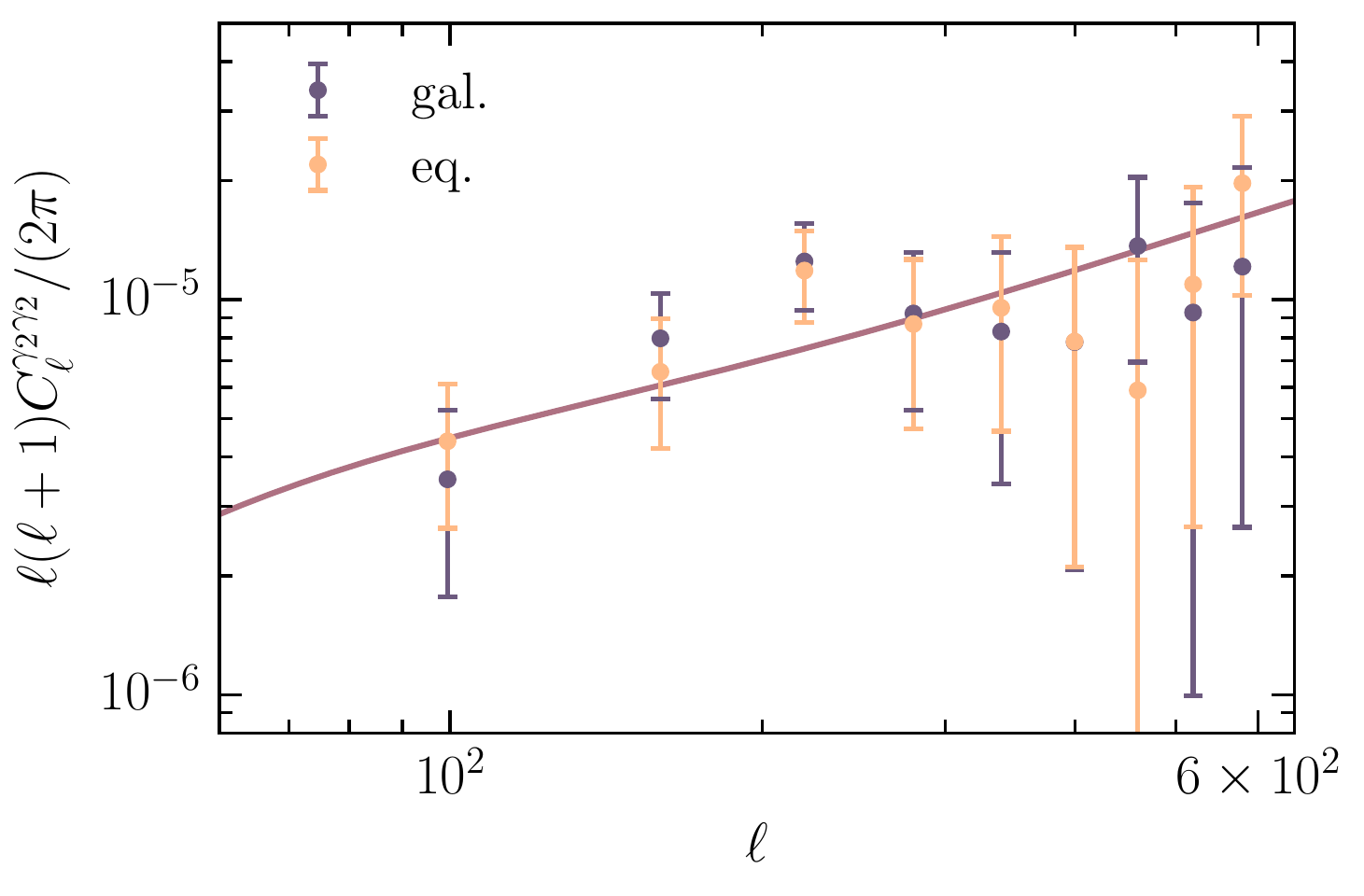}
\caption{Comparison of cosmic shear power spectrum from DES SV as measured using the maps in Galactic and equatorial coordinates. The theoretical prediction for the power spectrum has been convolved with the $\tt{PolSpice}$ kernels described in \ja{} and Section \ref{sec:cls}. The error bars are derived from the Gaussian simulations described in Section \ref{sec:covmat} and Appendix \ref{sec:ap-mocks}. The angular multipole range and binning scheme used is summarized in Tab.~\ref{tab:clparams}.}
\label{fig:descls_ecl_gal}
\end{center}
\end{figure}

We test that the measured power spectra are independent of the chosen coordinate system by comparing the results obtained from the maps in equatorial and Galactic coordinates. The only power spectrum that can be transformed between coordinate systems is the cosmic shear power spectrum for DES SV and the results are shown in Fig.~\ref{fig:descls_ecl_gal}. We find discrepancies between the two power spectra, which are comparable to the discrepancies found for SDSS Stripe 82 in \ja. This suggests that coordinate-dependent bias corrections are not the only explanation for differences between the power spectra but that these differences are partly also due to different shape noise properties in different coordinate systems. Since the discrepancies detected are well within our measurement uncertanties, we do not investigate this issue further and include the spherical harmonic power spectrum measured from the maps in Galactic coordinates into our analysis.

\subsection{Comparison between spherical harmonic power spectra measured from different foreground-reduced CMB temperature maps}

\begin{figure*}
\begin{center}
\includegraphics[scale=0.5]{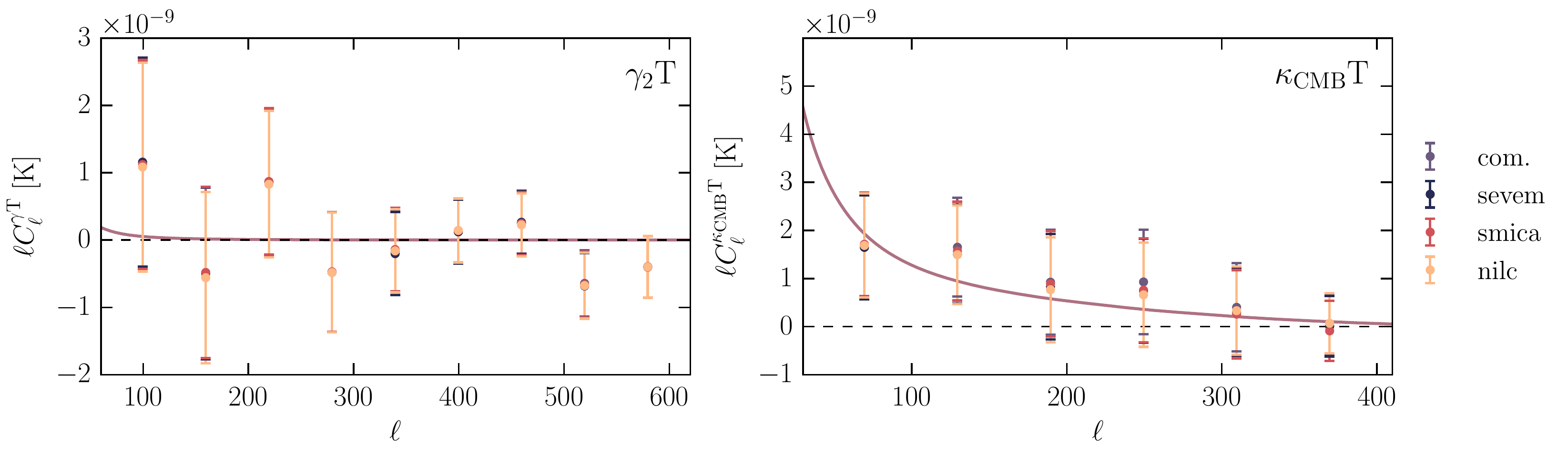}
\caption{Comparison of the spherical harmonic power spectra $C^{\gamma_{2} \mathrm{T}}_{\ell}, C_{\ell}^{\kappa_{\mathrm{CMB}}\mathrm{T}}$ obtained from the foreground-reduced CMB temperature anisotropy maps derived using the four different algorithms $\tt{Commander}$, $\tt{NILC}$, $\tt{SEVEM}$ and $\tt{SMICA}$. The theoretical predictions for the power spectra have been convolved with the $\tt{PolSpice}$ kernels described in \ja{} and Section \ref{sec:cls}. The error bars are derived from the Gaussian simulations described in Section \ref{sec:covmat} and Appendix \ref{sec:ap-mocks}. The angular multipole ranges and binning schemes used for all the power spectra are summarized in Tab.~\ref{tab:clparams}.}
\label{fig:cls_foreground_rem}
\end{center}
\end{figure*}

We investigate the impact of our choice of fiducial foreground-reduced CMB temperature anisotropy map by comparing the power spectra involving CMB temperature data obtained using the four foreground reduction algorithms employed by the Planck Collaboration \cite{Planck-Collaboration:2016ab}. These are $\tt{Commander}$, $\tt{NILC}$, $\tt{SEVEM}$ and $\tt{SMICA}$ and the results for $C_{\ell}^{\gamma_{2}\mathrm{T}}$ and $C_{\ell}^{\kappa_{\mathrm{CMB}}\mathrm{T}}$ are shown in Fig.~\ref{fig:cls_foreground_rem}. As can be seen, we find the derived power spectra to be well consistent with each other.

\begin{figure*}
\begin{center}
\includegraphics[scale=0.5]{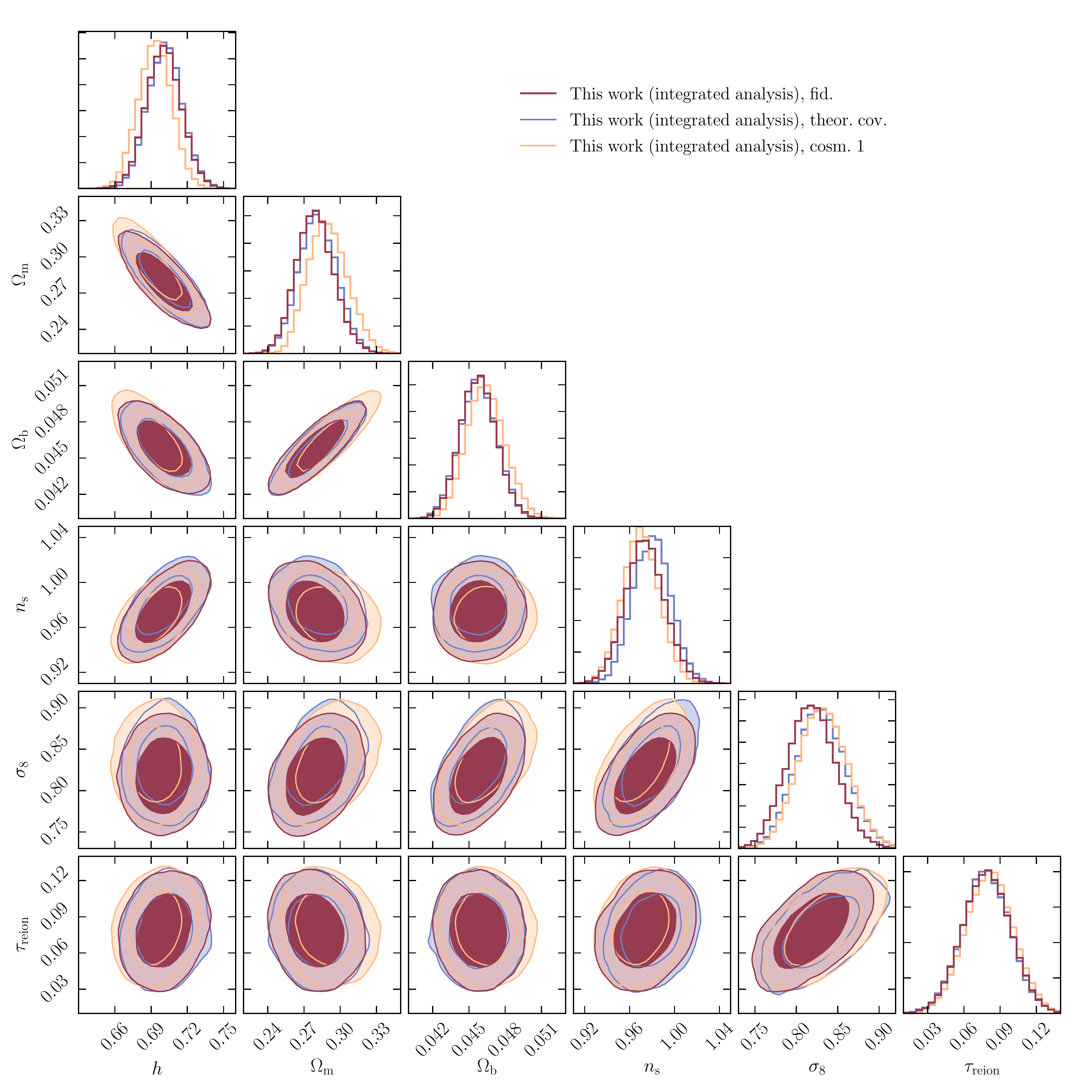}
\caption{Comparison of the constraints obtained from the fiducial joint analysis to the constraints obtained using the theoretical covariance matrix and the constraints obtained assuming a different cosmological model to estimate the empirical covariance. The constraints are marginalized over all nuisance parameters given in Tab.~\ref{tab:params}. In each case the inner (outer) contour shows the $68 \%$ c.l. ($95 \%$ c.l.).}
\label{fig:constraints-ja-diff-covs}
\end{center}
\end{figure*}

\section{Impact of unresolved foregrounds on CMB temperature anisotropies}\label{sec:ap-cmb-foregrounds}

\begin{figure*}
\begin{center}
\includegraphics[scale=0.5]{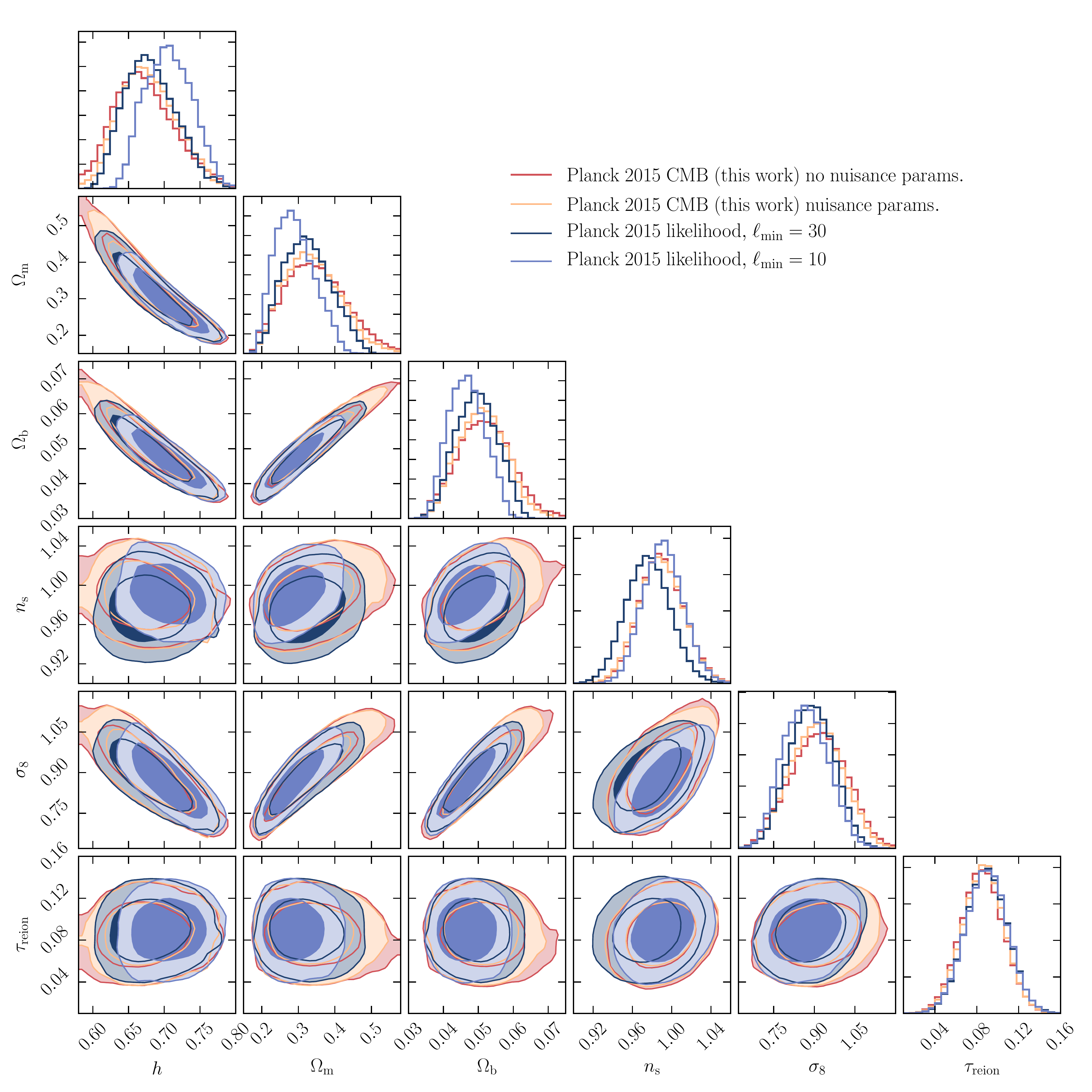}
\caption{Comparison of the constraints obtained from CMB temperature anisotropies for angular multipoles $\ell \in [10, 610]$ in this work both including the additional nuisance parameters $A_{\mathrm{ps}}$, $A_{\mathrm{cl}}$ and neglecting them. Also shown are the constraints obtained from the Planck likelihood with $\ell_{\mathrm{max}} \simeq 610$ and $\ell_{\mathrm{min}} = 10$ as well as $\ell_{\mathrm{min}} = 30$. In each case the inner (outer) contour shows the $68 \%$ c.l. ($95 \%$ c.l.).}
\label{fig:constraints-ja-cmb}
\end{center}
\end{figure*}

As described in Sec.~\ref{sec:systematics}, the foreground-reduced CMB temperature anisotropy maps are contaminated by unresolved extragalactic sources \cite{Planck-Collaboration:2014ac}. Following Ref.~\cite{Planck-Collaboration:2014af}, the power spectra of these foregrounds can be modelled using a contribution of an unclustered Poisson component $C_{\ell}^{\mathrm{ps}}$ and the contribution of a clustered component $C_{\ell}^{\mathrm{cl}}$ \cite{Planck-Collaboration:2014af}. These become significant at high angular multipoles. The two power spectra are defined in terms of their amplitudes $A_{\mathrm{ps}}$ and $A_{\mathrm{cl}}$ as:
\begin{equation}
\begin{aligned}
C_{\ell}^{\mathrm{ps}} &= 2 \pi \frac{A_{\mathrm{ps}}}{\ell_{\mathrm{p}}(\ell_{\mathrm{p}}+1)}, \\
C_{\ell}^{\mathrm{cl}} &= 2 \pi \frac{A_{\mathrm{cl}}}{\ell(\ell+1)} \left(\frac{\ell}{\ell_{\mathrm{p}}}\right)^{0.8},
\end{aligned}
\end{equation}
where the Pivot angular multipole is defined to be $\ell_{\mathrm{p}} = 3000$ and both amplitudes have units of $\mu$K$^{2}$. The normalization ensures that for $D_{\ell} = \frac{\ell(\ell+1)}{2 \pi} C_{\ell}$ we have $D_{\ell_{\mathrm{p}}}^{\mathrm{ps}} = A_{\mathrm{ps}}$ and $D_{\ell_{\mathrm{p}}}^{\mathrm{cl}} = A_{\mathrm{cl}}$. We investigate the impact of these residual foregrounds by comparing the cosmological parameter constraints obtained from CMB temperature data alone both including these two additional degrees of freedom and neglecting them. The constraints obtained in both cases are virtually the same as can be seen from Fig.~\ref{fig:constraints-ja-cmb}, which means that the low-$\ell$ temperature anisotropy power spectrum is insensitive to residual foregrounds. We therefore do not include these additional degrees of freedom in our fiducial analysis.

\begin{figure}
\begin{center}
\includegraphics[scale=0.5]{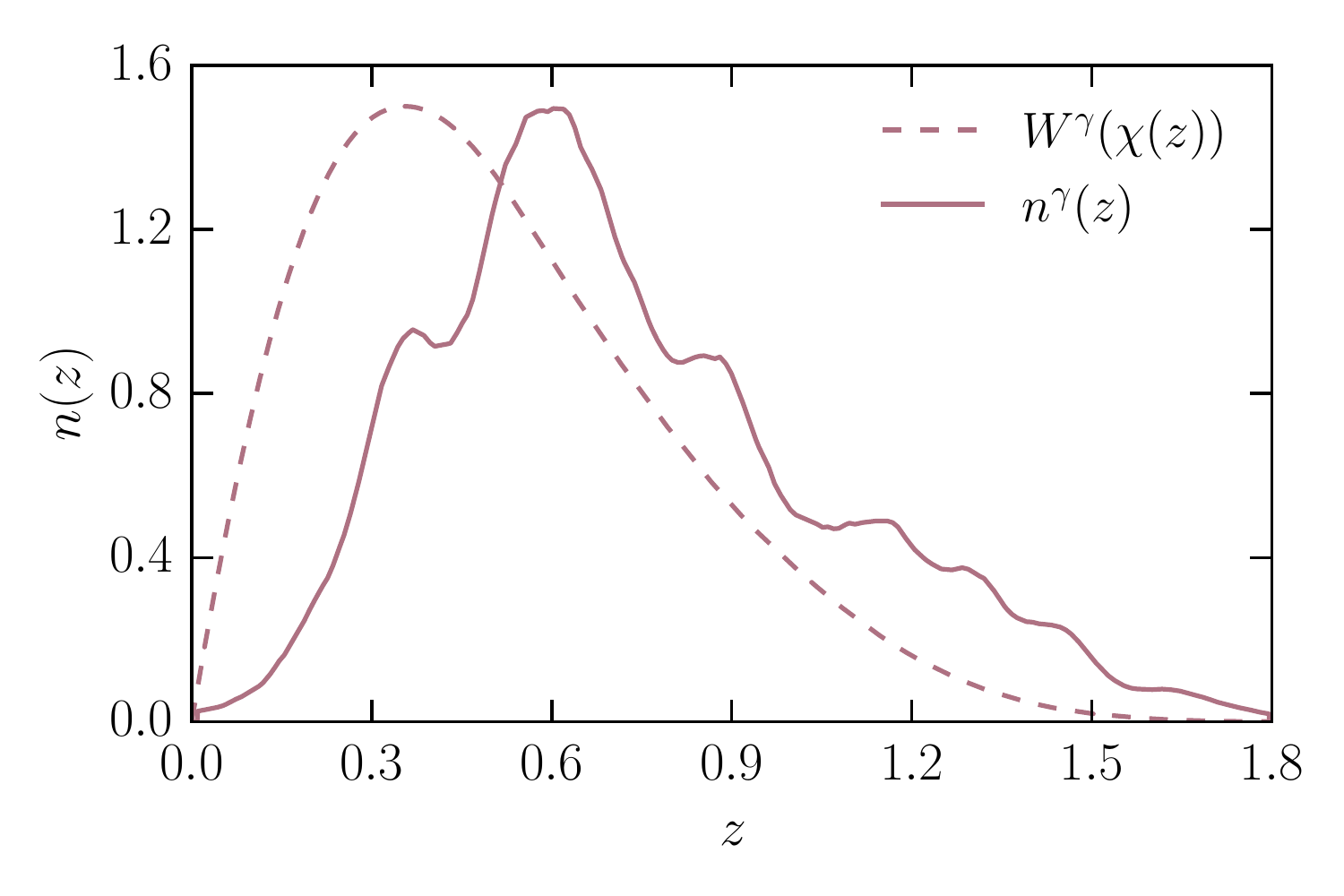}
\caption{Redshift selection function and weak lensing shear window function for the DES SV galaxies. The weak lensing window function has been rescaled relative to the redshift selection function. The window function has been derived using our best-fit cosmological parameters.} 
\label{fig:dessv-nz}
\end{center}
\end{figure}

\bibliography{main_text_incl_figs}

\begin{thebibliography}{98}%
\makeatletter
\providecommand \@ifxundefined [1]{%
 \@ifx{#1\undefined}
}%
\providecommand \@ifnum [1]{%
 \ifnum #1\expandafter \@firstoftwo
 \else \expandafter \@secondoftwo
 \fi
}%
\providecommand \@ifx [1]{%
 \ifx #1\expandafter \@firstoftwo
 \else \expandafter \@secondoftwo
 \fi
}%
\providecommand \natexlab [1]{#1}%
\providecommand \enquote  [1]{``#1''}%
\providecommand \bibnamefont  [1]{#1}%
\providecommand \bibfnamefont [1]{#1}%
\providecommand \citenamefont [1]{#1}%
\providecommand \href@noop [0]{\@secondoftwo}%
\providecommand \href [0]{\begingroup \@sanitize@url \@href}%
\providecommand \@href[1]{\@@startlink{#1}\@@href}%
\providecommand \@@href[1]{\endgroup#1\@@endlink}%
\providecommand \@sanitize@url [0]{\catcode `\\12\catcode `\$12\catcode
  `\&12\catcode `\#12\catcode `\^12\catcode `\_12\catcode `\%12\relax}%
\providecommand \@@startlink[1]{}%
\providecommand \@@endlink[0]{}%
\providecommand \url  [0]{\begingroup\@sanitize@url \@url }%
\providecommand \@url [1]{\endgroup\@href {#1}{\urlprefix }}%
\providecommand \urlprefix  [0]{URL }%
\providecommand \Eprint [0]{\href }%
\providecommand \doibase [0]{http://dx.doi.org/}%
\providecommand \selectlanguage [0]{\@gobble}%
\providecommand \bibinfo  [0]{\@secondoftwo}%
\providecommand \bibfield  [0]{\@secondoftwo}%
\providecommand \translation [1]{[#1]}%
\providecommand \BibitemOpen [0]{}%
\providecommand \bibitemStop [0]{}%
\providecommand \bibitemNoStop [0]{.\EOS\space}%
\providecommand \EOS [0]{\spacefactor3000\relax}%
\providecommand \BibitemShut  [1]{\csname bibitem#1\endcsname}%
\let\auto@bib@innerbib\@empty
%</preamble>
\bibitem [{\citenamefont {{Nicola}}\ \emph {et~al.}(2016)\citenamefont
  {{Nicola}}, \citenamefont {{Refregier}},\ and\ \citenamefont
  {{Amara}}}]{Nicola:2016}%
  \BibitemOpen
  \bibfield  {author} {\bibinfo {author} {\bibfnamefont {A.}~\bibnamefont
  {{Nicola}}}, \bibinfo {author} {\bibfnamefont {A.}~\bibnamefont
  {{Refregier}}}, \ and\ \bibinfo {author} {\bibfnamefont {A.}~\bibnamefont
  {{Amara}}},\ }\href {\doibase 10.1103/PhysRevD.94.083517} {\bibfield
  {journal} {\bibinfo  {journal} {\prd}\ }\textbf {\bibinfo {volume} {94}},\
  \bibinfo {eid} {083517} (\bibinfo {year} {2016})},\ \Eprint
  {http://arxiv.org/abs/1607.01014} {arXiv:1607.01014} \BibitemShut {NoStop}%
\bibitem [{\citenamefont {{Planck Collaboration}}\ \emph
  {et~al.}(2016{\natexlab{a}})\citenamefont {{Planck Collaboration}},
  \citenamefont {{Adam}}, \citenamefont {{Ade}}, \citenamefont {{Aghanim}},
  \citenamefont {{Akrami}}, \citenamefont {{Alves}}, \citenamefont
  {{Arg{\"u}eso}}, \citenamefont {{Arnaud}}, \citenamefont {{Arroja}},
  \citenamefont {{Ashdown}},\ and\ \citenamefont
  {et~al.}}]{Planck-Collaboration:2016ad}%
  \BibitemOpen
  \bibfield  {author} {\bibinfo {author} {\bibnamefont {{Planck
  Collaboration}}}, \bibinfo {author} {\bibfnamefont {R.}~\bibnamefont
  {{Adam}}}, \bibinfo {author} {\bibfnamefont {P.~A.~R.}\ \bibnamefont
  {{Ade}}}, \bibinfo {author} {\bibfnamefont {N.}~\bibnamefont {{Aghanim}}},
  \bibinfo {author} {\bibfnamefont {Y.}~\bibnamefont {{Akrami}}}, \bibinfo
  {author} {\bibfnamefont {M.~I.~R.}\ \bibnamefont {{Alves}}}, \bibinfo
  {author} {\bibfnamefont {F.}~\bibnamefont {{Arg{\"u}eso}}}, \bibinfo {author}
  {\bibfnamefont {M.}~\bibnamefont {{Arnaud}}}, \bibinfo {author}
  {\bibfnamefont {F.}~\bibnamefont {{Arroja}}}, \bibinfo {author}
  {\bibfnamefont {M.}~\bibnamefont {{Ashdown}}}, \ and\ \bibinfo {author}
  {\bibnamefont {et~al.}},\ }\href {\doibase 10.1051/0004-6361/201527101}
  {\bibfield  {journal} {\bibinfo  {journal} {\aap}\ }\textbf {\bibinfo
  {volume} {594}},\ \bibinfo {eid} {A1} (\bibinfo {year}
  {2016}{\natexlab{a}})},\ \Eprint {http://arxiv.org/abs/1502.01582}
  {arXiv:1502.01582} \BibitemShut {NoStop}%
\bibitem [{\citenamefont {{Aihara}}\ \emph {et~al.}(2011)\citenamefont
  {{Aihara}}, \citenamefont {{Allende Prieto}}, \citenamefont {{An}},
  \citenamefont {{Anderson}}, \citenamefont {{Aubourg}}, \citenamefont
  {{Balbinot}}, \citenamefont {{Beers}}, \citenamefont {{Berlind}},
  \citenamefont {{Bickerton}}, \citenamefont {{Bizyaev}}, \citenamefont
  {{Blanton}}, \citenamefont {{Bochanski}}, \citenamefont {{Bolton}},
  \citenamefont {{Bovy}}, \citenamefont {{Brandt}},\ and\ \citenamefont {{et
  al.}}}]{Aihara:2011}%
  \BibitemOpen
  \bibfield  {author} {\bibinfo {author} {\bibfnamefont {H.}~\bibnamefont
  {{Aihara}}}, \bibinfo {author} {\bibfnamefont {C.}~\bibnamefont {{Allende
  Prieto}}}, \bibinfo {author} {\bibfnamefont {D.}~\bibnamefont {{An}}},
  \bibinfo {author} {\bibfnamefont {S.~F.}\ \bibnamefont {{Anderson}}},
  \bibinfo {author} {\bibfnamefont {{\'E}.}~\bibnamefont {{Aubourg}}}, \bibinfo
  {author} {\bibfnamefont {E.}~\bibnamefont {{Balbinot}}}, \bibinfo {author}
  {\bibfnamefont {T.~C.}\ \bibnamefont {{Beers}}}, \bibinfo {author}
  {\bibfnamefont {A.~A.}\ \bibnamefont {{Berlind}}}, \bibinfo {author}
  {\bibfnamefont {S.~J.}\ \bibnamefont {{Bickerton}}}, \bibinfo {author}
  {\bibfnamefont {D.}~\bibnamefont {{Bizyaev}}}, \bibinfo {author}
  {\bibfnamefont {M.~R.}\ \bibnamefont {{Blanton}}}, \bibinfo {author}
  {\bibfnamefont {J.~J.}\ \bibnamefont {{Bochanski}}}, \bibinfo {author}
  {\bibfnamefont {A.~S.}\ \bibnamefont {{Bolton}}}, \bibinfo {author}
  {\bibfnamefont {J.}~\bibnamefont {{Bovy}}}, \bibinfo {author} {\bibfnamefont
  {W.~N.}\ \bibnamefont {{Brandt}}}, \ and\ \bibinfo {author} {\bibnamefont
  {{et al.}}},\ }\href {\doibase 10.1088/0067-0049/193/2/29} {\bibfield
  {journal} {\bibinfo  {journal} {\apjs}\ }\textbf {\bibinfo {volume} {193}},\
  \bibinfo {eid} {29} (\bibinfo {year} {2011})},\ \Eprint
  {http://arxiv.org/abs/1101.1559} {arXiv:1101.1559 [astro-ph.IM]} \BibitemShut
  {NoStop}%
\bibitem [{\citenamefont {{Annis}}\ \emph {et~al.}(2014)\citenamefont
  {{Annis}}, \citenamefont {{Soares-Santos}}, \citenamefont {{Strauss}},
  \citenamefont {{Becker}}, \citenamefont {{Dodelson}}, \citenamefont {{Fan}},
  \citenamefont {{Gunn}}, \citenamefont {{Hao}}, \citenamefont {{Ivezi{\'c}}},
  \citenamefont {{Jester}}, \citenamefont {{Jiang}}, \citenamefont
  {{Johnston}}, \citenamefont {{Kubo}}, \citenamefont {{Lampeitl}},
  \citenamefont {{Lin}}, \citenamefont {{Lupton}}, \citenamefont {{Miknaitis}},
  \citenamefont {{Seo}}, \citenamefont {{Simet}},\ and\ \citenamefont
  {{Yanny}}}]{Annis:2014}%
  \BibitemOpen
  \bibfield  {author} {\bibinfo {author} {\bibfnamefont {J.}~\bibnamefont
  {{Annis}}}, \bibinfo {author} {\bibfnamefont {M.}~\bibnamefont
  {{Soares-Santos}}}, \bibinfo {author} {\bibfnamefont {M.~A.}\ \bibnamefont
  {{Strauss}}}, \bibinfo {author} {\bibfnamefont {A.~C.}\ \bibnamefont
  {{Becker}}}, \bibinfo {author} {\bibfnamefont {S.}~\bibnamefont
  {{Dodelson}}}, \bibinfo {author} {\bibfnamefont {X.}~\bibnamefont {{Fan}}},
  \bibinfo {author} {\bibfnamefont {J.~E.}\ \bibnamefont {{Gunn}}}, \bibinfo
  {author} {\bibfnamefont {J.}~\bibnamefont {{Hao}}}, \bibinfo {author}
  {\bibfnamefont {{\v Z}.}~\bibnamefont {{Ivezi{\'c}}}}, \bibinfo {author}
  {\bibfnamefont {S.}~\bibnamefont {{Jester}}}, \bibinfo {author}
  {\bibfnamefont {L.}~\bibnamefont {{Jiang}}}, \bibinfo {author} {\bibfnamefont
  {D.~E.}\ \bibnamefont {{Johnston}}}, \bibinfo {author} {\bibfnamefont
  {J.~M.}\ \bibnamefont {{Kubo}}}, \bibinfo {author} {\bibfnamefont
  {H.}~\bibnamefont {{Lampeitl}}}, \bibinfo {author} {\bibfnamefont
  {H.}~\bibnamefont {{Lin}}}, \bibinfo {author} {\bibfnamefont {R.~H.}\
  \bibnamefont {{Lupton}}}, \bibinfo {author} {\bibfnamefont {G.}~\bibnamefont
  {{Miknaitis}}}, \bibinfo {author} {\bibfnamefont {H.-J.}\ \bibnamefont
  {{Seo}}}, \bibinfo {author} {\bibfnamefont {M.}~\bibnamefont {{Simet}}}, \
  and\ \bibinfo {author} {\bibfnamefont {B.}~\bibnamefont {{Yanny}}},\ }\href
  {\doibase 10.1088/0004-637X/794/2/120} {\bibfield  {journal} {\bibinfo
  {journal} {\apj}\ }\textbf {\bibinfo {volume} {794}},\ \bibinfo {eid} {120}
  (\bibinfo {year} {2014})},\ \Eprint {http://arxiv.org/abs/1111.6619}
  {arXiv:1111.6619} \BibitemShut {NoStop}%
\bibitem [{\citenamefont {{Planck Collaboration}}\ \emph
  {et~al.}(2016{\natexlab{b}})\citenamefont {{Planck Collaboration}},
  \citenamefont {{Ade}}, \citenamefont {{Aghanim}}, \citenamefont {{Arnaud}},
  \citenamefont {{Ashdown}}, \citenamefont {{Aumont}}, \citenamefont
  {{Baccigalupi}}, \citenamefont {{Banday}}, \citenamefont {{Barreiro}},
  \citenamefont {{Bartlett}},\ and\ \citenamefont
  {et~al.}}]{Planck-Collaboration:2016aa}%
  \BibitemOpen
  \bibfield  {author} {\bibinfo {author} {\bibnamefont {{Planck
  Collaboration}}}, \bibinfo {author} {\bibfnamefont {P.~A.~R.}\ \bibnamefont
  {{Ade}}}, \bibinfo {author} {\bibfnamefont {N.}~\bibnamefont {{Aghanim}}},
  \bibinfo {author} {\bibfnamefont {M.}~\bibnamefont {{Arnaud}}}, \bibinfo
  {author} {\bibfnamefont {M.}~\bibnamefont {{Ashdown}}}, \bibinfo {author}
  {\bibfnamefont {J.}~\bibnamefont {{Aumont}}}, \bibinfo {author}
  {\bibfnamefont {C.}~\bibnamefont {{Baccigalupi}}}, \bibinfo {author}
  {\bibfnamefont {A.~J.}\ \bibnamefont {{Banday}}}, \bibinfo {author}
  {\bibfnamefont {R.~B.}\ \bibnamefont {{Barreiro}}}, \bibinfo {author}
  {\bibfnamefont {J.~G.}\ \bibnamefont {{Bartlett}}}, \ and\ \bibinfo {author}
  {\bibnamefont {et~al.}},\ }\href {\doibase 10.1051/0004-6361/201525941}
  {\bibfield  {journal} {\bibinfo  {journal} {\aap}\ }\textbf {\bibinfo
  {volume} {594}},\ \bibinfo {eid} {A15} (\bibinfo {year}
  {2016}{\natexlab{b}})},\ \Eprint {http://arxiv.org/abs/1502.01591}
  {arXiv:1502.01591} \BibitemShut {NoStop}%
\bibitem [{\citenamefont {{Jarvis}}\ \emph {et~al.}(2016)\citenamefont
  {{Jarvis}}, \citenamefont {{Sheldon}}, \citenamefont {{Zuntz}}, \citenamefont
  {{Kacprzak}}, \citenamefont {{Bridle}}, \citenamefont {{Amara}},
  \citenamefont {{Armstrong}}, \citenamefont {{Becker}}, \citenamefont
  {{Bernstein}}, \citenamefont {{Bonnett}}, \citenamefont {{Chang}},
  \citenamefont {{Das}}, \citenamefont {{Dietrich}}, \citenamefont
  {{Drlica-Wagner}}, \citenamefont {{Eifler}},\ and\ \citenamefont {{et
  al.}}}]{Jarvis:2016}%
  \BibitemOpen
  \bibfield  {author} {\bibinfo {author} {\bibfnamefont {M.}~\bibnamefont
  {{Jarvis}}}, \bibinfo {author} {\bibfnamefont {E.}~\bibnamefont {{Sheldon}}},
  \bibinfo {author} {\bibfnamefont {J.}~\bibnamefont {{Zuntz}}}, \bibinfo
  {author} {\bibfnamefont {T.}~\bibnamefont {{Kacprzak}}}, \bibinfo {author}
  {\bibfnamefont {S.~L.}\ \bibnamefont {{Bridle}}}, \bibinfo {author}
  {\bibfnamefont {A.}~\bibnamefont {{Amara}}}, \bibinfo {author} {\bibfnamefont
  {R.}~\bibnamefont {{Armstrong}}}, \bibinfo {author} {\bibfnamefont {M.~R.}\
  \bibnamefont {{Becker}}}, \bibinfo {author} {\bibfnamefont {G.~M.}\
  \bibnamefont {{Bernstein}}}, \bibinfo {author} {\bibfnamefont
  {C.}~\bibnamefont {{Bonnett}}}, \bibinfo {author} {\bibfnamefont
  {C.}~\bibnamefont {{Chang}}}, \bibinfo {author} {\bibfnamefont
  {R.}~\bibnamefont {{Das}}}, \bibinfo {author} {\bibfnamefont {J.~P.}\
  \bibnamefont {{Dietrich}}}, \bibinfo {author} {\bibfnamefont
  {A.}~\bibnamefont {{Drlica-Wagner}}}, \bibinfo {author} {\bibfnamefont
  {T.~F.}\ \bibnamefont {{Eifler}}}, \ and\ \bibinfo {author} {\bibnamefont
  {{et al.}}},\ }\href {\doibase 10.1093/mnras/stw990} {\bibfield  {journal}
  {\bibinfo  {journal} {\mnras}\ }\textbf {\bibinfo {volume} {460}},\ \bibinfo
  {pages} {2245} (\bibinfo {year} {2016})},\ \Eprint
  {http://arxiv.org/abs/1507.05603} {arXiv:1507.05603 [astro-ph.IM]}
  \BibitemShut {NoStop}%
\bibitem [{\citenamefont {{Betoule}}\ \emph {et~al.}(2014)\citenamefont
  {{Betoule}}, \citenamefont {{Kessler}}, \citenamefont {{Guy}}, \citenamefont
  {{Mosher}}, \citenamefont {{Hardin}}, \citenamefont {{Biswas}}, \citenamefont
  {{Astier}}, \citenamefont {{El-Hage}}, \citenamefont {{Konig}}, \citenamefont
  {{Kuhlmann}}, \citenamefont {{Marriner}}, \citenamefont {{Pain}},
  \citenamefont {{Regnault}}, \citenamefont {{Balland}}, \citenamefont
  {{Bassett}},\ and\ \citenamefont {{et al.}}}]{Betoule:2014}%
  \BibitemOpen
  \bibfield  {author} {\bibinfo {author} {\bibfnamefont {M.}~\bibnamefont
  {{Betoule}}}, \bibinfo {author} {\bibfnamefont {R.}~\bibnamefont
  {{Kessler}}}, \bibinfo {author} {\bibfnamefont {J.}~\bibnamefont {{Guy}}},
  \bibinfo {author} {\bibfnamefont {J.}~\bibnamefont {{Mosher}}}, \bibinfo
  {author} {\bibfnamefont {D.}~\bibnamefont {{Hardin}}}, \bibinfo {author}
  {\bibfnamefont {R.}~\bibnamefont {{Biswas}}}, \bibinfo {author}
  {\bibfnamefont {P.}~\bibnamefont {{Astier}}}, \bibinfo {author}
  {\bibfnamefont {P.}~\bibnamefont {{El-Hage}}}, \bibinfo {author}
  {\bibfnamefont {M.}~\bibnamefont {{Konig}}}, \bibinfo {author} {\bibfnamefont
  {S.}~\bibnamefont {{Kuhlmann}}}, \bibinfo {author} {\bibfnamefont
  {J.}~\bibnamefont {{Marriner}}}, \bibinfo {author} {\bibfnamefont
  {R.}~\bibnamefont {{Pain}}}, \bibinfo {author} {\bibfnamefont
  {N.}~\bibnamefont {{Regnault}}}, \bibinfo {author} {\bibfnamefont
  {C.}~\bibnamefont {{Balland}}}, \bibinfo {author} {\bibfnamefont {B.~A.}\
  \bibnamefont {{Bassett}}}, \ and\ \bibinfo {author} {\bibnamefont {{et
  al.}}},\ }\href {\doibase 10.1051/0004-6361/201423413} {\bibfield  {journal}
  {\bibinfo  {journal} {\aap}\ }\textbf {\bibinfo {volume} {568}},\ \bibinfo
  {eid} {A22} (\bibinfo {year} {2014})},\ \Eprint
  {http://arxiv.org/abs/1401.4064} {arXiv:1401.4064} \BibitemShut {NoStop}%
\bibitem [{\citenamefont {{Riess}}\ \emph {et~al.}(2011)\citenamefont
  {{Riess}}, \citenamefont {{Macri}}, \citenamefont {{Casertano}},
  \citenamefont {{Lampeitl}}, \citenamefont {{Ferguson}}, \citenamefont
  {{Filippenko}}, \citenamefont {{Jha}}, \citenamefont {{Li}},\ and\
  \citenamefont {{Chornock}}}]{Riess:2011}%
  \BibitemOpen
  \bibfield  {author} {\bibinfo {author} {\bibfnamefont {A.~G.}\ \bibnamefont
  {{Riess}}}, \bibinfo {author} {\bibfnamefont {L.}~\bibnamefont {{Macri}}},
  \bibinfo {author} {\bibfnamefont {S.}~\bibnamefont {{Casertano}}}, \bibinfo
  {author} {\bibfnamefont {H.}~\bibnamefont {{Lampeitl}}}, \bibinfo {author}
  {\bibfnamefont {H.~C.}\ \bibnamefont {{Ferguson}}}, \bibinfo {author}
  {\bibfnamefont {A.~V.}\ \bibnamefont {{Filippenko}}}, \bibinfo {author}
  {\bibfnamefont {S.~W.}\ \bibnamefont {{Jha}}}, \bibinfo {author}
  {\bibfnamefont {W.}~\bibnamefont {{Li}}}, \ and\ \bibinfo {author}
  {\bibfnamefont {R.}~\bibnamefont {{Chornock}}},\ }\href {\doibase
  10.1088/0004-637X/730/2/119} {\bibfield  {journal} {\bibinfo  {journal}
  {\apj}\ }\textbf {\bibinfo {volume} {730}},\ \bibinfo {eid} {119} (\bibinfo
  {year} {2011})},\ \Eprint {http://arxiv.org/abs/1103.2976} {arXiv:1103.2976}
  \BibitemShut {NoStop}%
\bibitem [{\citenamefont {{Efstathiou}}(2014)}]{Efstathiou:2014}%
  \BibitemOpen
  \bibfield  {author} {\bibinfo {author} {\bibfnamefont {G.}~\bibnamefont
  {{Efstathiou}}},\ }\href {\doibase 10.1093/mnras/stu278} {\bibfield
  {journal} {\bibinfo  {journal} {\mnras}\ }\textbf {\bibinfo {volume} {440}},\
  \bibinfo {pages} {1138} (\bibinfo {year} {2014})},\ \Eprint
  {http://arxiv.org/abs/1311.3461} {arXiv:1311.3461} \BibitemShut {NoStop}%
\bibitem [{\citenamefont {{Giannantonio}}\ \emph {et~al.}(2014)\citenamefont
  {{Giannantonio}}, \citenamefont {{Ross}}, \citenamefont {{Percival}},
  \citenamefont {{Crittenden}}, \citenamefont {{Bacher}}, \citenamefont
  {{Kilbinger}}, \citenamefont {{Nichol}},\ and\ \citenamefont
  {{Weller}}}]{Giannantonio:2014aa}%
  \BibitemOpen
  \bibfield  {author} {\bibinfo {author} {\bibfnamefont {T.}~\bibnamefont
  {{Giannantonio}}}, \bibinfo {author} {\bibfnamefont {A.~J.}\ \bibnamefont
  {{Ross}}}, \bibinfo {author} {\bibfnamefont {W.~J.}\ \bibnamefont
  {{Percival}}}, \bibinfo {author} {\bibfnamefont {R.}~\bibnamefont
  {{Crittenden}}}, \bibinfo {author} {\bibfnamefont {D.}~\bibnamefont
  {{Bacher}}}, \bibinfo {author} {\bibfnamefont {M.}~\bibnamefont
  {{Kilbinger}}}, \bibinfo {author} {\bibfnamefont {R.}~\bibnamefont
  {{Nichol}}}, \ and\ \bibinfo {author} {\bibfnamefont {J.}~\bibnamefont
  {{Weller}}},\ }\href {\doibase 10.1103/PhysRevD.89.023511} {\bibfield
  {journal} {\bibinfo  {journal} {\prd}\ }\textbf {\bibinfo {volume} {89}},\
  \bibinfo {eid} {023511} (\bibinfo {year} {2014})},\ \Eprint
  {http://arxiv.org/abs/1303.1349} {arXiv:1303.1349 [astro-ph.CO]} \BibitemShut
  {NoStop}%
\bibitem [{\citenamefont {{Giannantonio}}\ and\ \citenamefont
  {{Percival}}(2014)}]{Giannantonio:2014ab}%
  \BibitemOpen
  \bibfield  {author} {\bibinfo {author} {\bibfnamefont {T.}~\bibnamefont
  {{Giannantonio}}}\ and\ \bibinfo {author} {\bibfnamefont {W.~J.}\
  \bibnamefont {{Percival}}},\ }\href {\doibase 10.1093/mnrasl/slu036}
  {\bibfield  {journal} {\bibinfo  {journal} {\mnras}\ }\textbf {\bibinfo
  {volume} {441}},\ \bibinfo {pages} {L16} (\bibinfo {year} {2014})},\ \Eprint
  {http://arxiv.org/abs/1312.5154} {arXiv:1312.5154} \BibitemShut {NoStop}%
\bibitem [{\citenamefont {{Soergel}}\ \emph {et~al.}(2015)\citenamefont
  {{Soergel}}, \citenamefont {{Giannantonio}}, \citenamefont {{Weller}},\ and\
  \citenamefont {{Battye}}}]{Soergel:2015}%
  \BibitemOpen
  \bibfield  {author} {\bibinfo {author} {\bibfnamefont {B.}~\bibnamefont
  {{Soergel}}}, \bibinfo {author} {\bibfnamefont {T.}~\bibnamefont
  {{Giannantonio}}}, \bibinfo {author} {\bibfnamefont {J.}~\bibnamefont
  {{Weller}}}, \ and\ \bibinfo {author} {\bibfnamefont {R.~A.}\ \bibnamefont
  {{Battye}}},\ }\href {\doibase 10.1088/1475-7516/2015/02/037} {\bibfield
  {journal} {\bibinfo  {journal} {\jcap}\ }\textbf {\bibinfo {volume} {2}},\
  \bibinfo {eid} {037} (\bibinfo {year} {2015})},\ \Eprint
  {http://arxiv.org/abs/1409.4540} {arXiv:1409.4540} \BibitemShut {NoStop}%
\bibitem [{\citenamefont {{Planck Collaboration}}\ \emph
  {et~al.}(2016{\natexlab{c}})\citenamefont {{Planck Collaboration}},
  \citenamefont {{Adam}}, \citenamefont {{Ade}}, \citenamefont {{Aghanim}},
  \citenamefont {{Arnaud}}, \citenamefont {{Ashdown}}, \citenamefont
  {{Aumont}}, \citenamefont {{Baccigalupi}}, \citenamefont {{Banday}},
  \citenamefont {{Barreiro}},\ and\ \citenamefont
  {et~al.}}]{Planck-Collaboration:2016ab}%
  \BibitemOpen
  \bibfield  {author} {\bibinfo {author} {\bibnamefont {{Planck
  Collaboration}}}, \bibinfo {author} {\bibfnamefont {R.}~\bibnamefont
  {{Adam}}}, \bibinfo {author} {\bibfnamefont {P.~A.~R.}\ \bibnamefont
  {{Ade}}}, \bibinfo {author} {\bibfnamefont {N.}~\bibnamefont {{Aghanim}}},
  \bibinfo {author} {\bibfnamefont {M.}~\bibnamefont {{Arnaud}}}, \bibinfo
  {author} {\bibfnamefont {M.}~\bibnamefont {{Ashdown}}}, \bibinfo {author}
  {\bibfnamefont {J.}~\bibnamefont {{Aumont}}}, \bibinfo {author}
  {\bibfnamefont {C.}~\bibnamefont {{Baccigalupi}}}, \bibinfo {author}
  {\bibfnamefont {A.~J.}\ \bibnamefont {{Banday}}}, \bibinfo {author}
  {\bibfnamefont {R.~B.}\ \bibnamefont {{Barreiro}}}, \ and\ \bibinfo {author}
  {\bibnamefont {et~al.}},\ }\href {\doibase 10.1051/0004-6361/201525936}
  {\bibfield  {journal} {\bibinfo  {journal} {\aap}\ }\textbf {\bibinfo
  {volume} {594}},\ \bibinfo {eid} {A9} (\bibinfo {year}
  {2016}{\natexlab{c}})},\ \Eprint {http://arxiv.org/abs/1502.05956}
  {arXiv:1502.05956} \BibitemShut {NoStop}%
\bibitem [{\citenamefont {{Ho}}\ \emph {et~al.}(2012)\citenamefont {{Ho}},
  \citenamefont {{Cuesta}}, \citenamefont {{Seo}}, \citenamefont {{de Putter}},
  \citenamefont {{Ross}}, \citenamefont {{White}}, \citenamefont
  {{Padmanabhan}}, \citenamefont {{Saito}}, \citenamefont {{Schlegel}},
  \citenamefont {{Schlafly}}, \citenamefont {{Seljak}}, \citenamefont
  {{Hern{\'a}ndez-Monteagudo}}, \citenamefont {{S{\'a}nchez}}, \citenamefont
  {{Percival}}, \citenamefont {{Blanton}},\ and\ \citenamefont {{et
  al.}}}]{Ho:2012}%
  \BibitemOpen
  \bibfield  {author} {\bibinfo {author} {\bibfnamefont {S.}~\bibnamefont
  {{Ho}}}, \bibinfo {author} {\bibfnamefont {A.}~\bibnamefont {{Cuesta}}},
  \bibinfo {author} {\bibfnamefont {H.-J.}\ \bibnamefont {{Seo}}}, \bibinfo
  {author} {\bibfnamefont {R.}~\bibnamefont {{de Putter}}}, \bibinfo {author}
  {\bibfnamefont {A.~J.}\ \bibnamefont {{Ross}}}, \bibinfo {author}
  {\bibfnamefont {M.}~\bibnamefont {{White}}}, \bibinfo {author} {\bibfnamefont
  {N.}~\bibnamefont {{Padmanabhan}}}, \bibinfo {author} {\bibfnamefont
  {S.}~\bibnamefont {{Saito}}}, \bibinfo {author} {\bibfnamefont {D.~J.}\
  \bibnamefont {{Schlegel}}}, \bibinfo {author} {\bibfnamefont
  {E.}~\bibnamefont {{Schlafly}}}, \bibinfo {author} {\bibfnamefont
  {U.}~\bibnamefont {{Seljak}}}, \bibinfo {author} {\bibfnamefont
  {C.}~\bibnamefont {{Hern{\'a}ndez-Monteagudo}}}, \bibinfo {author}
  {\bibfnamefont {A.~G.}\ \bibnamefont {{S{\'a}nchez}}}, \bibinfo {author}
  {\bibfnamefont {W.~J.}\ \bibnamefont {{Percival}}}, \bibinfo {author}
  {\bibfnamefont {M.}~\bibnamefont {{Blanton}}}, \ and\ \bibinfo {author}
  {\bibnamefont {{et al.}}},\ }\href {\doibase 10.1088/0004-637X/761/1/14}
  {\bibfield  {journal} {\bibinfo  {journal} {\apj}\ }\textbf {\bibinfo
  {volume} {761}},\ \bibinfo {eid} {14} (\bibinfo {year} {2012})},\ \Eprint
  {http://arxiv.org/abs/1201.2137} {arXiv:1201.2137} \BibitemShut {NoStop}%
\bibitem [{\citenamefont {{Ross}}\ \emph {et~al.}(2011)\citenamefont {{Ross}},
  \citenamefont {{Ho}}, \citenamefont {{Cuesta}}, \citenamefont {{Tojeiro}},
  \citenamefont {{Percival}}, \citenamefont {{Wake}}, \citenamefont
  {{Masters}}, \citenamefont {{Nichol}}, \citenamefont {{Myers}}, \citenamefont
  {{de Simoni}}, \citenamefont {{Seo}}, \citenamefont
  {{Hern{\'a}ndez-Monteagudo}}, \citenamefont {{Crittenden}}, \citenamefont
  {{Blanton}}, \citenamefont {{Brinkmann}},\ and\ \citenamefont {{et
  al.}}}]{Ross:2011}%
  \BibitemOpen
  \bibfield  {author} {\bibinfo {author} {\bibfnamefont {A.~J.}\ \bibnamefont
  {{Ross}}}, \bibinfo {author} {\bibfnamefont {S.}~\bibnamefont {{Ho}}},
  \bibinfo {author} {\bibfnamefont {A.~J.}\ \bibnamefont {{Cuesta}}}, \bibinfo
  {author} {\bibfnamefont {R.}~\bibnamefont {{Tojeiro}}}, \bibinfo {author}
  {\bibfnamefont {W.~J.}\ \bibnamefont {{Percival}}}, \bibinfo {author}
  {\bibfnamefont {D.}~\bibnamefont {{Wake}}}, \bibinfo {author} {\bibfnamefont
  {K.~L.}\ \bibnamefont {{Masters}}}, \bibinfo {author} {\bibfnamefont {R.~C.}\
  \bibnamefont {{Nichol}}}, \bibinfo {author} {\bibfnamefont {A.~D.}\
  \bibnamefont {{Myers}}}, \bibinfo {author} {\bibfnamefont {F.}~\bibnamefont
  {{de Simoni}}}, \bibinfo {author} {\bibfnamefont {H.~J.}\ \bibnamefont
  {{Seo}}}, \bibinfo {author} {\bibfnamefont {C.}~\bibnamefont
  {{Hern{\'a}ndez-Monteagudo}}}, \bibinfo {author} {\bibfnamefont
  {R.}~\bibnamefont {{Crittenden}}}, \bibinfo {author} {\bibfnamefont
  {M.}~\bibnamefont {{Blanton}}}, \bibinfo {author} {\bibfnamefont
  {J.}~\bibnamefont {{Brinkmann}}}, \ and\ \bibinfo {author} {\bibnamefont {{et
  al.}}},\ }\href {\doibase 10.1111/j.1365-2966.2011.19351.x} {\bibfield
  {journal} {\bibinfo  {journal} {\mnras}\ }\textbf {\bibinfo {volume} {417}},\
  \bibinfo {pages} {1350} (\bibinfo {year} {2011})},\ \Eprint
  {http://arxiv.org/abs/1105.2320} {arXiv:1105.2320} \BibitemShut {NoStop}%
\bibitem [{\citenamefont {{Lin}}\ \emph {et~al.}(2012)\citenamefont {{Lin}},
  \citenamefont {{Dodelson}}, \citenamefont {{Seo}}, \citenamefont
  {{Soares-Santos}}, \citenamefont {{Annis}}, \citenamefont {{Hao}},
  \citenamefont {{Johnston}}, \citenamefont {{Kubo}}, \citenamefont {{Reis}},\
  and\ \citenamefont {{Simet}}}]{Lin:2012}%
  \BibitemOpen
  \bibfield  {author} {\bibinfo {author} {\bibfnamefont {H.}~\bibnamefont
  {{Lin}}}, \bibinfo {author} {\bibfnamefont {S.}~\bibnamefont {{Dodelson}}},
  \bibinfo {author} {\bibfnamefont {H.-J.}\ \bibnamefont {{Seo}}}, \bibinfo
  {author} {\bibfnamefont {M.}~\bibnamefont {{Soares-Santos}}}, \bibinfo
  {author} {\bibfnamefont {J.}~\bibnamefont {{Annis}}}, \bibinfo {author}
  {\bibfnamefont {J.}~\bibnamefont {{Hao}}}, \bibinfo {author} {\bibfnamefont
  {D.}~\bibnamefont {{Johnston}}}, \bibinfo {author} {\bibfnamefont {J.~M.}\
  \bibnamefont {{Kubo}}}, \bibinfo {author} {\bibfnamefont {R.~R.~R.}\
  \bibnamefont {{Reis}}}, \ and\ \bibinfo {author} {\bibfnamefont
  {M.}~\bibnamefont {{Simet}}},\ }\href {\doibase 10.1088/0004-637X/761/1/15}
  {\bibfield  {journal} {\bibinfo  {journal} {\apj}\ }\textbf {\bibinfo
  {volume} {761}},\ \bibinfo {eid} {15} (\bibinfo {year} {2012})},\ \Eprint
  {http://arxiv.org/abs/1111.6622} {arXiv:1111.6622} \BibitemShut {NoStop}%
\bibitem [{\citenamefont {{Humphreys}}\ \emph {et~al.}(2013)\citenamefont
  {{Humphreys}}, \citenamefont {{Reid}}, \citenamefont {{Moran}}, \citenamefont
  {{Greenhill}},\ and\ \citenamefont {{Argon}}}]{Humphreys:2013}%
  \BibitemOpen
  \bibfield  {author} {\bibinfo {author} {\bibfnamefont {E.~M.~L.}\
  \bibnamefont {{Humphreys}}}, \bibinfo {author} {\bibfnamefont {M.~J.}\
  \bibnamefont {{Reid}}}, \bibinfo {author} {\bibfnamefont {J.~M.}\
  \bibnamefont {{Moran}}}, \bibinfo {author} {\bibfnamefont {L.~J.}\
  \bibnamefont {{Greenhill}}}, \ and\ \bibinfo {author} {\bibfnamefont {A.~L.}\
  \bibnamefont {{Argon}}},\ }\href {\doibase 10.1088/0004-637X/775/1/13}
  {\bibfield  {journal} {\bibinfo  {journal} {\apj}\ }\textbf {\bibinfo
  {volume} {775}},\ \bibinfo {eid} {13} (\bibinfo {year} {2013})},\ \Eprint
  {http://arxiv.org/abs/1307.6031} {arXiv:1307.6031} \BibitemShut {NoStop}%
\bibitem [{\citenamefont {{Flaugher}}\ \emph {et~al.}(2015)\citenamefont
  {{Flaugher}}, \citenamefont {{Diehl}}, \citenamefont {{Honscheid}},
  \citenamefont {{Abbott}}, \citenamefont {{Alvarez}}, \citenamefont
  {{Angstadt}}, \citenamefont {{Annis}}, \citenamefont {{Antonik}},
  \citenamefont {{Ballester}}, \citenamefont {{Beaufore}}, \citenamefont
  {{Bernstein}}, \citenamefont {{Bernstein}}, \citenamefont {{Bigelow}},
  \citenamefont {{Bonati}}, \citenamefont {{Boprie}},\ and\ \citenamefont {{et
  al.}}}]{Flaugher:2015}%
  \BibitemOpen
  \bibfield  {author} {\bibinfo {author} {\bibfnamefont {B.}~\bibnamefont
  {{Flaugher}}}, \bibinfo {author} {\bibfnamefont {H.~T.}\ \bibnamefont
  {{Diehl}}}, \bibinfo {author} {\bibfnamefont {K.}~\bibnamefont
  {{Honscheid}}}, \bibinfo {author} {\bibfnamefont {T.~M.~C.}\ \bibnamefont
  {{Abbott}}}, \bibinfo {author} {\bibfnamefont {O.}~\bibnamefont {{Alvarez}}},
  \bibinfo {author} {\bibfnamefont {R.}~\bibnamefont {{Angstadt}}}, \bibinfo
  {author} {\bibfnamefont {J.~T.}\ \bibnamefont {{Annis}}}, \bibinfo {author}
  {\bibfnamefont {M.}~\bibnamefont {{Antonik}}}, \bibinfo {author}
  {\bibfnamefont {O.}~\bibnamefont {{Ballester}}}, \bibinfo {author}
  {\bibfnamefont {L.}~\bibnamefont {{Beaufore}}}, \bibinfo {author}
  {\bibfnamefont {G.~M.}\ \bibnamefont {{Bernstein}}}, \bibinfo {author}
  {\bibfnamefont {R.~A.}\ \bibnamefont {{Bernstein}}}, \bibinfo {author}
  {\bibfnamefont {B.}~\bibnamefont {{Bigelow}}}, \bibinfo {author}
  {\bibfnamefont {M.}~\bibnamefont {{Bonati}}}, \bibinfo {author}
  {\bibfnamefont {D.}~\bibnamefont {{Boprie}}}, \ and\ \bibinfo {author}
  {\bibnamefont {{et al.}}},\ }\href {\doibase 10.1088/0004-6256/150/5/150}
  {\bibfield  {journal} {\bibinfo  {journal} {\aj}\ }\textbf {\bibinfo {volume}
  {150}},\ \bibinfo {eid} {150} (\bibinfo {year} {2015})},\ \Eprint
  {http://arxiv.org/abs/1504.02900} {arXiv:1504.02900 [astro-ph.IM]}
  \BibitemShut {NoStop}%
\bibitem [{\citenamefont {{Dark Energy Survey Collaboration}}\ \emph
  {et~al.}(2016)\citenamefont {{Dark Energy Survey Collaboration}},
  \citenamefont {{Abbott}}, \citenamefont {{Abdalla}}, \citenamefont
  {{Aleksi{\'c}}}, \citenamefont {{Allam}}, \citenamefont {{Amara}},
  \citenamefont {{Bacon}}, \citenamefont {{Balbinot}}, \citenamefont
  {{Banerji}}, \citenamefont {{Bechtol}}, \citenamefont {{Benoit-L{\'e}vy}},
  \citenamefont {{Bernstein}}, \citenamefont {{Bertin}}, \citenamefont
  {{Blazek}}, \citenamefont {{Bonnett}}, \citenamefont {{Bridle}},\ and\
  \citenamefont {{et al.}}}]{DESCollaboration:2016aa}%
  \BibitemOpen
  \bibfield  {author} {\bibinfo {author} {\bibnamefont {{Dark Energy Survey
  Collaboration}}}, \bibinfo {author} {\bibfnamefont {T.}~\bibnamefont
  {{Abbott}}}, \bibinfo {author} {\bibfnamefont {F.~B.}\ \bibnamefont
  {{Abdalla}}}, \bibinfo {author} {\bibfnamefont {J.}~\bibnamefont
  {{Aleksi{\'c}}}}, \bibinfo {author} {\bibfnamefont {S.}~\bibnamefont
  {{Allam}}}, \bibinfo {author} {\bibfnamefont {A.}~\bibnamefont {{Amara}}},
  \bibinfo {author} {\bibfnamefont {D.}~\bibnamefont {{Bacon}}}, \bibinfo
  {author} {\bibfnamefont {E.}~\bibnamefont {{Balbinot}}}, \bibinfo {author}
  {\bibfnamefont {M.}~\bibnamefont {{Banerji}}}, \bibinfo {author}
  {\bibfnamefont {K.}~\bibnamefont {{Bechtol}}}, \bibinfo {author}
  {\bibfnamefont {A.}~\bibnamefont {{Benoit-L{\'e}vy}}}, \bibinfo {author}
  {\bibfnamefont {G.~M.}\ \bibnamefont {{Bernstein}}}, \bibinfo {author}
  {\bibfnamefont {E.}~\bibnamefont {{Bertin}}}, \bibinfo {author}
  {\bibfnamefont {J.}~\bibnamefont {{Blazek}}}, \bibinfo {author}
  {\bibfnamefont {C.}~\bibnamefont {{Bonnett}}}, \bibinfo {author}
  {\bibfnamefont {S.}~\bibnamefont {{Bridle}}}, \ and\ \bibinfo {author}
  {\bibnamefont {{et al.}}},\ }\href {\doibase 10.1093/mnras/stw641} {\bibfield
   {journal} {\bibinfo  {journal} {\mnras}\ }\textbf {\bibinfo {volume}
  {460}},\ \bibinfo {pages} {1270} (\bibinfo {year} {2016})},\ \Eprint
  {http://arxiv.org/abs/1601.00329} {arXiv:1601.00329} \BibitemShut {NoStop}%
\bibitem [{\citenamefont {{Sheldon}}(2014)}]{Sheldon:2014}%
  \BibitemOpen
  \bibfield  {author} {\bibinfo {author} {\bibfnamefont {E.~S.}\ \bibnamefont
  {{Sheldon}}},\ }\href {\doibase 10.1093/mnrasl/slu104} {\bibfield  {journal}
  {\bibinfo  {journal} {\mnras}\ }\textbf {\bibinfo {volume} {444}},\ \bibinfo
  {pages} {L25} (\bibinfo {year} {2014})},\ \Eprint
  {http://arxiv.org/abs/1403.7669} {arXiv:1403.7669} \BibitemShut {NoStop}%
\bibitem [{\citenamefont {{Zuntz}}\ \emph {et~al.}(2013)\citenamefont
  {{Zuntz}}, \citenamefont {{Kacprzak}}, \citenamefont {{Voigt}}, \citenamefont
  {{Hirsch}}, \citenamefont {{Rowe}},\ and\ \citenamefont
  {{Bridle}}}]{Zuntz:2013}%
  \BibitemOpen
  \bibfield  {author} {\bibinfo {author} {\bibfnamefont {J.}~\bibnamefont
  {{Zuntz}}}, \bibinfo {author} {\bibfnamefont {T.}~\bibnamefont {{Kacprzak}}},
  \bibinfo {author} {\bibfnamefont {L.}~\bibnamefont {{Voigt}}}, \bibinfo
  {author} {\bibfnamefont {M.}~\bibnamefont {{Hirsch}}}, \bibinfo {author}
  {\bibfnamefont {B.}~\bibnamefont {{Rowe}}}, \ and\ \bibinfo {author}
  {\bibfnamefont {S.}~\bibnamefont {{Bridle}}},\ }\href {\doibase
  10.1093/mnras/stt1125} {\bibfield  {journal} {\bibinfo  {journal} {\mnras}\
  }\textbf {\bibinfo {volume} {434}},\ \bibinfo {pages} {1604} (\bibinfo {year}
  {2013})},\ \Eprint {http://arxiv.org/abs/1302.0183} {arXiv:1302.0183}
  \BibitemShut {NoStop}%
\bibitem [{\citenamefont {{Bonnett}}\ \emph {et~al.}(2016)\citenamefont
  {{Bonnett}}, \citenamefont {{Troxel}}, \citenamefont {{Hartley}},
  \citenamefont {{Amara}}, \citenamefont {{Leistedt}}, \citenamefont
  {{Becker}}, \citenamefont {{Bernstein}}, \citenamefont {{Bridle}},
  \citenamefont {{Bruderer}}, \citenamefont {{Busha}}, \citenamefont {{Carrasco
  Kind}}, \citenamefont {{Childress}}, \citenamefont {{Castander}},
  \citenamefont {{Chang}}, \citenamefont {{Crocce}},\ and\ \citenamefont {{et
  al.}}}]{Bonnett:2016}%
  \BibitemOpen
  \bibfield  {author} {\bibinfo {author} {\bibfnamefont {C.}~\bibnamefont
  {{Bonnett}}}, \bibinfo {author} {\bibfnamefont {M.~A.}\ \bibnamefont
  {{Troxel}}}, \bibinfo {author} {\bibfnamefont {W.}~\bibnamefont {{Hartley}}},
  \bibinfo {author} {\bibfnamefont {A.}~\bibnamefont {{Amara}}}, \bibinfo
  {author} {\bibfnamefont {B.}~\bibnamefont {{Leistedt}}}, \bibinfo {author}
  {\bibfnamefont {M.~R.}\ \bibnamefont {{Becker}}}, \bibinfo {author}
  {\bibfnamefont {G.~M.}\ \bibnamefont {{Bernstein}}}, \bibinfo {author}
  {\bibfnamefont {S.~L.}\ \bibnamefont {{Bridle}}}, \bibinfo {author}
  {\bibfnamefont {C.}~\bibnamefont {{Bruderer}}}, \bibinfo {author}
  {\bibfnamefont {M.~T.}\ \bibnamefont {{Busha}}}, \bibinfo {author}
  {\bibfnamefont {M.}~\bibnamefont {{Carrasco Kind}}}, \bibinfo {author}
  {\bibfnamefont {M.~J.}\ \bibnamefont {{Childress}}}, \bibinfo {author}
  {\bibfnamefont {F.~J.}\ \bibnamefont {{Castander}}}, \bibinfo {author}
  {\bibfnamefont {C.}~\bibnamefont {{Chang}}}, \bibinfo {author} {\bibfnamefont
  {M.}~\bibnamefont {{Crocce}}}, \ and\ \bibinfo {author} {\bibnamefont {{et
  al.}}},\ }\href {\doibase 10.1103/PhysRevD.94.042005} {\bibfield  {journal}
  {\bibinfo  {journal} {\prd}\ }\textbf {\bibinfo {volume} {94}},\ \bibinfo
  {eid} {042005} (\bibinfo {year} {2016})},\ \Eprint
  {http://arxiv.org/abs/1507.05909} {arXiv:1507.05909} \BibitemShut {NoStop}%
\bibitem [{\citenamefont {{Becker}}\ \emph {et~al.}(2016)\citenamefont
  {{Becker}}, \citenamefont {{Troxel}}, \citenamefont {{MacCrann}},
  \citenamefont {{Krause}}, \citenamefont {{Eifler}}, \citenamefont
  {{Friedrich}}, \citenamefont {{Nicola}}, \citenamefont {{Refregier}},
  \citenamefont {{Amara}}, \citenamefont {{Bacon}}, \citenamefont
  {{Bernstein}}, \citenamefont {{Bonnett}}, \citenamefont {{Bridle}},
  \citenamefont {{Busha}}, \citenamefont {{Chang}},\ and\ \citenamefont {{et
  al.}}}]{Becker:2016}%
  \BibitemOpen
  \bibfield  {author} {\bibinfo {author} {\bibfnamefont {M.~R.}\ \bibnamefont
  {{Becker}}}, \bibinfo {author} {\bibfnamefont {M.~A.}\ \bibnamefont
  {{Troxel}}}, \bibinfo {author} {\bibfnamefont {N.}~\bibnamefont
  {{MacCrann}}}, \bibinfo {author} {\bibfnamefont {E.}~\bibnamefont
  {{Krause}}}, \bibinfo {author} {\bibfnamefont {T.~F.}\ \bibnamefont
  {{Eifler}}}, \bibinfo {author} {\bibfnamefont {O.}~\bibnamefont
  {{Friedrich}}}, \bibinfo {author} {\bibfnamefont {A.}~\bibnamefont
  {{Nicola}}}, \bibinfo {author} {\bibfnamefont {A.}~\bibnamefont
  {{Refregier}}}, \bibinfo {author} {\bibfnamefont {A.}~\bibnamefont
  {{Amara}}}, \bibinfo {author} {\bibfnamefont {D.}~\bibnamefont {{Bacon}}},
  \bibinfo {author} {\bibfnamefont {G.~M.}\ \bibnamefont {{Bernstein}}},
  \bibinfo {author} {\bibfnamefont {C.}~\bibnamefont {{Bonnett}}}, \bibinfo
  {author} {\bibfnamefont {S.~L.}\ \bibnamefont {{Bridle}}}, \bibinfo {author}
  {\bibfnamefont {M.~T.}\ \bibnamefont {{Busha}}}, \bibinfo {author}
  {\bibfnamefont {C.}~\bibnamefont {{Chang}}}, \ and\ \bibinfo {author}
  {\bibnamefont {{et al.}}},\ }\href {\doibase 10.1103/PhysRevD.94.022002}
  {\bibfield  {journal} {\bibinfo  {journal} {\prd}\ }\textbf {\bibinfo
  {volume} {94}},\ \bibinfo {eid} {022002} (\bibinfo {year} {2016})},\ \Eprint
  {http://arxiv.org/abs/1507.05598} {arXiv:1507.05598} \BibitemShut {NoStop}%
\bibitem [{\citenamefont {{The Dark Energy Survey Collaboration}}\ \emph
  {et~al.}(2015)\citenamefont {{The Dark Energy Survey Collaboration}},
  \citenamefont {{Abbott}}, \citenamefont {{Abdalla}}, \citenamefont {{Allam}},
  \citenamefont {{Amara}}, \citenamefont {{Annis}}, \citenamefont
  {{Armstrong}}, \citenamefont {{Bacon}}, \citenamefont {{Banerji}},
  \citenamefont {{Bauer}}, \citenamefont {{Baxter}}, \citenamefont {{Becker}},
  \citenamefont {{Benoit-L{\'e}vy}}, \citenamefont {{Bernstein}}, \citenamefont
  {{Bernstein}},\ and\ \citenamefont {et~al.}}]{DES-Collaboration:2015}%
  \BibitemOpen
  \bibfield  {author} {\bibinfo {author} {\bibnamefont {{The Dark Energy Survey
  Collaboration}}}, \bibinfo {author} {\bibfnamefont {T.}~\bibnamefont
  {{Abbott}}}, \bibinfo {author} {\bibfnamefont {F.~B.}\ \bibnamefont
  {{Abdalla}}}, \bibinfo {author} {\bibfnamefont {S.}~\bibnamefont {{Allam}}},
  \bibinfo {author} {\bibfnamefont {A.}~\bibnamefont {{Amara}}}, \bibinfo
  {author} {\bibfnamefont {J.}~\bibnamefont {{Annis}}}, \bibinfo {author}
  {\bibfnamefont {R.}~\bibnamefont {{Armstrong}}}, \bibinfo {author}
  {\bibfnamefont {D.}~\bibnamefont {{Bacon}}}, \bibinfo {author} {\bibfnamefont
  {M.}~\bibnamefont {{Banerji}}}, \bibinfo {author} {\bibfnamefont {A.~H.}\
  \bibnamefont {{Bauer}}}, \bibinfo {author} {\bibfnamefont {E.}~\bibnamefont
  {{Baxter}}}, \bibinfo {author} {\bibfnamefont {M.~R.}\ \bibnamefont
  {{Becker}}}, \bibinfo {author} {\bibfnamefont {A.}~\bibnamefont
  {{Benoit-L{\'e}vy}}}, \bibinfo {author} {\bibfnamefont {R.~A.}\ \bibnamefont
  {{Bernstein}}}, \bibinfo {author} {\bibfnamefont {G.~M.}\ \bibnamefont
  {{Bernstein}}}, \ and\ \bibinfo {author} {\bibnamefont {et~al.}},\
  }\href@noop {} {\bibfield  {journal} {\bibinfo  {journal} {ArXiv e-prints}\ }
  (\bibinfo {year} {2015})},\ \Eprint {http://arxiv.org/abs/1507.05552}
  {arXiv:1507.05552} \BibitemShut {NoStop}%
\bibitem [{\citenamefont {{G{\'o}rski}}\ \emph {et~al.}(2005)\citenamefont
  {{G{\'o}rski}}, \citenamefont {{Hivon}}, \citenamefont {{Banday}},
  \citenamefont {{Wandelt}}, \citenamefont {{Hansen}}, \citenamefont
  {{Reinecke}},\ and\ \citenamefont {{Bartelmann}}}]{Gorski:2005}%
  \BibitemOpen
  \bibfield  {author} {\bibinfo {author} {\bibfnamefont {K.~M.}\ \bibnamefont
  {{G{\'o}rski}}}, \bibinfo {author} {\bibfnamefont {E.}~\bibnamefont
  {{Hivon}}}, \bibinfo {author} {\bibfnamefont {A.~J.}\ \bibnamefont
  {{Banday}}}, \bibinfo {author} {\bibfnamefont {B.~D.}\ \bibnamefont
  {{Wandelt}}}, \bibinfo {author} {\bibfnamefont {F.~K.}\ \bibnamefont
  {{Hansen}}}, \bibinfo {author} {\bibfnamefont {M.}~\bibnamefont
  {{Reinecke}}}, \ and\ \bibinfo {author} {\bibfnamefont {M.}~\bibnamefont
  {{Bartelmann}}},\ }\href {\doibase 10.1086/427976} {\bibfield  {journal}
  {\bibinfo  {journal} {\apj}\ }\textbf {\bibinfo {volume} {622}},\ \bibinfo
  {pages} {759} (\bibinfo {year} {2005})},\ \Eprint
  {http://arxiv.org/abs/astro-ph/0409513} {astro-ph/0409513} \BibitemShut
  {NoStop}%
\bibitem [{\citenamefont {{Okamoto}}\ and\ \citenamefont
  {{Hu}}(2003)}]{Okamoto:2003}%
  \BibitemOpen
  \bibfield  {author} {\bibinfo {author} {\bibfnamefont {T.}~\bibnamefont
  {{Okamoto}}}\ and\ \bibinfo {author} {\bibfnamefont {W.}~\bibnamefont
  {{Hu}}},\ }\href {\doibase 10.1103/PhysRevD.67.083002} {\bibfield  {journal}
  {\bibinfo  {journal} {\prd}\ }\textbf {\bibinfo {volume} {67}},\ \bibinfo
  {eid} {083002} (\bibinfo {year} {2003})},\ \Eprint
  {http://arxiv.org/abs/astro-ph/0301031} {astro-ph/0301031} \BibitemShut
  {NoStop}%
\bibitem [{\citenamefont {{Frieman}}\ \emph {et~al.}(2008)\citenamefont
  {{Frieman}}, \citenamefont {{Bassett}}, \citenamefont {{Becker}},
  \citenamefont {{Choi}}, \citenamefont {{Cinabro}}, \citenamefont {{DeJongh}},
  \citenamefont {{Depoy}}, \citenamefont {{Dilday}}, \citenamefont {{Doi}},
  \citenamefont {{Garnavich}}, \citenamefont {{Hogan}}, \citenamefont
  {{Holtzman}}, \citenamefont {{Im}}, \citenamefont {{Jha}}, \citenamefont
  {{Kessler}},\ and\ \citenamefont {{et al.}}}]{Frieman:2008}%
  \BibitemOpen
  \bibfield  {author} {\bibinfo {author} {\bibfnamefont {J.~A.}\ \bibnamefont
  {{Frieman}}}, \bibinfo {author} {\bibfnamefont {B.}~\bibnamefont
  {{Bassett}}}, \bibinfo {author} {\bibfnamefont {A.}~\bibnamefont {{Becker}}},
  \bibinfo {author} {\bibfnamefont {C.}~\bibnamefont {{Choi}}}, \bibinfo
  {author} {\bibfnamefont {D.}~\bibnamefont {{Cinabro}}}, \bibinfo {author}
  {\bibfnamefont {F.}~\bibnamefont {{DeJongh}}}, \bibinfo {author}
  {\bibfnamefont {D.~L.}\ \bibnamefont {{Depoy}}}, \bibinfo {author}
  {\bibfnamefont {B.}~\bibnamefont {{Dilday}}}, \bibinfo {author}
  {\bibfnamefont {M.}~\bibnamefont {{Doi}}}, \bibinfo {author} {\bibfnamefont
  {P.~M.}\ \bibnamefont {{Garnavich}}}, \bibinfo {author} {\bibfnamefont
  {C.~J.}\ \bibnamefont {{Hogan}}}, \bibinfo {author} {\bibfnamefont
  {J.}~\bibnamefont {{Holtzman}}}, \bibinfo {author} {\bibfnamefont
  {M.}~\bibnamefont {{Im}}}, \bibinfo {author} {\bibfnamefont {S.}~\bibnamefont
  {{Jha}}}, \bibinfo {author} {\bibfnamefont {R.}~\bibnamefont {{Kessler}}}, \
  and\ \bibinfo {author} {\bibnamefont {{et al.}}},\ }\href {\doibase
  10.1088/0004-6256/135/1/338} {\bibfield  {journal} {\bibinfo  {journal}
  {\aj}\ }\textbf {\bibinfo {volume} {135}},\ \bibinfo {pages} {338} (\bibinfo
  {year} {2008})},\ \Eprint {http://arxiv.org/abs/0708.2749} {arXiv:0708.2749}
  \BibitemShut {NoStop}%
\bibitem [{\citenamefont {{Kessler}}\ \emph {et~al.}(2009)\citenamefont
  {{Kessler}}, \citenamefont {{Becker}}, \citenamefont {{Cinabro}},
  \citenamefont {{Vanderplas}}, \citenamefont {{Frieman}}, \citenamefont
  {{Marriner}}, \citenamefont {{Davis}}, \citenamefont {{Dilday}},
  \citenamefont {{Holtzman}}, \citenamefont {{Jha}}, \citenamefont
  {{Lampeitl}}, \citenamefont {{Sako}}, \citenamefont {{Smith}}, \citenamefont
  {{Zheng}}, \citenamefont {{Nichol}},\ and\ \citenamefont {{et
  al.}}}]{Kessler:2009}%
  \BibitemOpen
  \bibfield  {author} {\bibinfo {author} {\bibfnamefont {R.}~\bibnamefont
  {{Kessler}}}, \bibinfo {author} {\bibfnamefont {A.~C.}\ \bibnamefont
  {{Becker}}}, \bibinfo {author} {\bibfnamefont {D.}~\bibnamefont {{Cinabro}}},
  \bibinfo {author} {\bibfnamefont {J.}~\bibnamefont {{Vanderplas}}}, \bibinfo
  {author} {\bibfnamefont {J.~A.}\ \bibnamefont {{Frieman}}}, \bibinfo {author}
  {\bibfnamefont {J.}~\bibnamefont {{Marriner}}}, \bibinfo {author}
  {\bibfnamefont {T.~M.}\ \bibnamefont {{Davis}}}, \bibinfo {author}
  {\bibfnamefont {B.}~\bibnamefont {{Dilday}}}, \bibinfo {author}
  {\bibfnamefont {J.}~\bibnamefont {{Holtzman}}}, \bibinfo {author}
  {\bibfnamefont {S.~W.}\ \bibnamefont {{Jha}}}, \bibinfo {author}
  {\bibfnamefont {H.}~\bibnamefont {{Lampeitl}}}, \bibinfo {author}
  {\bibfnamefont {M.}~\bibnamefont {{Sako}}}, \bibinfo {author} {\bibfnamefont
  {M.}~\bibnamefont {{Smith}}}, \bibinfo {author} {\bibfnamefont
  {C.}~\bibnamefont {{Zheng}}}, \bibinfo {author} {\bibfnamefont {R.~C.}\
  \bibnamefont {{Nichol}}}, \ and\ \bibinfo {author} {\bibnamefont {{et
  al.}}},\ }\href {\doibase 10.1088/0067-0049/185/1/32} {\bibfield  {journal}
  {\bibinfo  {journal} {\apjs}\ }\textbf {\bibinfo {volume} {185}},\ \bibinfo
  {pages} {32} (\bibinfo {year} {2009})},\ \Eprint
  {http://arxiv.org/abs/0908.4274} {arXiv:0908.4274} \BibitemShut {NoStop}%
\bibitem [{\citenamefont {{Sollerman}}\ \emph {et~al.}(2009)\citenamefont
  {{Sollerman}}, \citenamefont {{M{\"o}rtsell}}, \citenamefont {{Davis}},
  \citenamefont {{Blomqvist}}, \citenamefont {{Bassett}}, \citenamefont
  {{Becker}}, \citenamefont {{Cinabro}}, \citenamefont {{Filippenko}},
  \citenamefont {{Foley}}, \citenamefont {{Frieman}}, \citenamefont
  {{Garnavich}}, \citenamefont {{Lampeitl}}, \citenamefont {{Marriner}},
  \citenamefont {{Miquel}}, \citenamefont {{Nichol}},\ and\ \citenamefont {{et
  al.}}}]{Sollerman:2009}%
  \BibitemOpen
  \bibfield  {author} {\bibinfo {author} {\bibfnamefont {J.}~\bibnamefont
  {{Sollerman}}}, \bibinfo {author} {\bibfnamefont {E.}~\bibnamefont
  {{M{\"o}rtsell}}}, \bibinfo {author} {\bibfnamefont {T.~M.}\ \bibnamefont
  {{Davis}}}, \bibinfo {author} {\bibfnamefont {M.}~\bibnamefont
  {{Blomqvist}}}, \bibinfo {author} {\bibfnamefont {B.}~\bibnamefont
  {{Bassett}}}, \bibinfo {author} {\bibfnamefont {A.~C.}\ \bibnamefont
  {{Becker}}}, \bibinfo {author} {\bibfnamefont {D.}~\bibnamefont {{Cinabro}}},
  \bibinfo {author} {\bibfnamefont {A.~V.}\ \bibnamefont {{Filippenko}}},
  \bibinfo {author} {\bibfnamefont {R.~J.}\ \bibnamefont {{Foley}}}, \bibinfo
  {author} {\bibfnamefont {J.}~\bibnamefont {{Frieman}}}, \bibinfo {author}
  {\bibfnamefont {P.}~\bibnamefont {{Garnavich}}}, \bibinfo {author}
  {\bibfnamefont {H.}~\bibnamefont {{Lampeitl}}}, \bibinfo {author}
  {\bibfnamefont {J.}~\bibnamefont {{Marriner}}}, \bibinfo {author}
  {\bibfnamefont {R.}~\bibnamefont {{Miquel}}}, \bibinfo {author}
  {\bibfnamefont {R.~C.}\ \bibnamefont {{Nichol}}}, \ and\ \bibinfo {author}
  {\bibnamefont {{et al.}}},\ }\href {\doibase 10.1088/0004-637X/703/2/1374}
  {\bibfield  {journal} {\bibinfo  {journal} {\apj}\ }\textbf {\bibinfo
  {volume} {703}},\ \bibinfo {pages} {1374} (\bibinfo {year} {2009})},\ \Eprint
  {http://arxiv.org/abs/0908.4276} {arXiv:0908.4276} \BibitemShut {NoStop}%
\bibitem [{\citenamefont {{Lampeitl}}\ \emph {et~al.}(2010)\citenamefont
  {{Lampeitl}}, \citenamefont {{Nichol}}, \citenamefont {{Seo}}, \citenamefont
  {{Giannantonio}}, \citenamefont {{Shapiro}}, \citenamefont {{Bassett}},
  \citenamefont {{Percival}}, \citenamefont {{Davis}}, \citenamefont
  {{Dilday}}, \citenamefont {{Frieman}}, \citenamefont {{Garnavich}},
  \citenamefont {{Sako}}, \citenamefont {{Smith}}, \citenamefont {{Sollerman}},
  \citenamefont {{Becker}},\ and\ \citenamefont {{et al.}}}]{Lampeitl:2010}%
  \BibitemOpen
  \bibfield  {author} {\bibinfo {author} {\bibfnamefont {H.}~\bibnamefont
  {{Lampeitl}}}, \bibinfo {author} {\bibfnamefont {R.~C.}\ \bibnamefont
  {{Nichol}}}, \bibinfo {author} {\bibfnamefont {H.-J.}\ \bibnamefont {{Seo}}},
  \bibinfo {author} {\bibfnamefont {T.}~\bibnamefont {{Giannantonio}}},
  \bibinfo {author} {\bibfnamefont {C.}~\bibnamefont {{Shapiro}}}, \bibinfo
  {author} {\bibfnamefont {B.}~\bibnamefont {{Bassett}}}, \bibinfo {author}
  {\bibfnamefont {W.~J.}\ \bibnamefont {{Percival}}}, \bibinfo {author}
  {\bibfnamefont {T.~M.}\ \bibnamefont {{Davis}}}, \bibinfo {author}
  {\bibfnamefont {B.}~\bibnamefont {{Dilday}}}, \bibinfo {author}
  {\bibfnamefont {J.}~\bibnamefont {{Frieman}}}, \bibinfo {author}
  {\bibfnamefont {P.}~\bibnamefont {{Garnavich}}}, \bibinfo {author}
  {\bibfnamefont {M.}~\bibnamefont {{Sako}}}, \bibinfo {author} {\bibfnamefont
  {M.}~\bibnamefont {{Smith}}}, \bibinfo {author} {\bibfnamefont
  {J.}~\bibnamefont {{Sollerman}}}, \bibinfo {author} {\bibfnamefont {A.~C.}\
  \bibnamefont {{Becker}}}, \ and\ \bibinfo {author} {\bibnamefont {{et
  al.}}},\ }\href {\doibase 10.1111/j.1365-2966.2009.15851.x} {\bibfield
  {journal} {\bibinfo  {journal} {\mnras}\ }\textbf {\bibinfo {volume} {401}},\
  \bibinfo {pages} {2331} (\bibinfo {year} {2010})},\ \Eprint
  {http://arxiv.org/abs/0910.2193} {arXiv:0910.2193 [astro-ph.CO]} \BibitemShut
  {NoStop}%
\bibitem [{\citenamefont {{Campbell}}\ \emph {et~al.}(2013)\citenamefont
  {{Campbell}}, \citenamefont {{D'Andrea}}, \citenamefont {{Nichol}},
  \citenamefont {{Sako}}, \citenamefont {{Smith}}, \citenamefont {{Lampeitl}},
  \citenamefont {{Olmstead}}, \citenamefont {{Bassett}}, \citenamefont
  {{Biswas}}, \citenamefont {{Brown}}, \citenamefont {{Cinabro}}, \citenamefont
  {{Dawson}}, \citenamefont {{Dilday}}, \citenamefont {{Foley}}, \citenamefont
  {{Frieman}},\ and\ \citenamefont {{et al.}}}]{Campbell:2013}%
  \BibitemOpen
  \bibfield  {author} {\bibinfo {author} {\bibfnamefont {H.}~\bibnamefont
  {{Campbell}}}, \bibinfo {author} {\bibfnamefont {C.~B.}\ \bibnamefont
  {{D'Andrea}}}, \bibinfo {author} {\bibfnamefont {R.~C.}\ \bibnamefont
  {{Nichol}}}, \bibinfo {author} {\bibfnamefont {M.}~\bibnamefont {{Sako}}},
  \bibinfo {author} {\bibfnamefont {M.}~\bibnamefont {{Smith}}}, \bibinfo
  {author} {\bibfnamefont {H.}~\bibnamefont {{Lampeitl}}}, \bibinfo {author}
  {\bibfnamefont {M.~D.}\ \bibnamefont {{Olmstead}}}, \bibinfo {author}
  {\bibfnamefont {B.}~\bibnamefont {{Bassett}}}, \bibinfo {author}
  {\bibfnamefont {R.}~\bibnamefont {{Biswas}}}, \bibinfo {author}
  {\bibfnamefont {P.}~\bibnamefont {{Brown}}}, \bibinfo {author} {\bibfnamefont
  {D.}~\bibnamefont {{Cinabro}}}, \bibinfo {author} {\bibfnamefont {K.~S.}\
  \bibnamefont {{Dawson}}}, \bibinfo {author} {\bibfnamefont {B.}~\bibnamefont
  {{Dilday}}}, \bibinfo {author} {\bibfnamefont {R.~J.}\ \bibnamefont
  {{Foley}}}, \bibinfo {author} {\bibfnamefont {J.~A.}\ \bibnamefont
  {{Frieman}}}, \ and\ \bibinfo {author} {\bibnamefont {{et al.}}},\ }\href
  {\doibase 10.1088/0004-637X/763/2/88} {\bibfield  {journal} {\bibinfo
  {journal} {\apj}\ }\textbf {\bibinfo {volume} {763}},\ \bibinfo {eid} {88}
  (\bibinfo {year} {2013})},\ \Eprint {http://arxiv.org/abs/1211.4480}
  {arXiv:1211.4480 [astro-ph.CO]} \BibitemShut {NoStop}%
\bibitem [{\citenamefont {{Astier}}\ \emph {et~al.}(2006)\citenamefont
  {{Astier}}, \citenamefont {{Guy}}, \citenamefont {{Regnault}}, \citenamefont
  {{Pain}}, \citenamefont {{Aubourg}}, \citenamefont {{Balam}}, \citenamefont
  {{Basa}}, \citenamefont {{Carlberg}}, \citenamefont {{Fabbro}}, \citenamefont
  {{Fouchez}}, \citenamefont {{Hook}}, \citenamefont {{Howell}}, \citenamefont
  {{Lafoux}}, \citenamefont {{Neill}}, \citenamefont
  {{Palanque-Delabrouille}},\ and\ \citenamefont {{et al.}}}]{Astier:2006}%
  \BibitemOpen
  \bibfield  {author} {\bibinfo {author} {\bibfnamefont {P.}~\bibnamefont
  {{Astier}}}, \bibinfo {author} {\bibfnamefont {J.}~\bibnamefont {{Guy}}},
  \bibinfo {author} {\bibfnamefont {N.}~\bibnamefont {{Regnault}}}, \bibinfo
  {author} {\bibfnamefont {R.}~\bibnamefont {{Pain}}}, \bibinfo {author}
  {\bibfnamefont {E.}~\bibnamefont {{Aubourg}}}, \bibinfo {author}
  {\bibfnamefont {D.}~\bibnamefont {{Balam}}}, \bibinfo {author} {\bibfnamefont
  {S.}~\bibnamefont {{Basa}}}, \bibinfo {author} {\bibfnamefont {R.~G.}\
  \bibnamefont {{Carlberg}}}, \bibinfo {author} {\bibfnamefont
  {S.}~\bibnamefont {{Fabbro}}}, \bibinfo {author} {\bibfnamefont
  {D.}~\bibnamefont {{Fouchez}}}, \bibinfo {author} {\bibfnamefont {I.~M.}\
  \bibnamefont {{Hook}}}, \bibinfo {author} {\bibfnamefont {D.~A.}\
  \bibnamefont {{Howell}}}, \bibinfo {author} {\bibfnamefont {H.}~\bibnamefont
  {{Lafoux}}}, \bibinfo {author} {\bibfnamefont {J.~D.}\ \bibnamefont
  {{Neill}}}, \bibinfo {author} {\bibfnamefont {N.}~\bibnamefont
  {{Palanque-Delabrouille}}}, \ and\ \bibinfo {author} {\bibnamefont {{et
  al.}}},\ }\href {\doibase 10.1051/0004-6361:20054185} {\bibfield  {journal}
  {\bibinfo  {journal} {\aap}\ }\textbf {\bibinfo {volume} {447}},\ \bibinfo
  {pages} {31} (\bibinfo {year} {2006})},\ \Eprint
  {http://arxiv.org/abs/astro-ph/0510447} {astro-ph/0510447} \BibitemShut
  {NoStop}%
\bibitem [{\citenamefont {{Sullivan}}\ \emph {et~al.}(2011)\citenamefont
  {{Sullivan}}, \citenamefont {{Guy}}, \citenamefont {{Conley}}, \citenamefont
  {{Regnault}}, \citenamefont {{Astier}}, \citenamefont {{Balland}},
  \citenamefont {{Basa}}, \citenamefont {{Carlberg}}, \citenamefont
  {{Fouchez}}, \citenamefont {{Hardin}}, \citenamefont {{Hook}}, \citenamefont
  {{Howell}}, \citenamefont {{Pain}}, \citenamefont {{Palanque-Delabrouille}},
  \citenamefont {{Perrett}},\ and\ \citenamefont {{et al.}}}]{Sullivan:2011}%
  \BibitemOpen
  \bibfield  {author} {\bibinfo {author} {\bibfnamefont {M.}~\bibnamefont
  {{Sullivan}}}, \bibinfo {author} {\bibfnamefont {J.}~\bibnamefont {{Guy}}},
  \bibinfo {author} {\bibfnamefont {A.}~\bibnamefont {{Conley}}}, \bibinfo
  {author} {\bibfnamefont {N.}~\bibnamefont {{Regnault}}}, \bibinfo {author}
  {\bibfnamefont {P.}~\bibnamefont {{Astier}}}, \bibinfo {author}
  {\bibfnamefont {C.}~\bibnamefont {{Balland}}}, \bibinfo {author}
  {\bibfnamefont {S.}~\bibnamefont {{Basa}}}, \bibinfo {author} {\bibfnamefont
  {R.~G.}\ \bibnamefont {{Carlberg}}}, \bibinfo {author} {\bibfnamefont
  {D.}~\bibnamefont {{Fouchez}}}, \bibinfo {author} {\bibfnamefont
  {D.}~\bibnamefont {{Hardin}}}, \bibinfo {author} {\bibfnamefont {I.~M.}\
  \bibnamefont {{Hook}}}, \bibinfo {author} {\bibfnamefont {D.~A.}\
  \bibnamefont {{Howell}}}, \bibinfo {author} {\bibfnamefont {R.}~\bibnamefont
  {{Pain}}}, \bibinfo {author} {\bibfnamefont {N.}~\bibnamefont
  {{Palanque-Delabrouille}}}, \bibinfo {author} {\bibfnamefont {K.~M.}\
  \bibnamefont {{Perrett}}}, \ and\ \bibinfo {author} {\bibnamefont {{et
  al.}}},\ }\href {\doibase 10.1088/0004-637X/737/2/102} {\bibfield  {journal}
  {\bibinfo  {journal} {\apj}\ }\textbf {\bibinfo {volume} {737}},\ \bibinfo
  {eid} {102} (\bibinfo {year} {2011})},\ \Eprint
  {http://arxiv.org/abs/1104.1444} {arXiv:1104.1444 [astro-ph.CO]} \BibitemShut
  {NoStop}%
\bibitem [{\citenamefont {{Riess}}\ \emph {et~al.}(2007)\citenamefont
  {{Riess}}, \citenamefont {{Strolger}}, \citenamefont {{Casertano}},
  \citenamefont {{Ferguson}}, \citenamefont {{Mobasher}}, \citenamefont
  {{Gold}}, \citenamefont {{Challis}}, \citenamefont {{Filippenko}},
  \citenamefont {{Jha}}, \citenamefont {{Li}}, \citenamefont {{Tonry}},
  \citenamefont {{Foley}}, \citenamefont {{Kirshner}}, \citenamefont
  {{Dickinson}}, \citenamefont {{MacDonald}},\ and\ \citenamefont {{et
  al.}}}]{Riess:2007}%
  \BibitemOpen
  \bibfield  {author} {\bibinfo {author} {\bibfnamefont {A.~G.}\ \bibnamefont
  {{Riess}}}, \bibinfo {author} {\bibfnamefont {L.-G.}\ \bibnamefont
  {{Strolger}}}, \bibinfo {author} {\bibfnamefont {S.}~\bibnamefont
  {{Casertano}}}, \bibinfo {author} {\bibfnamefont {H.~C.}\ \bibnamefont
  {{Ferguson}}}, \bibinfo {author} {\bibfnamefont {B.}~\bibnamefont
  {{Mobasher}}}, \bibinfo {author} {\bibfnamefont {B.}~\bibnamefont {{Gold}}},
  \bibinfo {author} {\bibfnamefont {P.~J.}\ \bibnamefont {{Challis}}}, \bibinfo
  {author} {\bibfnamefont {A.~V.}\ \bibnamefont {{Filippenko}}}, \bibinfo
  {author} {\bibfnamefont {S.}~\bibnamefont {{Jha}}}, \bibinfo {author}
  {\bibfnamefont {W.}~\bibnamefont {{Li}}}, \bibinfo {author} {\bibfnamefont
  {J.}~\bibnamefont {{Tonry}}}, \bibinfo {author} {\bibfnamefont
  {R.}~\bibnamefont {{Foley}}}, \bibinfo {author} {\bibfnamefont {R.~P.}\
  \bibnamefont {{Kirshner}}}, \bibinfo {author} {\bibfnamefont
  {M.}~\bibnamefont {{Dickinson}}}, \bibinfo {author} {\bibfnamefont
  {E.}~\bibnamefont {{MacDonald}}}, \ and\ \bibinfo {author} {\bibnamefont {{et
  al.}}},\ }\href {\doibase 10.1086/510378} {\bibfield  {journal} {\bibinfo
  {journal} {\apj}\ }\textbf {\bibinfo {volume} {659}},\ \bibinfo {pages} {98}
  (\bibinfo {year} {2007})},\ \Eprint {http://arxiv.org/abs/astro-ph/0611572}
  {astro-ph/0611572} \BibitemShut {NoStop}%
\bibitem [{\citenamefont {{Suzuki}}\ \emph {et~al.}(2012)\citenamefont
  {{Suzuki}}, \citenamefont {{Rubin}}, \citenamefont {{Lidman}}, \citenamefont
  {{Aldering}}, \citenamefont {{Amanullah}}, \citenamefont {{Barbary}},
  \citenamefont {{Barrientos}}, \citenamefont {{Botyanszki}}, \citenamefont
  {{Brodwin}}, \citenamefont {{Connolly}}, \citenamefont {{Dawson}},
  \citenamefont {{Dey}}, \citenamefont {{Doi}}, \citenamefont {{Donahue}},
  \citenamefont {{Deustua}},\ and\ \citenamefont {{et al.}}}]{Suzuki:2012}%
  \BibitemOpen
  \bibfield  {author} {\bibinfo {author} {\bibfnamefont {N.}~\bibnamefont
  {{Suzuki}}}, \bibinfo {author} {\bibfnamefont {D.}~\bibnamefont {{Rubin}}},
  \bibinfo {author} {\bibfnamefont {C.}~\bibnamefont {{Lidman}}}, \bibinfo
  {author} {\bibfnamefont {G.}~\bibnamefont {{Aldering}}}, \bibinfo {author}
  {\bibfnamefont {R.}~\bibnamefont {{Amanullah}}}, \bibinfo {author}
  {\bibfnamefont {K.}~\bibnamefont {{Barbary}}}, \bibinfo {author}
  {\bibfnamefont {L.~F.}\ \bibnamefont {{Barrientos}}}, \bibinfo {author}
  {\bibfnamefont {J.}~\bibnamefont {{Botyanszki}}}, \bibinfo {author}
  {\bibfnamefont {M.}~\bibnamefont {{Brodwin}}}, \bibinfo {author}
  {\bibfnamefont {N.}~\bibnamefont {{Connolly}}}, \bibinfo {author}
  {\bibfnamefont {K.~S.}\ \bibnamefont {{Dawson}}}, \bibinfo {author}
  {\bibfnamefont {A.}~\bibnamefont {{Dey}}}, \bibinfo {author} {\bibfnamefont
  {M.}~\bibnamefont {{Doi}}}, \bibinfo {author} {\bibfnamefont
  {M.}~\bibnamefont {{Donahue}}}, \bibinfo {author} {\bibfnamefont
  {S.}~\bibnamefont {{Deustua}}}, \ and\ \bibinfo {author} {\bibnamefont {{et
  al.}}},\ }\href {\doibase 10.1088/0004-637X/746/1/85} {\bibfield  {journal}
  {\bibinfo  {journal} {\apj}\ }\textbf {\bibinfo {volume} {746}},\ \bibinfo
  {eid} {85} (\bibinfo {year} {2012})},\ \Eprint
  {http://arxiv.org/abs/1105.3470} {arXiv:1105.3470 [astro-ph.CO]} \BibitemShut
  {NoStop}%
\bibitem [{\citenamefont {{Limber}}(1953)}]{Limber:1953}%
  \BibitemOpen
  \bibfield  {author} {\bibinfo {author} {\bibfnamefont {D.~N.}\ \bibnamefont
  {{Limber}}},\ }\href {\doibase 10.1086/145672} {\bibfield  {journal}
  {\bibinfo  {journal} {\apj}\ }\textbf {\bibinfo {volume} {117}},\ \bibinfo
  {pages} {134} (\bibinfo {year} {1953})}\BibitemShut {NoStop}%
\bibitem [{\citenamefont {{Kaiser}}(1992)}]{Kaiser:1992}%
  \BibitemOpen
  \bibfield  {author} {\bibinfo {author} {\bibfnamefont {N.}~\bibnamefont
  {{Kaiser}}},\ }\href {\doibase 10.1086/171151} {\bibfield  {journal}
  {\bibinfo  {journal} {\apj}\ }\textbf {\bibinfo {volume} {388}},\ \bibinfo
  {pages} {272} (\bibinfo {year} {1992})}\BibitemShut {NoStop}%
\bibitem [{\citenamefont {{Kaiser}}(1998)}]{Kaiser:1998}%
  \BibitemOpen
  \bibfield  {author} {\bibinfo {author} {\bibfnamefont {N.}~\bibnamefont
  {{Kaiser}}},\ }\href {\doibase 10.1086/305515} {\bibfield  {journal}
  {\bibinfo  {journal} {\apj}\ }\textbf {\bibinfo {volume} {498}},\ \bibinfo
  {pages} {26} (\bibinfo {year} {1998})},\ \Eprint
  {http://arxiv.org/abs/astro-ph/9610120} {astro-ph/9610120} \BibitemShut
  {NoStop}%
\bibitem [{\citenamefont {{Sachs}}\ and\ \citenamefont
  {{Wolfe}}(1967)}]{Sachs:1967}%
  \BibitemOpen
  \bibfield  {author} {\bibinfo {author} {\bibfnamefont {R.~K.}\ \bibnamefont
  {{Sachs}}}\ and\ \bibinfo {author} {\bibfnamefont {A.~M.}\ \bibnamefont
  {{Wolfe}}},\ }\href {\doibase 10.1086/148982} {\bibfield  {journal} {\bibinfo
   {journal} {\apj}\ }\textbf {\bibinfo {volume} {147}},\ \bibinfo {pages} {73}
  (\bibinfo {year} {1967})}\BibitemShut {NoStop}%
\bibitem [{\citenamefont {{Sunyaev}}\ and\ \citenamefont
  {{Zeldovich}}(1980)}]{Sunyaev:1980}%
  \BibitemOpen
  \bibfield  {author} {\bibinfo {author} {\bibfnamefont {R.~A.}\ \bibnamefont
  {{Sunyaev}}}\ and\ \bibinfo {author} {\bibfnamefont {I.~B.}\ \bibnamefont
  {{Zeldovich}}},\ }\href {\doibase 10.1093/mnras/190.3.413} {\bibfield
  {journal} {\bibinfo  {journal} {\mnras}\ }\textbf {\bibinfo {volume} {190}},\
  \bibinfo {pages} {413} (\bibinfo {year} {1980})}\BibitemShut {NoStop}%
\bibitem [{\citenamefont {{Goldberg}}\ and\ \citenamefont
  {{Spergel}}(1999)}]{Goldberg:1999}%
  \BibitemOpen
  \bibfield  {author} {\bibinfo {author} {\bibfnamefont {D.~M.}\ \bibnamefont
  {{Goldberg}}}\ and\ \bibinfo {author} {\bibfnamefont {D.~N.}\ \bibnamefont
  {{Spergel}}},\ }\href {\doibase 10.1103/PhysRevD.59.103002} {\bibfield
  {journal} {\bibinfo  {journal} {\prd}\ }\textbf {\bibinfo {volume} {59}},\
  \bibinfo {eid} {103002} (\bibinfo {year} {1999})},\ \Eprint
  {http://arxiv.org/abs/astro-ph/9811251} {astro-ph/9811251} \BibitemShut
  {NoStop}%
\bibitem [{\citenamefont {{Cooray}}\ and\ \citenamefont
  {{Hu}}(2000)}]{Cooray:2000}%
  \BibitemOpen
  \bibfield  {author} {\bibinfo {author} {\bibfnamefont {A.}~\bibnamefont
  {{Cooray}}}\ and\ \bibinfo {author} {\bibfnamefont {W.}~\bibnamefont
  {{Hu}}},\ }\href {\doibase 10.1086/308799} {\bibfield  {journal} {\bibinfo
  {journal} {\apj}\ }\textbf {\bibinfo {volume} {534}},\ \bibinfo {pages} {533}
  (\bibinfo {year} {2000})},\ \Eprint {http://arxiv.org/abs/astro-ph/9910397}
  {astro-ph/9910397} \BibitemShut {NoStop}%
\bibitem [{\citenamefont {{Lesgourgues}}(2011)}]{Lesgourgues:2011}%
  \BibitemOpen
  \bibfield  {author} {\bibinfo {author} {\bibfnamefont {J.}~\bibnamefont
  {{Lesgourgues}}},\ }\href@noop {} {\bibfield  {journal} {\bibinfo  {journal}
  {ArXiv e-prints}\ } (\bibinfo {year} {2011})},\ \Eprint
  {http://arxiv.org/abs/1104.2932} {arXiv:1104.2932 [astro-ph.IM]} \BibitemShut
  {NoStop}%
\bibitem [{\citenamefont {{Refregier}}\ \emph {et~al.}(2016)\citenamefont
  {{Refregier}}, \citenamefont {{Gamper}},\ and\ \citenamefont {{et
  al.}}}]{Refregier:2016}%
  \BibitemOpen
  \bibfield  {author} {\bibinfo {author} {\bibfnamefont {A.}~\bibnamefont
  {{Refregier}}}, \bibinfo {author} {\bibfnamefont {L.}~\bibnamefont
  {{Gamper}}}, \ and\ \bibinfo {author} {\bibnamefont {{et al.}}},\ }\href@noop
  {} {\bibfield  {journal} {\bibinfo  {journal} {in prep.}\ } (\bibinfo {year}
  {2016})}\BibitemShut {NoStop}%
\bibitem [{\citenamefont {{Eisenstein}}\ and\ \citenamefont
  {{Hu}}(1998)}]{Eisenstein:1998}%
  \BibitemOpen
  \bibfield  {author} {\bibinfo {author} {\bibfnamefont {D.~J.}\ \bibnamefont
  {{Eisenstein}}}\ and\ \bibinfo {author} {\bibfnamefont {W.}~\bibnamefont
  {{Hu}}},\ }\href {\doibase 10.1086/305424} {\bibfield  {journal} {\bibinfo
  {journal} {\apj}\ }\textbf {\bibinfo {volume} {496}},\ \bibinfo {pages} {605}
  (\bibinfo {year} {1998})},\ \Eprint {http://arxiv.org/abs/astro-ph/9709112}
  {astro-ph/9709112} \BibitemShut {NoStop}%
\bibitem [{\citenamefont {{Smith}}\ \emph {et~al.}(2003)\citenamefont
  {{Smith}}, \citenamefont {{Peacock}}, \citenamefont {{Jenkins}},
  \citenamefont {{White}}, \citenamefont {{Frenk}}, \citenamefont {{Pearce}},
  \citenamefont {{Thomas}}, \citenamefont {{Efstathiou}},\ and\ \citenamefont
  {{Couchman}}}]{Smith:2003}%
  \BibitemOpen
  \bibfield  {author} {\bibinfo {author} {\bibfnamefont {R.~E.}\ \bibnamefont
  {{Smith}}}, \bibinfo {author} {\bibfnamefont {J.~A.}\ \bibnamefont
  {{Peacock}}}, \bibinfo {author} {\bibfnamefont {A.}~\bibnamefont
  {{Jenkins}}}, \bibinfo {author} {\bibfnamefont {S.~D.~M.}\ \bibnamefont
  {{White}}}, \bibinfo {author} {\bibfnamefont {C.~S.}\ \bibnamefont
  {{Frenk}}}, \bibinfo {author} {\bibfnamefont {F.~R.}\ \bibnamefont
  {{Pearce}}}, \bibinfo {author} {\bibfnamefont {P.~A.}\ \bibnamefont
  {{Thomas}}}, \bibinfo {author} {\bibfnamefont {G.}~\bibnamefont
  {{Efstathiou}}}, \ and\ \bibinfo {author} {\bibfnamefont {H.~M.~P.}\
  \bibnamefont {{Couchman}}},\ }\href {\doibase
  10.1046/j.1365-8711.2003.06503.x} {\bibfield  {journal} {\bibinfo  {journal}
  {\mnras}\ }\textbf {\bibinfo {volume} {341}},\ \bibinfo {pages} {1311}
  (\bibinfo {year} {2003})},\ \Eprint {http://arxiv.org/abs/astro-ph/0207664}
  {astro-ph/0207664} \BibitemShut {NoStop}%
\bibitem [{\citenamefont {{Takahashi}}\ \emph {et~al.}(2012)\citenamefont
  {{Takahashi}}, \citenamefont {{Sato}}, \citenamefont {{Nishimichi}},
  \citenamefont {{Taruya}},\ and\ \citenamefont {{Oguri}}}]{Takahashi:2012}%
  \BibitemOpen
  \bibfield  {author} {\bibinfo {author} {\bibfnamefont {R.}~\bibnamefont
  {{Takahashi}}}, \bibinfo {author} {\bibfnamefont {M.}~\bibnamefont {{Sato}}},
  \bibinfo {author} {\bibfnamefont {T.}~\bibnamefont {{Nishimichi}}}, \bibinfo
  {author} {\bibfnamefont {A.}~\bibnamefont {{Taruya}}}, \ and\ \bibinfo
  {author} {\bibfnamefont {M.}~\bibnamefont {{Oguri}}},\ }\href {\doibase
  10.1088/0004-637X/761/2/152} {\bibfield  {journal} {\bibinfo  {journal}
  {\apj}\ }\textbf {\bibinfo {volume} {761}},\ \bibinfo {eid} {152} (\bibinfo
  {year} {2012})},\ \Eprint {http://arxiv.org/abs/1208.2701} {arXiv:1208.2701}
  \BibitemShut {NoStop}%
\bibitem [{\citenamefont {{Szapudi}}\ \emph
  {et~al.}(2001{\natexlab{a}})\citenamefont {{Szapudi}}, \citenamefont
  {{Prunet}}, \citenamefont {{Pogosyan}}, \citenamefont {{Szalay}},\ and\
  \citenamefont {{Bond}}}]{Szapudi:2001}%
  \BibitemOpen
  \bibfield  {author} {\bibinfo {author} {\bibfnamefont {I.}~\bibnamefont
  {{Szapudi}}}, \bibinfo {author} {\bibfnamefont {S.}~\bibnamefont {{Prunet}}},
  \bibinfo {author} {\bibfnamefont {D.}~\bibnamefont {{Pogosyan}}}, \bibinfo
  {author} {\bibfnamefont {A.~S.}\ \bibnamefont {{Szalay}}}, \ and\ \bibinfo
  {author} {\bibfnamefont {J.~R.}\ \bibnamefont {{Bond}}},\ }\href {\doibase
  10.1086/319105} {\bibfield  {journal} {\bibinfo  {journal} {\apjl}\ }\textbf
  {\bibinfo {volume} {548}},\ \bibinfo {pages} {L115} (\bibinfo {year}
  {2001}{\natexlab{a}})}\BibitemShut {NoStop}%
\bibitem [{\citenamefont {{Szapudi}}\ \emph
  {et~al.}(2001{\natexlab{b}})\citenamefont {{Szapudi}}, \citenamefont
  {{Prunet}},\ and\ \citenamefont {{Colombi}}}]{Szapudi:2001ab}%
  \BibitemOpen
  \bibfield  {author} {\bibinfo {author} {\bibfnamefont {I.}~\bibnamefont
  {{Szapudi}}}, \bibinfo {author} {\bibfnamefont {S.}~\bibnamefont {{Prunet}}},
  \ and\ \bibinfo {author} {\bibfnamefont {S.}~\bibnamefont {{Colombi}}},\
  }\href {\doibase 10.1086/324312} {\bibfield  {journal} {\bibinfo  {journal}
  {\apjl}\ }\textbf {\bibinfo {volume} {561}},\ \bibinfo {pages} {L11}
  (\bibinfo {year} {2001}{\natexlab{b}})}\BibitemShut {NoStop}%
\bibitem [{\citenamefont {{Chon}}\ \emph {et~al.}(2004)\citenamefont {{Chon}},
  \citenamefont {{Challinor}}, \citenamefont {{Prunet}}, \citenamefont
  {{Hivon}},\ and\ \citenamefont {{Szapudi}}}]{Chon:2004}%
  \BibitemOpen
  \bibfield  {author} {\bibinfo {author} {\bibfnamefont {G.}~\bibnamefont
  {{Chon}}}, \bibinfo {author} {\bibfnamefont {A.}~\bibnamefont {{Challinor}}},
  \bibinfo {author} {\bibfnamefont {S.}~\bibnamefont {{Prunet}}}, \bibinfo
  {author} {\bibfnamefont {E.}~\bibnamefont {{Hivon}}}, \ and\ \bibinfo
  {author} {\bibfnamefont {I.}~\bibnamefont {{Szapudi}}},\ }\href {\doibase
  10.1111/j.1365-2966.2004.07737.x} {\bibfield  {journal} {\bibinfo  {journal}
  {\mnras}\ }\textbf {\bibinfo {volume} {350}},\ \bibinfo {pages} {914}
  (\bibinfo {year} {2004})},\ \Eprint {http://arxiv.org/abs/astro-ph/0303414}
  {astro-ph/0303414} \BibitemShut {NoStop}%
\bibitem [{\citenamefont {{Planck Collaboration}}\ \emph
  {et~al.}(2014{\natexlab{a}})\citenamefont {{Planck Collaboration}},
  \citenamefont {{Ade}}, \citenamefont {{Aghanim}}, \citenamefont
  {{Armitage-Caplan}}, \citenamefont {{Arnaud}}, \citenamefont {{Ashdown}},
  \citenamefont {{Atrio-Barandela}}, \citenamefont {{Aumont}}, \citenamefont
  {{Baccigalupi}}, \citenamefont {{Banday}},\ and\ \citenamefont
  {et~al.}}]{Planck-Collaboration:2014aa}%
  \BibitemOpen
  \bibfield  {author} {\bibinfo {author} {\bibnamefont {{Planck
  Collaboration}}}, \bibinfo {author} {\bibfnamefont {P.~A.~R.}\ \bibnamefont
  {{Ade}}}, \bibinfo {author} {\bibfnamefont {N.}~\bibnamefont {{Aghanim}}},
  \bibinfo {author} {\bibfnamefont {C.}~\bibnamefont {{Armitage-Caplan}}},
  \bibinfo {author} {\bibfnamefont {M.}~\bibnamefont {{Arnaud}}}, \bibinfo
  {author} {\bibfnamefont {M.}~\bibnamefont {{Ashdown}}}, \bibinfo {author}
  {\bibfnamefont {F.}~\bibnamefont {{Atrio-Barandela}}}, \bibinfo {author}
  {\bibfnamefont {J.}~\bibnamefont {{Aumont}}}, \bibinfo {author}
  {\bibfnamefont {C.}~\bibnamefont {{Baccigalupi}}}, \bibinfo {author}
  {\bibfnamefont {A.~J.}\ \bibnamefont {{Banday}}}, \ and\ \bibinfo {author}
  {\bibnamefont {et~al.}},\ }\href {\doibase 10.1051/0004-6361/201321543}
  {\bibfield  {journal} {\bibinfo  {journal} {\aap}\ }\textbf {\bibinfo
  {volume} {571}},\ \bibinfo {eid} {A17} (\bibinfo {year}
  {2014}{\natexlab{a}})},\ \Eprint {http://arxiv.org/abs/1303.5077}
  {arXiv:1303.5077} \BibitemShut {NoStop}%
\bibitem [{\citenamefont {{Planck Collaboration}}\ \emph
  {et~al.}(2014{\natexlab{b}})\citenamefont {{Planck Collaboration}},
  \citenamefont {{Ade}}, \citenamefont {{Aghanim}}, \citenamefont
  {{Armitage-Caplan}}, \citenamefont {{Arnaud}}, \citenamefont {{Ashdown}},
  \citenamefont {{Atrio-Barandela}}, \citenamefont {{Aumont}}, \citenamefont
  {{Baccigalupi}}, \citenamefont {{Banday}},\ and\ \citenamefont
  {et~al.}}]{Planck-Collaboration:2014ac}%
  \BibitemOpen
  \bibfield  {author} {\bibinfo {author} {\bibnamefont {{Planck
  Collaboration}}}, \bibinfo {author} {\bibfnamefont {P.~A.~R.}\ \bibnamefont
  {{Ade}}}, \bibinfo {author} {\bibfnamefont {N.}~\bibnamefont {{Aghanim}}},
  \bibinfo {author} {\bibfnamefont {C.}~\bibnamefont {{Armitage-Caplan}}},
  \bibinfo {author} {\bibfnamefont {M.}~\bibnamefont {{Arnaud}}}, \bibinfo
  {author} {\bibfnamefont {M.}~\bibnamefont {{Ashdown}}}, \bibinfo {author}
  {\bibfnamefont {F.}~\bibnamefont {{Atrio-Barandela}}}, \bibinfo {author}
  {\bibfnamefont {J.}~\bibnamefont {{Aumont}}}, \bibinfo {author}
  {\bibfnamefont {C.}~\bibnamefont {{Baccigalupi}}}, \bibinfo {author}
  {\bibfnamefont {A.~J.}\ \bibnamefont {{Banday}}}, \ and\ \bibinfo {author}
  {\bibnamefont {et~al.}},\ }\href {\doibase 10.1051/0004-6361/201321573}
  {\bibfield  {journal} {\bibinfo  {journal} {\aap}\ }\textbf {\bibinfo
  {volume} {571}},\ \bibinfo {eid} {A15} (\bibinfo {year}
  {2014}{\natexlab{b}})},\ \Eprint {http://arxiv.org/abs/1303.5075}
  {arXiv:1303.5075} \BibitemShut {NoStop}%
\bibitem [{\citenamefont {{Planck Collaboration}}\ \emph
  {et~al.}(2014{\natexlab{c}})\citenamefont {{Planck Collaboration}},
  \citenamefont {{Ade}}, \citenamefont {{Aghanim}}, \citenamefont
  {{Armitage-Caplan}}, \citenamefont {{Arnaud}}, \citenamefont {{Ashdown}},
  \citenamefont {{Atrio-Barandela}}, \citenamefont {{Aumont}}, \citenamefont
  {{Baccigalupi}}, \citenamefont {{Banday}},\ and\ \citenamefont
  {et~al.}}]{Planck-Collaboration:2014af}%
  \BibitemOpen
  \bibfield  {author} {\bibinfo {author} {\bibnamefont {{Planck
  Collaboration}}}, \bibinfo {author} {\bibfnamefont {P.~A.~R.}\ \bibnamefont
  {{Ade}}}, \bibinfo {author} {\bibfnamefont {N.}~\bibnamefont {{Aghanim}}},
  \bibinfo {author} {\bibfnamefont {C.}~\bibnamefont {{Armitage-Caplan}}},
  \bibinfo {author} {\bibfnamefont {M.}~\bibnamefont {{Arnaud}}}, \bibinfo
  {author} {\bibfnamefont {M.}~\bibnamefont {{Ashdown}}}, \bibinfo {author}
  {\bibfnamefont {F.}~\bibnamefont {{Atrio-Barandela}}}, \bibinfo {author}
  {\bibfnamefont {J.}~\bibnamefont {{Aumont}}}, \bibinfo {author}
  {\bibfnamefont {C.}~\bibnamefont {{Baccigalupi}}}, \bibinfo {author}
  {\bibfnamefont {A.~J.}\ \bibnamefont {{Banday}}}, \ and\ \bibinfo {author}
  {\bibnamefont {et~al.}},\ }\href {\doibase 10.1051/0004-6361/201321580}
  {\bibfield  {journal} {\bibinfo  {journal} {\aap}\ }\textbf {\bibinfo
  {volume} {571}},\ \bibinfo {eid} {A12} (\bibinfo {year}
  {2014}{\natexlab{c}})},\ \Eprint {http://arxiv.org/abs/1303.5072}
  {arXiv:1303.5072} \BibitemShut {NoStop}%
\bibitem [{\citenamefont {{Schlegel}}\ \emph {et~al.}(1998)\citenamefont
  {{Schlegel}}, \citenamefont {{Finkbeiner}},\ and\ \citenamefont
  {{Davis}}}]{Schlegel:1998}%
  \BibitemOpen
  \bibfield  {author} {\bibinfo {author} {\bibfnamefont {D.~J.}\ \bibnamefont
  {{Schlegel}}}, \bibinfo {author} {\bibfnamefont {D.~P.}\ \bibnamefont
  {{Finkbeiner}}}, \ and\ \bibinfo {author} {\bibfnamefont {M.}~\bibnamefont
  {{Davis}}},\ }\href {\doibase 10.1086/305772} {\bibfield  {journal} {\bibinfo
   {journal} {\apj}\ }\textbf {\bibinfo {volume} {500}},\ \bibinfo {pages}
  {525} (\bibinfo {year} {1998})},\ \Eprint
  {http://arxiv.org/abs/astro-ph/9710327} {astro-ph/9710327} \BibitemShut
  {NoStop}%
\bibitem [{\citenamefont {{Joachimi}}\ \emph {et~al.}(2015)\citenamefont
  {{Joachimi}}, \citenamefont {{Cacciato}}, \citenamefont {{Kitching}},
  \citenamefont {{Leonard}}, \citenamefont {{Mandelbaum}}, \citenamefont
  {{Sch{\"a}fer}}, \citenamefont {{Sif{\'o}n}}, \citenamefont {{Hoekstra}},
  \citenamefont {{Kiessling}}, \citenamefont {{Kirk}},\ and\ \citenamefont
  {{Rassat}}}]{Joachimi:2015}%
  \BibitemOpen
  \bibfield  {author} {\bibinfo {author} {\bibfnamefont {B.}~\bibnamefont
  {{Joachimi}}}, \bibinfo {author} {\bibfnamefont {M.}~\bibnamefont
  {{Cacciato}}}, \bibinfo {author} {\bibfnamefont {T.~D.}\ \bibnamefont
  {{Kitching}}}, \bibinfo {author} {\bibfnamefont {A.}~\bibnamefont
  {{Leonard}}}, \bibinfo {author} {\bibfnamefont {R.}~\bibnamefont
  {{Mandelbaum}}}, \bibinfo {author} {\bibfnamefont {B.~M.}\ \bibnamefont
  {{Sch{\"a}fer}}}, \bibinfo {author} {\bibfnamefont {C.}~\bibnamefont
  {{Sif{\'o}n}}}, \bibinfo {author} {\bibfnamefont {H.}~\bibnamefont
  {{Hoekstra}}}, \bibinfo {author} {\bibfnamefont {A.}~\bibnamefont
  {{Kiessling}}}, \bibinfo {author} {\bibfnamefont {D.}~\bibnamefont {{Kirk}}},
  \ and\ \bibinfo {author} {\bibfnamefont {A.}~\bibnamefont {{Rassat}}},\
  }\href {\doibase 10.1007/s11214-015-0177-4} {\bibfield  {journal} {\bibinfo
  {journal} {\ssr}\ }\textbf {\bibinfo {volume} {193}},\ \bibinfo {pages} {1}
  (\bibinfo {year} {2015})},\ \Eprint {http://arxiv.org/abs/1504.05456}
  {arXiv:1504.05456} \BibitemShut {NoStop}%
\bibitem [{\citenamefont {{Troxel}}\ and\ \citenamefont
  {{Ishak}}(2015)}]{Troxel:2015}%
  \BibitemOpen
  \bibfield  {author} {\bibinfo {author} {\bibfnamefont {M.~A.}\ \bibnamefont
  {{Troxel}}}\ and\ \bibinfo {author} {\bibfnamefont {M.}~\bibnamefont
  {{Ishak}}},\ }\href {\doibase 10.1016/j.physrep.2014.11.001} {\bibfield
  {journal} {\bibinfo  {journal} {\physrep}\ }\textbf {\bibinfo {volume}
  {558}},\ \bibinfo {pages} {1} (\bibinfo {year} {2015})},\ \Eprint
  {http://arxiv.org/abs/1407.6990} {arXiv:1407.6990} \BibitemShut {NoStop}%
\bibitem [{\citenamefont {{Heymans}}\ \emph {et~al.}(2006)\citenamefont
  {{Heymans}}, \citenamefont {{White}}, \citenamefont {{Heavens}},
  \citenamefont {{Vale}},\ and\ \citenamefont {{van Waerbeke}}}]{Heymans:2006}%
  \BibitemOpen
  \bibfield  {author} {\bibinfo {author} {\bibfnamefont {C.}~\bibnamefont
  {{Heymans}}}, \bibinfo {author} {\bibfnamefont {M.}~\bibnamefont {{White}}},
  \bibinfo {author} {\bibfnamefont {A.}~\bibnamefont {{Heavens}}}, \bibinfo
  {author} {\bibfnamefont {C.}~\bibnamefont {{Vale}}}, \ and\ \bibinfo {author}
  {\bibfnamefont {L.}~\bibnamefont {{van Waerbeke}}},\ }\href {\doibase
  10.1111/j.1365-2966.2006.10705.x} {\bibfield  {journal} {\bibinfo  {journal}
  {\mnras}\ }\textbf {\bibinfo {volume} {371}},\ \bibinfo {pages} {750}
  (\bibinfo {year} {2006})},\ \Eprint {http://arxiv.org/abs/astro-ph/0604001}
  {astro-ph/0604001} \BibitemShut {NoStop}%
\bibitem [{\citenamefont {{Mandelbaum}}\ \emph {et~al.}(2006)\citenamefont
  {{Mandelbaum}}, \citenamefont {{Hirata}}, \citenamefont {{Ishak}},
  \citenamefont {{Seljak}},\ and\ \citenamefont
  {{Brinkmann}}}]{Mandelbaum:2006}%
  \BibitemOpen
  \bibfield  {author} {\bibinfo {author} {\bibfnamefont {R.}~\bibnamefont
  {{Mandelbaum}}}, \bibinfo {author} {\bibfnamefont {C.~M.}\ \bibnamefont
  {{Hirata}}}, \bibinfo {author} {\bibfnamefont {M.}~\bibnamefont {{Ishak}}},
  \bibinfo {author} {\bibfnamefont {U.}~\bibnamefont {{Seljak}}}, \ and\
  \bibinfo {author} {\bibfnamefont {J.}~\bibnamefont {{Brinkmann}}},\ }\href
  {\doibase 10.1111/j.1365-2966.2005.09946.x} {\bibfield  {journal} {\bibinfo
  {journal} {\mnras}\ }\textbf {\bibinfo {volume} {367}},\ \bibinfo {pages}
  {611} (\bibinfo {year} {2006})},\ \Eprint
  {http://arxiv.org/abs/astro-ph/0509026} {astro-ph/0509026} \BibitemShut
  {NoStop}%
\bibitem [{\citenamefont {{Hirata}}\ \emph {et~al.}(2007)\citenamefont
  {{Hirata}}, \citenamefont {{Mandelbaum}}, \citenamefont {{Ishak}},
  \citenamefont {{Seljak}}, \citenamefont {{Nichol}}, \citenamefont
  {{Pimbblet}}, \citenamefont {{Ross}},\ and\ \citenamefont
  {{Wake}}}]{Hirata:2007}%
  \BibitemOpen
  \bibfield  {author} {\bibinfo {author} {\bibfnamefont {C.~M.}\ \bibnamefont
  {{Hirata}}}, \bibinfo {author} {\bibfnamefont {R.}~\bibnamefont
  {{Mandelbaum}}}, \bibinfo {author} {\bibfnamefont {M.}~\bibnamefont
  {{Ishak}}}, \bibinfo {author} {\bibfnamefont {U.}~\bibnamefont {{Seljak}}},
  \bibinfo {author} {\bibfnamefont {R.}~\bibnamefont {{Nichol}}}, \bibinfo
  {author} {\bibfnamefont {K.~A.}\ \bibnamefont {{Pimbblet}}}, \bibinfo
  {author} {\bibfnamefont {N.~P.}\ \bibnamefont {{Ross}}}, \ and\ \bibinfo
  {author} {\bibfnamefont {D.}~\bibnamefont {{Wake}}},\ }\href {\doibase
  10.1111/j.1365-2966.2007.12312.x} {\bibfield  {journal} {\bibinfo  {journal}
  {\mnras}\ }\textbf {\bibinfo {volume} {381}},\ \bibinfo {pages} {1197}
  (\bibinfo {year} {2007})},\ \Eprint {http://arxiv.org/abs/astro-ph/0701671}
  {astro-ph/0701671} \BibitemShut {NoStop}%
\bibitem [{\citenamefont {{Faltenbacher}}\ \emph {et~al.}(2009)\citenamefont
  {{Faltenbacher}}, \citenamefont {{Li}}, \citenamefont {{White}},
  \citenamefont {{Jing}}, \citenamefont {{Shu-DeMao}},\ and\ \citenamefont
  {{Wang}}}]{Faltenbacher:2009}%
  \BibitemOpen
  \bibfield  {author} {\bibinfo {author} {\bibfnamefont {A.}~\bibnamefont
  {{Faltenbacher}}}, \bibinfo {author} {\bibfnamefont {C.}~\bibnamefont
  {{Li}}}, \bibinfo {author} {\bibfnamefont {S.~D.~M.}\ \bibnamefont
  {{White}}}, \bibinfo {author} {\bibfnamefont {Y.-P.}\ \bibnamefont {{Jing}}},
  \bibinfo {author} {\bibnamefont {{Shu-DeMao}}}, \ and\ \bibinfo {author}
  {\bibfnamefont {J.}~\bibnamefont {{Wang}}},\ }\href {\doibase
  10.1088/1674-4527/9/1/004} {\bibfield  {journal} {\bibinfo  {journal}
  {Research in Astronomy and Astrophysics}\ }\textbf {\bibinfo {volume} {9}},\
  \bibinfo {pages} {41} (\bibinfo {year} {2009})},\ \Eprint
  {http://arxiv.org/abs/0811.1995} {arXiv:0811.1995} \BibitemShut {NoStop}%
\bibitem [{\citenamefont {{Okumura}}\ and\ \citenamefont
  {{Jing}}(2009)}]{Okumura:2009}%
  \BibitemOpen
  \bibfield  {author} {\bibinfo {author} {\bibfnamefont {T.}~\bibnamefont
  {{Okumura}}}\ and\ \bibinfo {author} {\bibfnamefont {Y.~P.}\ \bibnamefont
  {{Jing}}},\ }\href {\doibase 10.1088/0004-637X/694/1/L83} {\bibfield
  {journal} {\bibinfo  {journal} {\apjl}\ }\textbf {\bibinfo {volume} {694}},\
  \bibinfo {pages} {L83} (\bibinfo {year} {2009})},\ \Eprint
  {http://arxiv.org/abs/0812.2935} {arXiv:0812.2935} \BibitemShut {NoStop}%
\bibitem [{\citenamefont {{Joachimi}}\ \emph {et~al.}(2011)\citenamefont
  {{Joachimi}}, \citenamefont {{Mandelbaum}}, \citenamefont {{Abdalla}},\ and\
  \citenamefont {{Bridle}}}]{Joachimi:2011}%
  \BibitemOpen
  \bibfield  {author} {\bibinfo {author} {\bibfnamefont {B.}~\bibnamefont
  {{Joachimi}}}, \bibinfo {author} {\bibfnamefont {R.}~\bibnamefont
  {{Mandelbaum}}}, \bibinfo {author} {\bibfnamefont {F.~B.}\ \bibnamefont
  {{Abdalla}}}, \ and\ \bibinfo {author} {\bibfnamefont {S.~L.}\ \bibnamefont
  {{Bridle}}},\ }\href {\doibase 10.1051/0004-6361/201015621} {\bibfield
  {journal} {\bibinfo  {journal} {\aap}\ }\textbf {\bibinfo {volume} {527}},\
  \bibinfo {eid} {A26} (\bibinfo {year} {2011})},\ \Eprint
  {http://arxiv.org/abs/1008.3491} {arXiv:1008.3491 [astro-ph.CO]} \BibitemShut
  {NoStop}%
\bibitem [{\citenamefont {{Singh}}\ \emph {et~al.}(2015)\citenamefont
  {{Singh}}, \citenamefont {{Mandelbaum}},\ and\ \citenamefont
  {{More}}}]{Singh:2015}%
  \BibitemOpen
  \bibfield  {author} {\bibinfo {author} {\bibfnamefont {S.}~\bibnamefont
  {{Singh}}}, \bibinfo {author} {\bibfnamefont {R.}~\bibnamefont
  {{Mandelbaum}}}, \ and\ \bibinfo {author} {\bibfnamefont {S.}~\bibnamefont
  {{More}}},\ }\href {\doibase 10.1093/mnras/stv778} {\bibfield  {journal}
  {\bibinfo  {journal} {\mnras}\ }\textbf {\bibinfo {volume} {450}},\ \bibinfo
  {pages} {2195} (\bibinfo {year} {2015})},\ \Eprint
  {http://arxiv.org/abs/1411.1755} {arXiv:1411.1755} \BibitemShut {NoStop}%
\bibitem [{\citenamefont {{Catelan}}\ \emph {et~al.}(2001)\citenamefont
  {{Catelan}}, \citenamefont {{Kamionkowski}},\ and\ \citenamefont
  {{Blandford}}}]{Catelan:2001}%
  \BibitemOpen
  \bibfield  {author} {\bibinfo {author} {\bibfnamefont {P.}~\bibnamefont
  {{Catelan}}}, \bibinfo {author} {\bibfnamefont {M.}~\bibnamefont
  {{Kamionkowski}}}, \ and\ \bibinfo {author} {\bibfnamefont {R.~D.}\
  \bibnamefont {{Blandford}}},\ }\href {\doibase
  10.1046/j.1365-8711.2001.04105.x} {\bibfield  {journal} {\bibinfo  {journal}
  {\mnras}\ }\textbf {\bibinfo {volume} {320}},\ \bibinfo {pages} {L7}
  (\bibinfo {year} {2001})},\ \Eprint {http://arxiv.org/abs/astro-ph/0005470}
  {astro-ph/0005470} \BibitemShut {NoStop}%
\bibitem [{\citenamefont {{Hirata}}\ and\ \citenamefont
  {{Seljak}}(2004)}]{Hirata:2004aa}%
  \BibitemOpen
  \bibfield  {author} {\bibinfo {author} {\bibfnamefont {C.~M.}\ \bibnamefont
  {{Hirata}}}\ and\ \bibinfo {author} {\bibfnamefont {U.}~\bibnamefont
  {{Seljak}}},\ }\href {\doibase 10.1103/PhysRevD.70.063526} {\bibfield
  {journal} {\bibinfo  {journal} {\prd}\ }\textbf {\bibinfo {volume} {70}},\
  \bibinfo {eid} {063526} (\bibinfo {year} {2004})},\ \Eprint
  {http://arxiv.org/abs/astro-ph/0406275} {astro-ph/0406275} \BibitemShut
  {NoStop}%
\bibitem [{\citenamefont {{Bridle}}\ and\ \citenamefont
  {{King}}(2007)}]{Bridle:2007}%
  \BibitemOpen
  \bibfield  {author} {\bibinfo {author} {\bibfnamefont {S.}~\bibnamefont
  {{Bridle}}}\ and\ \bibinfo {author} {\bibfnamefont {L.}~\bibnamefont
  {{King}}},\ }\href {\doibase 10.1088/1367-2630/9/12/444} {\bibfield
  {journal} {\bibinfo  {journal} {New Journal of Physics}\ }\textbf {\bibinfo
  {volume} {9}},\ \bibinfo {pages} {444} (\bibinfo {year} {2007})},\ \Eprint
  {http://arxiv.org/abs/0705.0166} {arXiv:0705.0166} \BibitemShut {NoStop}%
\bibitem [{\citenamefont {{Mead}}\ \emph {et~al.}(2015)\citenamefont {{Mead}},
  \citenamefont {{Peacock}}, \citenamefont {{Heymans}}, \citenamefont
  {{Joudaki}},\ and\ \citenamefont {{Heavens}}}]{Mead:2015}%
  \BibitemOpen
  \bibfield  {author} {\bibinfo {author} {\bibfnamefont {A.~J.}\ \bibnamefont
  {{Mead}}}, \bibinfo {author} {\bibfnamefont {J.~A.}\ \bibnamefont
  {{Peacock}}}, \bibinfo {author} {\bibfnamefont {C.}~\bibnamefont
  {{Heymans}}}, \bibinfo {author} {\bibfnamefont {S.}~\bibnamefont
  {{Joudaki}}}, \ and\ \bibinfo {author} {\bibfnamefont {A.~F.}\ \bibnamefont
  {{Heavens}}},\ }\href {\doibase 10.1093/mnras/stv2036} {\bibfield  {journal}
  {\bibinfo  {journal} {\mnras}\ }\textbf {\bibinfo {volume} {454}},\ \bibinfo
  {pages} {1958} (\bibinfo {year} {2015})},\ \Eprint
  {http://arxiv.org/abs/1505.07833} {arXiv:1505.07833} \BibitemShut {NoStop}%
\bibitem [{\citenamefont {{Sullivan}}\ \emph {et~al.}(2010)\citenamefont
  {{Sullivan}}, \citenamefont {{Conley}}, \citenamefont {{Howell}},
  \citenamefont {{Neill}}, \citenamefont {{Astier}}, \citenamefont {{Balland}},
  \citenamefont {{Basa}}, \citenamefont {{Carlberg}}, \citenamefont
  {{Fouchez}}, \citenamefont {{Guy}}, \citenamefont {{Hardin}}, \citenamefont
  {{Hook}}, \citenamefont {{Pain}}, \citenamefont {{Palanque-Delabrouille}},
  \citenamefont {{Perrett}},\ and\ \citenamefont {{et al.}}}]{Sullivan:2010}%
  \BibitemOpen
  \bibfield  {author} {\bibinfo {author} {\bibfnamefont {M.}~\bibnamefont
  {{Sullivan}}}, \bibinfo {author} {\bibfnamefont {A.}~\bibnamefont
  {{Conley}}}, \bibinfo {author} {\bibfnamefont {D.~A.}\ \bibnamefont
  {{Howell}}}, \bibinfo {author} {\bibfnamefont {J.~D.}\ \bibnamefont
  {{Neill}}}, \bibinfo {author} {\bibfnamefont {P.}~\bibnamefont {{Astier}}},
  \bibinfo {author} {\bibfnamefont {C.}~\bibnamefont {{Balland}}}, \bibinfo
  {author} {\bibfnamefont {S.}~\bibnamefont {{Basa}}}, \bibinfo {author}
  {\bibfnamefont {R.~G.}\ \bibnamefont {{Carlberg}}}, \bibinfo {author}
  {\bibfnamefont {D.}~\bibnamefont {{Fouchez}}}, \bibinfo {author}
  {\bibfnamefont {J.}~\bibnamefont {{Guy}}}, \bibinfo {author} {\bibfnamefont
  {D.}~\bibnamefont {{Hardin}}}, \bibinfo {author} {\bibfnamefont {I.~M.}\
  \bibnamefont {{Hook}}}, \bibinfo {author} {\bibfnamefont {R.}~\bibnamefont
  {{Pain}}}, \bibinfo {author} {\bibfnamefont {N.}~\bibnamefont
  {{Palanque-Delabrouille}}}, \bibinfo {author} {\bibfnamefont {K.~M.}\
  \bibnamefont {{Perrett}}}, \ and\ \bibinfo {author} {\bibnamefont {{et
  al.}}},\ }\href {\doibase 10.1111/j.1365-2966.2010.16731.x} {\bibfield
  {journal} {\bibinfo  {journal} {\mnras}\ }\textbf {\bibinfo {volume} {406}},\
  \bibinfo {pages} {782} (\bibinfo {year} {2010})},\ \Eprint
  {http://arxiv.org/abs/1003.5119} {arXiv:1003.5119} \BibitemShut {NoStop}%
\bibitem [{\citenamefont {{Johansson}}\ \emph {et~al.}(2013)\citenamefont
  {{Johansson}}, \citenamefont {{Thomas}}, \citenamefont {{Pforr}},
  \citenamefont {{Maraston}}, \citenamefont {{Nichol}}, \citenamefont
  {{Smith}}, \citenamefont {{Lampeitl}}, \citenamefont {{Beifiori}},
  \citenamefont {{Gupta}},\ and\ \citenamefont {{Schneider}}}]{Johansson:2013}%
  \BibitemOpen
  \bibfield  {author} {\bibinfo {author} {\bibfnamefont {J.}~\bibnamefont
  {{Johansson}}}, \bibinfo {author} {\bibfnamefont {D.}~\bibnamefont
  {{Thomas}}}, \bibinfo {author} {\bibfnamefont {J.}~\bibnamefont {{Pforr}}},
  \bibinfo {author} {\bibfnamefont {C.}~\bibnamefont {{Maraston}}}, \bibinfo
  {author} {\bibfnamefont {R.~C.}\ \bibnamefont {{Nichol}}}, \bibinfo {author}
  {\bibfnamefont {M.}~\bibnamefont {{Smith}}}, \bibinfo {author} {\bibfnamefont
  {H.}~\bibnamefont {{Lampeitl}}}, \bibinfo {author} {\bibfnamefont
  {A.}~\bibnamefont {{Beifiori}}}, \bibinfo {author} {\bibfnamefont {R.~R.}\
  \bibnamefont {{Gupta}}}, \ and\ \bibinfo {author} {\bibfnamefont {D.~P.}\
  \bibnamefont {{Schneider}}},\ }\href {\doibase 10.1093/mnras/stt1408}
  {\bibfield  {journal} {\bibinfo  {journal} {\mnras}\ }\textbf {\bibinfo
  {volume} {435}},\ \bibinfo {pages} {1680} (\bibinfo {year} {2013})},\ \Eprint
  {http://arxiv.org/abs/1211.1386} {arXiv:1211.1386} \BibitemShut {NoStop}%
\bibitem [{\citenamefont {{van Engelen}}\ \emph {et~al.}(2012)\citenamefont
  {{van Engelen}}, \citenamefont {{Keisler}}, \citenamefont {{Zahn}},
  \citenamefont {{Aird}}, \citenamefont {{Benson}}, \citenamefont {{Bleem}},
  \citenamefont {{Carlstrom}}, \citenamefont {{Chang}}, \citenamefont {{Cho}},
  \citenamefont {{Crawford}}, \citenamefont {{Crites}}, \citenamefont {{de
  Haan}}, \citenamefont {{Dobbs}}, \citenamefont {{Dudley}}, \citenamefont
  {{George}},\ and\ \citenamefont {{et al.}}}]{van-Engelen:2012}%
  \BibitemOpen
  \bibfield  {author} {\bibinfo {author} {\bibfnamefont {A.}~\bibnamefont {{van
  Engelen}}}, \bibinfo {author} {\bibfnamefont {R.}~\bibnamefont {{Keisler}}},
  \bibinfo {author} {\bibfnamefont {O.}~\bibnamefont {{Zahn}}}, \bibinfo
  {author} {\bibfnamefont {K.~A.}\ \bibnamefont {{Aird}}}, \bibinfo {author}
  {\bibfnamefont {B.~A.}\ \bibnamefont {{Benson}}}, \bibinfo {author}
  {\bibfnamefont {L.~E.}\ \bibnamefont {{Bleem}}}, \bibinfo {author}
  {\bibfnamefont {J.~E.}\ \bibnamefont {{Carlstrom}}}, \bibinfo {author}
  {\bibfnamefont {C.~L.}\ \bibnamefont {{Chang}}}, \bibinfo {author}
  {\bibfnamefont {H.~M.}\ \bibnamefont {{Cho}}}, \bibinfo {author}
  {\bibfnamefont {T.~M.}\ \bibnamefont {{Crawford}}}, \bibinfo {author}
  {\bibfnamefont {A.~T.}\ \bibnamefont {{Crites}}}, \bibinfo {author}
  {\bibfnamefont {T.}~\bibnamefont {{de Haan}}}, \bibinfo {author}
  {\bibfnamefont {M.~A.}\ \bibnamefont {{Dobbs}}}, \bibinfo {author}
  {\bibfnamefont {J.}~\bibnamefont {{Dudley}}}, \bibinfo {author}
  {\bibfnamefont {E.~M.}\ \bibnamefont {{George}}}, \ and\ \bibinfo {author}
  {\bibnamefont {{et al.}}},\ }\href {\doibase 10.1088/0004-637X/756/2/142}
  {\bibfield  {journal} {\bibinfo  {journal} {\apj}\ }\textbf {\bibinfo
  {volume} {756}},\ \bibinfo {eid} {142} (\bibinfo {year} {2012})},\ \Eprint
  {http://arxiv.org/abs/1202.0546} {arXiv:1202.0546} \BibitemShut {NoStop}%
\bibitem [{\citenamefont {{Kaufman}}(1967)}]{Kaufman:1967}%
  \BibitemOpen
  \bibfield  {author} {\bibinfo {author} {\bibfnamefont {G.~M.}\ \bibnamefont
  {{Kaufman}}},\ }\href@noop {} {\bibfield  {journal} {\bibinfo  {journal}
  {Report No. 6710, Centre for Operations Research and Econometrics, Catholic
  University of Louvain, Heverlee, Belgium}\ } (\bibinfo {year}
  {1967})}\BibitemShut {NoStop}%
\bibitem [{\citenamefont {Anderson}(2003)}]{Anderson:2003}%
  \BibitemOpen
  \bibfield  {author} {\bibinfo {author} {\bibfnamefont {T.}~\bibnamefont
  {Anderson}},\ }\href {https://books.google.ch/books?id=Cmm9QgAACAAJ} {\emph
  {\bibinfo {title} {An Introduction to Multivariate Statistical Analysis}}},\
  Wiley Series in Probability and Statistics\ (\bibinfo  {publisher} {Wiley},\
  \bibinfo {year} {2003})\BibitemShut {NoStop}%
\bibitem [{\citenamefont {{Hartlap}}\ \emph {et~al.}(2007)\citenamefont
  {{Hartlap}}, \citenamefont {{Simon}},\ and\ \citenamefont
  {{Schneider}}}]{Hartlap:2007}%
  \BibitemOpen
  \bibfield  {author} {\bibinfo {author} {\bibfnamefont {J.}~\bibnamefont
  {{Hartlap}}}, \bibinfo {author} {\bibfnamefont {P.}~\bibnamefont {{Simon}}},
  \ and\ \bibinfo {author} {\bibfnamefont {P.}~\bibnamefont {{Schneider}}},\
  }\href {\doibase 10.1051/0004-6361:20066170} {\bibfield  {journal} {\bibinfo
  {journal} {\aap}\ }\textbf {\bibinfo {volume} {464}},\ \bibinfo {pages} {399}
  (\bibinfo {year} {2007})},\ \Eprint {http://arxiv.org/abs/astro-ph/0608064}
  {astro-ph/0608064} \BibitemShut {NoStop}%
\bibitem [{\citenamefont {{Scovacricchi}}\ \emph {et~al.}(2016)\citenamefont
  {{Scovacricchi}}, \citenamefont {{Nichol}}, \citenamefont {{Macaulay}},\ and\
  \citenamefont {{Bacon}}}]{Scovacricchi:2016}%
  \BibitemOpen
  \bibfield  {author} {\bibinfo {author} {\bibfnamefont {D.}~\bibnamefont
  {{Scovacricchi}}}, \bibinfo {author} {\bibfnamefont {R.~C.}\ \bibnamefont
  {{Nichol}}}, \bibinfo {author} {\bibfnamefont {E.}~\bibnamefont
  {{Macaulay}}}, \ and\ \bibinfo {author} {\bibfnamefont {D.}~\bibnamefont
  {{Bacon}}},\ }\href@noop {} {\bibfield  {journal} {\bibinfo  {journal} {ArXiv
  e-prints}\ } (\bibinfo {year} {2016})},\ \Eprint
  {http://arxiv.org/abs/1611.01315} {arXiv:1611.01315} \BibitemShut {NoStop}%
\bibitem [{\citenamefont {{Akeret}}\ \emph {et~al.}(2013)\citenamefont
  {{Akeret}}, \citenamefont {{Seehars}}, \citenamefont {{Amara}}, \citenamefont
  {{Refregier}},\ and\ \citenamefont {{Csillaghy}}}]{Akeret:2012}%
  \BibitemOpen
  \bibfield  {author} {\bibinfo {author} {\bibfnamefont {J.}~\bibnamefont
  {{Akeret}}}, \bibinfo {author} {\bibfnamefont {S.}~\bibnamefont {{Seehars}}},
  \bibinfo {author} {\bibfnamefont {A.}~\bibnamefont {{Amara}}}, \bibinfo
  {author} {\bibfnamefont {A.}~\bibnamefont {{Refregier}}}, \ and\ \bibinfo
  {author} {\bibfnamefont {A.}~\bibnamefont {{Csillaghy}}},\ }\href@noop {}
  {\bibfield  {journal} {\bibinfo  {journal} {Astronomy and Computing}\
  }\textbf {\bibinfo {volume} {2}},\ \bibinfo {pages} {27} (\bibinfo {year}
  {2013})},\ \Eprint {http://arxiv.org/abs/1212.1721} {arXiv:1212.1721
  [astro-ph.CO]} \BibitemShut {NoStop}%
\bibitem [{\citenamefont {{Pen}}(1998)}]{Pen:1998}%
  \BibitemOpen
  \bibfield  {author} {\bibinfo {author} {\bibfnamefont {U.-L.}\ \bibnamefont
  {{Pen}}},\ }\href {\doibase 10.1086/306098} {\bibfield  {journal} {\bibinfo
  {journal} {\apj}\ }\textbf {\bibinfo {volume} {504}},\ \bibinfo {pages} {601}
  (\bibinfo {year} {1998})},\ \Eprint {http://arxiv.org/abs/astro-ph/9711180}
  {astro-ph/9711180} \BibitemShut {NoStop}%
\bibitem [{\citenamefont {{Tegmark}}\ and\ \citenamefont
  {{Peebles}}(1998)}]{Tegmark:1998}%
  \BibitemOpen
  \bibfield  {author} {\bibinfo {author} {\bibfnamefont {M.}~\bibnamefont
  {{Tegmark}}}\ and\ \bibinfo {author} {\bibfnamefont {P.~J.~E.}\ \bibnamefont
  {{Peebles}}},\ }\href {\doibase 10.1086/311426} {\bibfield  {journal}
  {\bibinfo  {journal} {\apjl}\ }\textbf {\bibinfo {volume} {500}},\ \bibinfo
  {pages} {L79} (\bibinfo {year} {1998})},\ \Eprint
  {http://arxiv.org/abs/astro-ph/9804067} {astro-ph/9804067} \BibitemShut
  {NoStop}%
\bibitem [{\citenamefont {{Dekel}}\ and\ \citenamefont
  {{Lahav}}(1999)}]{Dekel:1999}%
  \BibitemOpen
  \bibfield  {author} {\bibinfo {author} {\bibfnamefont {A.}~\bibnamefont
  {{Dekel}}}\ and\ \bibinfo {author} {\bibfnamefont {O.}~\bibnamefont
  {{Lahav}}},\ }\href {\doibase 10.1086/307428} {\bibfield  {journal} {\bibinfo
   {journal} {\apj}\ }\textbf {\bibinfo {volume} {520}},\ \bibinfo {pages} {24}
  (\bibinfo {year} {1999})},\ \Eprint {http://arxiv.org/abs/astro-ph/9806193}
  {astro-ph/9806193} \BibitemShut {NoStop}%
\bibitem [{\citenamefont {{Hinshaw}}\ \emph {et~al.}(2013)\citenamefont
  {{Hinshaw}}, \citenamefont {{Larson}}, \citenamefont {{Komatsu}},
  \citenamefont {{Spergel}}, \citenamefont {{Bennett}}, \citenamefont
  {{Dunkley}}, \citenamefont {{Nolta}}, \citenamefont {{Halpern}},
  \citenamefont {{Hill}}, \citenamefont {{Odegard}}, \citenamefont {{Page}},
  \citenamefont {{Smith}}, \citenamefont {{Weiland}}, \citenamefont {{Gold}},
  \citenamefont {{Jarosik}},\ and\ \citenamefont {et~al.}}]{Hinshaw:2013}%
  \BibitemOpen
  \bibfield  {author} {\bibinfo {author} {\bibfnamefont {G.}~\bibnamefont
  {{Hinshaw}}}, \bibinfo {author} {\bibfnamefont {D.}~\bibnamefont {{Larson}}},
  \bibinfo {author} {\bibfnamefont {E.}~\bibnamefont {{Komatsu}}}, \bibinfo
  {author} {\bibfnamefont {D.~N.}\ \bibnamefont {{Spergel}}}, \bibinfo {author}
  {\bibfnamefont {C.~L.}\ \bibnamefont {{Bennett}}}, \bibinfo {author}
  {\bibfnamefont {J.}~\bibnamefont {{Dunkley}}}, \bibinfo {author}
  {\bibfnamefont {M.~R.}\ \bibnamefont {{Nolta}}}, \bibinfo {author}
  {\bibfnamefont {M.}~\bibnamefont {{Halpern}}}, \bibinfo {author}
  {\bibfnamefont {R.~S.}\ \bibnamefont {{Hill}}}, \bibinfo {author}
  {\bibfnamefont {N.}~\bibnamefont {{Odegard}}}, \bibinfo {author}
  {\bibfnamefont {L.}~\bibnamefont {{Page}}}, \bibinfo {author} {\bibfnamefont
  {K.~M.}\ \bibnamefont {{Smith}}}, \bibinfo {author} {\bibfnamefont {J.~L.}\
  \bibnamefont {{Weiland}}}, \bibinfo {author} {\bibfnamefont {B.}~\bibnamefont
  {{Gold}}}, \bibinfo {author} {\bibfnamefont {N.}~\bibnamefont {{Jarosik}}}, \
  and\ \bibinfo {author} {\bibnamefont {et~al.}},\ }\href {\doibase
  10.1088/0067-0049/208/2/19} {\bibfield  {journal} {\bibinfo  {journal}
  {\apjs}\ }\textbf {\bibinfo {volume} {208}},\ \bibinfo {eid} {19} (\bibinfo
  {year} {2013})},\ \Eprint {http://arxiv.org/abs/1212.5226} {arXiv:1212.5226}
  \BibitemShut {NoStop}%
\bibitem [{\citenamefont {{Hirata}}\ and\ \citenamefont
  {{Seljak}}(2003)}]{Hirata:2003}%
  \BibitemOpen
  \bibfield  {author} {\bibinfo {author} {\bibfnamefont {C.}~\bibnamefont
  {{Hirata}}}\ and\ \bibinfo {author} {\bibfnamefont {U.}~\bibnamefont
  {{Seljak}}},\ }\href {\doibase 10.1046/j.1365-8711.2003.06683.x} {\bibfield
  {journal} {\bibinfo  {journal} {\mnras}\ }\textbf {\bibinfo {volume} {343}},\
  \bibinfo {pages} {459} (\bibinfo {year} {2003})},\ \Eprint
  {http://arxiv.org/abs/astro-ph/0301054} {astro-ph/0301054} \BibitemShut
  {NoStop}%
\bibitem [{\citenamefont {{Planck Collaboration}}\ \emph
  {et~al.}(2016{\natexlab{d}})\citenamefont {{Planck Collaboration}},
  \citenamefont {{Ade}}, \citenamefont {{Aghanim}}, \citenamefont {{Arnaud}},
  \citenamefont {{Ashdown}}, \citenamefont {{Aumont}}, \citenamefont
  {{Baccigalupi}}, \citenamefont {{Banday}}, \citenamefont {{Barreiro}},
  \citenamefont {{Bartlett}},\ and\ \citenamefont
  {et~al.}}]{Planck-Collaboration:2016ae}%
  \BibitemOpen
  \bibfield  {author} {\bibinfo {author} {\bibnamefont {{Planck
  Collaboration}}}, \bibinfo {author} {\bibfnamefont {P.~A.~R.}\ \bibnamefont
  {{Ade}}}, \bibinfo {author} {\bibfnamefont {N.}~\bibnamefont {{Aghanim}}},
  \bibinfo {author} {\bibfnamefont {M.}~\bibnamefont {{Arnaud}}}, \bibinfo
  {author} {\bibfnamefont {M.}~\bibnamefont {{Ashdown}}}, \bibinfo {author}
  {\bibfnamefont {J.}~\bibnamefont {{Aumont}}}, \bibinfo {author}
  {\bibfnamefont {C.}~\bibnamefont {{Baccigalupi}}}, \bibinfo {author}
  {\bibfnamefont {A.~J.}\ \bibnamefont {{Banday}}}, \bibinfo {author}
  {\bibfnamefont {R.~B.}\ \bibnamefont {{Barreiro}}}, \bibinfo {author}
  {\bibfnamefont {J.~G.}\ \bibnamefont {{Bartlett}}}, \ and\ \bibinfo {author}
  {\bibnamefont {et~al.}},\ }\href {\doibase 10.1051/0004-6361/201525830}
  {\bibfield  {journal} {\bibinfo  {journal} {\aap}\ }\textbf {\bibinfo
  {volume} {594}},\ \bibinfo {eid} {A13} (\bibinfo {year}
  {2016}{\natexlab{d}})},\ \Eprint {http://arxiv.org/abs/1502.01589}
  {arXiv:1502.01589} \BibitemShut {NoStop}%
\bibitem [{\citenamefont {{Hildebrandt}}\ \emph {et~al.}(2016)\citenamefont
  {{Hildebrandt}}, \citenamefont {{Viola}}, \citenamefont {{Heymans}},
  \citenamefont {{Joudaki}}, \citenamefont {{Kuijken}}, \citenamefont
  {{Blake}}, \citenamefont {{Erben}}, \citenamefont {{Joachimi}}, \citenamefont
  {{Klaes}}, \citenamefont {{Miller}}, \citenamefont {{Morrison}},
  \citenamefont {{Nakajima}}, \citenamefont {{Verdoes Kleijn}}, \citenamefont
  {{Amon}},\ and\ \citenamefont {{et al.}}}]{Hildebrandt:2016}%
  \BibitemOpen
  \bibfield  {author} {\bibinfo {author} {\bibfnamefont {H.}~\bibnamefont
  {{Hildebrandt}}}, \bibinfo {author} {\bibfnamefont {M.}~\bibnamefont
  {{Viola}}}, \bibinfo {author} {\bibfnamefont {C.}~\bibnamefont {{Heymans}}},
  \bibinfo {author} {\bibfnamefont {S.}~\bibnamefont {{Joudaki}}}, \bibinfo
  {author} {\bibfnamefont {K.}~\bibnamefont {{Kuijken}}}, \bibinfo {author}
  {\bibfnamefont {C.}~\bibnamefont {{Blake}}}, \bibinfo {author} {\bibfnamefont
  {T.}~\bibnamefont {{Erben}}}, \bibinfo {author} {\bibfnamefont
  {B.}~\bibnamefont {{Joachimi}}}, \bibinfo {author} {\bibfnamefont
  {D.}~\bibnamefont {{Klaes}}}, \bibinfo {author} {\bibfnamefont
  {L.}~\bibnamefont {{Miller}}}, \bibinfo {author} {\bibfnamefont {C.~B.}\
  \bibnamefont {{Morrison}}}, \bibinfo {author} {\bibfnamefont
  {R.}~\bibnamefont {{Nakajima}}}, \bibinfo {author} {\bibfnamefont
  {G.}~\bibnamefont {{Verdoes Kleijn}}}, \bibinfo {author} {\bibfnamefont
  {A.}~\bibnamefont {{Amon}}}, \ and\ \bibinfo {author} {\bibnamefont {{et
  al.}}},\ }\href@noop {} {\bibfield  {journal} {\bibinfo  {journal} {ArXiv
  e-prints}\ } (\bibinfo {year} {2016})},\ \Eprint
  {http://arxiv.org/abs/1606.05338} {arXiv:1606.05338} \BibitemShut {NoStop}%
\bibitem [{\citenamefont {{Planck Collaboration}}\ \emph
  {et~al.}(2016{\natexlab{e}})\citenamefont {{Planck Collaboration}},
  \citenamefont {{Aghanim}}, \citenamefont {{Akrami}}, \citenamefont
  {{Ashdown}}, \citenamefont {{Aumont}}, \citenamefont {{Ballardini}},
  \citenamefont {{Banday}}, \citenamefont {{Barreiro}}, \citenamefont
  {{Bartolo}}, \citenamefont {{Basak}}, \citenamefont {{Benabed}},
  \citenamefont {{Bersanelli}}, \citenamefont {{Bielewicz}}, \citenamefont
  {{Bonaldi}}, \citenamefont {{Bonavera}}, \citenamefont {{Bond}},\ and\
  \citenamefont {{et al.}}}]{Planck-Collaboration:2016ac}%
  \BibitemOpen
  \bibfield  {author} {\bibinfo {author} {\bibnamefont {{Planck
  Collaboration}}}, \bibinfo {author} {\bibfnamefont {N.}~\bibnamefont
  {{Aghanim}}}, \bibinfo {author} {\bibfnamefont {Y.}~\bibnamefont {{Akrami}}},
  \bibinfo {author} {\bibfnamefont {M.}~\bibnamefont {{Ashdown}}}, \bibinfo
  {author} {\bibfnamefont {J.}~\bibnamefont {{Aumont}}}, \bibinfo {author}
  {\bibfnamefont {M.}~\bibnamefont {{Ballardini}}}, \bibinfo {author}
  {\bibfnamefont {A.~J.}\ \bibnamefont {{Banday}}}, \bibinfo {author}
  {\bibfnamefont {R.~B.}\ \bibnamefont {{Barreiro}}}, \bibinfo {author}
  {\bibfnamefont {N.}~\bibnamefont {{Bartolo}}}, \bibinfo {author}
  {\bibfnamefont {S.}~\bibnamefont {{Basak}}}, \bibinfo {author} {\bibfnamefont
  {K.}~\bibnamefont {{Benabed}}}, \bibinfo {author} {\bibfnamefont
  {M.}~\bibnamefont {{Bersanelli}}}, \bibinfo {author} {\bibfnamefont
  {P.}~\bibnamefont {{Bielewicz}}}, \bibinfo {author} {\bibfnamefont
  {A.}~\bibnamefont {{Bonaldi}}}, \bibinfo {author} {\bibfnamefont
  {L.}~\bibnamefont {{Bonavera}}}, \bibinfo {author} {\bibfnamefont {J.~R.}\
  \bibnamefont {{Bond}}}, \ and\ \bibinfo {author} {\bibnamefont {{et al.}}},\
  }\href@noop {} {\bibfield  {journal} {\bibinfo  {journal} {ArXiv e-prints}\ }
  (\bibinfo {year} {2016}{\natexlab{e}})},\ \Eprint
  {http://arxiv.org/abs/1608.02487} {arXiv:1608.02487} \BibitemShut {NoStop}%
\bibitem [{\citenamefont {{Addison}}\ \emph {et~al.}(2016)\citenamefont
  {{Addison}}, \citenamefont {{Huang}}, \citenamefont {{Watts}}, \citenamefont
  {{Bennett}}, \citenamefont {{Halpern}}, \citenamefont {{Hinshaw}},\ and\
  \citenamefont {{Weiland}}}]{Addison:2016}%
  \BibitemOpen
  \bibfield  {author} {\bibinfo {author} {\bibfnamefont {G.~E.}\ \bibnamefont
  {{Addison}}}, \bibinfo {author} {\bibfnamefont {Y.}~\bibnamefont {{Huang}}},
  \bibinfo {author} {\bibfnamefont {D.~J.}\ \bibnamefont {{Watts}}}, \bibinfo
  {author} {\bibfnamefont {C.~L.}\ \bibnamefont {{Bennett}}}, \bibinfo {author}
  {\bibfnamefont {M.}~\bibnamefont {{Halpern}}}, \bibinfo {author}
  {\bibfnamefont {G.}~\bibnamefont {{Hinshaw}}}, \ and\ \bibinfo {author}
  {\bibfnamefont {J.~L.}\ \bibnamefont {{Weiland}}},\ }\href {\doibase
  10.3847/0004-637X/818/2/132} {\bibfield  {journal} {\bibinfo  {journal}
  {\apj}\ }\textbf {\bibinfo {volume} {818}},\ \bibinfo {eid} {132} (\bibinfo
  {year} {2016})},\ \Eprint {http://arxiv.org/abs/1511.00055}
  {arXiv:1511.00055} \BibitemShut {NoStop}%
\bibitem [{\citenamefont {{Fowler}}\ \emph {et~al.}(2010)\citenamefont
  {{Fowler}}, \citenamefont {{Acquaviva}}, \citenamefont {{Ade}}, \citenamefont
  {{Aguirre}}, \citenamefont {{Amiri}}, \citenamefont {{Appel}}, \citenamefont
  {{Barrientos}}, \citenamefont {{Battistelli}}, \citenamefont {{Bond}},
  \citenamefont {{Brown}}, \citenamefont {{Burger}}, \citenamefont
  {{Chervenak}}, \citenamefont {{Das}}, \citenamefont {{Devlin}}, \citenamefont
  {{Dicker}},\ and\ \citenamefont {{et al.}}}]{Fowler:2010}%
  \BibitemOpen
  \bibfield  {author} {\bibinfo {author} {\bibfnamefont {J.~W.}\ \bibnamefont
  {{Fowler}}}, \bibinfo {author} {\bibfnamefont {V.}~\bibnamefont
  {{Acquaviva}}}, \bibinfo {author} {\bibfnamefont {P.~A.~R.}\ \bibnamefont
  {{Ade}}}, \bibinfo {author} {\bibfnamefont {P.}~\bibnamefont {{Aguirre}}},
  \bibinfo {author} {\bibfnamefont {M.}~\bibnamefont {{Amiri}}}, \bibinfo
  {author} {\bibfnamefont {J.~W.}\ \bibnamefont {{Appel}}}, \bibinfo {author}
  {\bibfnamefont {L.~F.}\ \bibnamefont {{Barrientos}}}, \bibinfo {author}
  {\bibfnamefont {E.~S.}\ \bibnamefont {{Battistelli}}}, \bibinfo {author}
  {\bibfnamefont {J.~R.}\ \bibnamefont {{Bond}}}, \bibinfo {author}
  {\bibfnamefont {B.}~\bibnamefont {{Brown}}}, \bibinfo {author} {\bibfnamefont
  {B.}~\bibnamefont {{Burger}}}, \bibinfo {author} {\bibfnamefont
  {J.}~\bibnamefont {{Chervenak}}}, \bibinfo {author} {\bibfnamefont
  {S.}~\bibnamefont {{Das}}}, \bibinfo {author} {\bibfnamefont {M.~J.}\
  \bibnamefont {{Devlin}}}, \bibinfo {author} {\bibfnamefont {S.~R.}\
  \bibnamefont {{Dicker}}}, \ and\ \bibinfo {author} {\bibnamefont {{et
  al.}}},\ }\href {\doibase 10.1088/0004-637X/722/2/1148} {\bibfield  {journal}
  {\bibinfo  {journal} {\apj}\ }\textbf {\bibinfo {volume} {722}},\ \bibinfo
  {pages} {1148} (\bibinfo {year} {2010})},\ \Eprint
  {http://arxiv.org/abs/1001.2934} {arXiv:1001.2934 [astro-ph.CO]} \BibitemShut
  {NoStop}%
\bibitem [{\citenamefont {{Das}}\ \emph {et~al.}(2011)\citenamefont {{Das}},
  \citenamefont {{Marriage}}, \citenamefont {{Ade}}, \citenamefont {{Aguirre}},
  \citenamefont {{Amiri}}, \citenamefont {{Appel}}, \citenamefont
  {{Barrientos}}, \citenamefont {{Battistelli}}, \citenamefont {{Bond}},
  \citenamefont {{Brown}}, \citenamefont {{Burger}}, \citenamefont
  {{Chervenak}}, \citenamefont {{Devlin}}, \citenamefont {{Dicker}},
  \citenamefont {{Bertrand Doriese}},\ and\ \citenamefont {{et
  al.}}}]{Das:2011}%
  \BibitemOpen
  \bibfield  {author} {\bibinfo {author} {\bibfnamefont {S.}~\bibnamefont
  {{Das}}}, \bibinfo {author} {\bibfnamefont {T.~A.}\ \bibnamefont
  {{Marriage}}}, \bibinfo {author} {\bibfnamefont {P.~A.~R.}\ \bibnamefont
  {{Ade}}}, \bibinfo {author} {\bibfnamefont {P.}~\bibnamefont {{Aguirre}}},
  \bibinfo {author} {\bibfnamefont {M.}~\bibnamefont {{Amiri}}}, \bibinfo
  {author} {\bibfnamefont {J.~W.}\ \bibnamefont {{Appel}}}, \bibinfo {author}
  {\bibfnamefont {L.~F.}\ \bibnamefont {{Barrientos}}}, \bibinfo {author}
  {\bibfnamefont {E.~S.}\ \bibnamefont {{Battistelli}}}, \bibinfo {author}
  {\bibfnamefont {J.~R.}\ \bibnamefont {{Bond}}}, \bibinfo {author}
  {\bibfnamefont {B.}~\bibnamefont {{Brown}}}, \bibinfo {author} {\bibfnamefont
  {B.}~\bibnamefont {{Burger}}}, \bibinfo {author} {\bibfnamefont
  {J.}~\bibnamefont {{Chervenak}}}, \bibinfo {author} {\bibfnamefont {M.~J.}\
  \bibnamefont {{Devlin}}}, \bibinfo {author} {\bibfnamefont {S.~R.}\
  \bibnamefont {{Dicker}}}, \bibinfo {author} {\bibfnamefont {W.}~\bibnamefont
  {{Bertrand Doriese}}}, \ and\ \bibinfo {author} {\bibnamefont {{et al.}}},\
  }\href {\doibase 10.1088/0004-637X/729/1/62} {\bibfield  {journal} {\bibinfo
  {journal} {\apj}\ }\textbf {\bibinfo {volume} {729}},\ \bibinfo {eid} {62}
  (\bibinfo {year} {2011})},\ \Eprint {http://arxiv.org/abs/1009.0847}
  {arXiv:1009.0847} \BibitemShut {NoStop}%
\bibitem [{\citenamefont {{Keisler}}\ \emph {et~al.}(2011)\citenamefont
  {{Keisler}}, \citenamefont {{Reichardt}}, \citenamefont {{Aird}},
  \citenamefont {{Benson}}, \citenamefont {{Bleem}}, \citenamefont
  {{Carlstrom}}, \citenamefont {{Chang}}, \citenamefont {{Cho}}, \citenamefont
  {{Crawford}}, \citenamefont {{Crites}}, \citenamefont {{de Haan}},
  \citenamefont {{Dobbs}}, \citenamefont {{Dudley}}, \citenamefont {{George}},
  \citenamefont {{Halverson}},\ and\ \citenamefont {{et al.}}}]{Keisler:2011}%
  \BibitemOpen
  \bibfield  {author} {\bibinfo {author} {\bibfnamefont {R.}~\bibnamefont
  {{Keisler}}}, \bibinfo {author} {\bibfnamefont {C.~L.}\ \bibnamefont
  {{Reichardt}}}, \bibinfo {author} {\bibfnamefont {K.~A.}\ \bibnamefont
  {{Aird}}}, \bibinfo {author} {\bibfnamefont {B.~A.}\ \bibnamefont
  {{Benson}}}, \bibinfo {author} {\bibfnamefont {L.~E.}\ \bibnamefont
  {{Bleem}}}, \bibinfo {author} {\bibfnamefont {J.~E.}\ \bibnamefont
  {{Carlstrom}}}, \bibinfo {author} {\bibfnamefont {C.~L.}\ \bibnamefont
  {{Chang}}}, \bibinfo {author} {\bibfnamefont {H.~M.}\ \bibnamefont {{Cho}}},
  \bibinfo {author} {\bibfnamefont {T.~M.}\ \bibnamefont {{Crawford}}},
  \bibinfo {author} {\bibfnamefont {A.~T.}\ \bibnamefont {{Crites}}}, \bibinfo
  {author} {\bibfnamefont {T.}~\bibnamefont {{de Haan}}}, \bibinfo {author}
  {\bibfnamefont {M.~A.}\ \bibnamefont {{Dobbs}}}, \bibinfo {author}
  {\bibfnamefont {J.}~\bibnamefont {{Dudley}}}, \bibinfo {author}
  {\bibfnamefont {E.~M.}\ \bibnamefont {{George}}}, \bibinfo {author}
  {\bibfnamefont {N.~W.}\ \bibnamefont {{Halverson}}}, \ and\ \bibinfo {author}
  {\bibnamefont {{et al.}}},\ }\href {\doibase 10.1088/0004-637X/743/1/28}
  {\bibfield  {journal} {\bibinfo  {journal} {\apj}\ }\textbf {\bibinfo
  {volume} {743}},\ \bibinfo {eid} {28} (\bibinfo {year} {2011})},\ \Eprint
  {http://arxiv.org/abs/1105.3182} {arXiv:1105.3182} \BibitemShut {NoStop}%
\bibitem [{\citenamefont {{Reichardt}}\ \emph {et~al.}(2012)\citenamefont
  {{Reichardt}}, \citenamefont {{Shaw}}, \citenamefont {{Zahn}}, \citenamefont
  {{Aird}}, \citenamefont {{Benson}}, \citenamefont {{Bleem}}, \citenamefont
  {{Carlstrom}}, \citenamefont {{Chang}}, \citenamefont {{Cho}}, \citenamefont
  {{Crawford}}, \citenamefont {{Crites}}, \citenamefont {{de Haan}},
  \citenamefont {{Dobbs}}, \citenamefont {{Dudley}}, \citenamefont {{George}},\
  and\ \citenamefont {{et al.}}}]{Reichardt:2012}%
  \BibitemOpen
  \bibfield  {author} {\bibinfo {author} {\bibfnamefont {C.~L.}\ \bibnamefont
  {{Reichardt}}}, \bibinfo {author} {\bibfnamefont {L.}~\bibnamefont {{Shaw}}},
  \bibinfo {author} {\bibfnamefont {O.}~\bibnamefont {{Zahn}}}, \bibinfo
  {author} {\bibfnamefont {K.~A.}\ \bibnamefont {{Aird}}}, \bibinfo {author}
  {\bibfnamefont {B.~A.}\ \bibnamefont {{Benson}}}, \bibinfo {author}
  {\bibfnamefont {L.~E.}\ \bibnamefont {{Bleem}}}, \bibinfo {author}
  {\bibfnamefont {J.~E.}\ \bibnamefont {{Carlstrom}}}, \bibinfo {author}
  {\bibfnamefont {C.~L.}\ \bibnamefont {{Chang}}}, \bibinfo {author}
  {\bibfnamefont {H.~M.}\ \bibnamefont {{Cho}}}, \bibinfo {author}
  {\bibfnamefont {T.~M.}\ \bibnamefont {{Crawford}}}, \bibinfo {author}
  {\bibfnamefont {A.~T.}\ \bibnamefont {{Crites}}}, \bibinfo {author}
  {\bibfnamefont {T.}~\bibnamefont {{de Haan}}}, \bibinfo {author}
  {\bibfnamefont {M.~A.}\ \bibnamefont {{Dobbs}}}, \bibinfo {author}
  {\bibfnamefont {J.}~\bibnamefont {{Dudley}}}, \bibinfo {author}
  {\bibfnamefont {E.~M.}\ \bibnamefont {{George}}}, \ and\ \bibinfo {author}
  {\bibnamefont {{et al.}}},\ }\href {\doibase 10.1088/0004-637X/755/1/70}
  {\bibfield  {journal} {\bibinfo  {journal} {\apj}\ }\textbf {\bibinfo
  {volume} {755}},\ \bibinfo {eid} {70} (\bibinfo {year} {2012})},\ \Eprint
  {http://arxiv.org/abs/1111.0932} {arXiv:1111.0932 [astro-ph.CO]} \BibitemShut
  {NoStop}%
\bibitem [{\citenamefont {{Foreman-Mackey}}(2016)}]{ForemanMackey:2016}%
  \BibitemOpen
  \bibfield  {author} {\bibinfo {author} {\bibfnamefont {D.}~\bibnamefont
  {{Foreman-Mackey}}},\ }\href@noop {} {\bibfield  {journal} {\bibinfo
  {journal} {The Journal of Open Source Software}\ }\textbf {\bibinfo {volume}
  {24}} (\bibinfo {year} {2016})}\BibitemShut {NoStop}%
\bibitem [{\citenamefont {{Joachimi}}\ and\ \citenamefont
  {{Bridle}}(2010)}]{Joachimi:2010}%
  \BibitemOpen
  \bibfield  {author} {\bibinfo {author} {\bibfnamefont {B.}~\bibnamefont
  {{Joachimi}}}\ and\ \bibinfo {author} {\bibfnamefont {S.~L.}\ \bibnamefont
  {{Bridle}}},\ }\href {\doibase 10.1051/0004-6361/200913657} {\bibfield
  {journal} {\bibinfo  {journal} {\aap}\ }\textbf {\bibinfo {volume} {523}},\
  \bibinfo {eid} {A1} (\bibinfo {year} {2010})},\ \Eprint
  {http://arxiv.org/abs/0911.2454} {arXiv:0911.2454} \BibitemShut {NoStop}%
\bibitem [{\citenamefont {{Hall}}\ and\ \citenamefont
  {{Taylor}}(2014)}]{Hall:2014}%
  \BibitemOpen
  \bibfield  {author} {\bibinfo {author} {\bibfnamefont {A.}~\bibnamefont
  {{Hall}}}\ and\ \bibinfo {author} {\bibfnamefont {A.}~\bibnamefont
  {{Taylor}}},\ }\href {\doibase 10.1093/mnrasl/slu094} {\bibfield  {journal}
  {\bibinfo  {journal} {\mnras}\ }\textbf {\bibinfo {volume} {443}},\ \bibinfo
  {pages} {L119} (\bibinfo {year} {2014})},\ \Eprint
  {http://arxiv.org/abs/1401.6018} {arXiv:1401.6018} \BibitemShut {NoStop}%
\bibitem [{\citenamefont {{Troxel}}\ and\ \citenamefont
  {{Ishak}}(2014)}]{Troxel:2014}%
  \BibitemOpen
  \bibfield  {author} {\bibinfo {author} {\bibfnamefont {M.~A.}\ \bibnamefont
  {{Troxel}}}\ and\ \bibinfo {author} {\bibfnamefont {M.}~\bibnamefont
  {{Ishak}}},\ }\href {\doibase 10.1103/PhysRevD.89.063528} {\bibfield
  {journal} {\bibinfo  {journal} {\prd}\ }\textbf {\bibinfo {volume} {89}},\
  \bibinfo {eid} {063528} (\bibinfo {year} {2014})},\ \Eprint
  {http://arxiv.org/abs/1401.7051} {arXiv:1401.7051} \BibitemShut {NoStop}%
\bibitem [{\citenamefont {{van Daalen}}\ \emph {et~al.}(2011)\citenamefont
  {{van Daalen}}, \citenamefont {{Schaye}}, \citenamefont {{Booth}},\ and\
  \citenamefont {{Dalla Vecchia}}}]{van-Daalen:2011}%
  \BibitemOpen
  \bibfield  {author} {\bibinfo {author} {\bibfnamefont {M.~P.}\ \bibnamefont
  {{van Daalen}}}, \bibinfo {author} {\bibfnamefont {J.}~\bibnamefont
  {{Schaye}}}, \bibinfo {author} {\bibfnamefont {C.~M.}\ \bibnamefont
  {{Booth}}}, \ and\ \bibinfo {author} {\bibfnamefont {C.}~\bibnamefont {{Dalla
  Vecchia}}},\ }\href {\doibase 10.1111/j.1365-2966.2011.18981.x} {\bibfield
  {journal} {\bibinfo  {journal} {\mnras}\ }\textbf {\bibinfo {volume} {415}},\
  \bibinfo {pages} {3649} (\bibinfo {year} {2011})},\ \Eprint
  {http://arxiv.org/abs/1104.1174} {arXiv:1104.1174 [astro-ph.CO]} \BibitemShut
  {NoStop}%
\bibitem [{\citenamefont {{Mead}}\ \emph {et~al.}(2016)\citenamefont {{Mead}},
  \citenamefont {{Heymans}}, \citenamefont {{Lombriser}}, \citenamefont
  {{Peacock}}, \citenamefont {{Steele}},\ and\ \citenamefont
  {{Winther}}}]{Mead:2016}%
  \BibitemOpen
  \bibfield  {author} {\bibinfo {author} {\bibfnamefont {A.~J.}\ \bibnamefont
  {{Mead}}}, \bibinfo {author} {\bibfnamefont {C.}~\bibnamefont {{Heymans}}},
  \bibinfo {author} {\bibfnamefont {L.}~\bibnamefont {{Lombriser}}}, \bibinfo
  {author} {\bibfnamefont {J.~A.}\ \bibnamefont {{Peacock}}}, \bibinfo {author}
  {\bibfnamefont {O.~I.}\ \bibnamefont {{Steele}}}, \ and\ \bibinfo {author}
  {\bibfnamefont {H.~A.}\ \bibnamefont {{Winther}}},\ }\href {\doibase
  10.1093/mnras/stw681} {\bibfield  {journal} {\bibinfo  {journal} {\mnras}\
  }\textbf {\bibinfo {volume} {459}},\ \bibinfo {pages} {1468} (\bibinfo {year}
  {2016})},\ \Eprint {http://arxiv.org/abs/1602.02154} {arXiv:1602.02154}
  \BibitemShut {NoStop}%
\bibitem [{\citenamefont {{Heitmann}}\ \emph {et~al.}(2010)\citenamefont
  {{Heitmann}}, \citenamefont {{White}}, \citenamefont {{Wagner}},
  \citenamefont {{Habib}},\ and\ \citenamefont {{Higdon}}}]{Heitmann:2010}%
  \BibitemOpen
  \bibfield  {author} {\bibinfo {author} {\bibfnamefont {K.}~\bibnamefont
  {{Heitmann}}}, \bibinfo {author} {\bibfnamefont {M.}~\bibnamefont {{White}}},
  \bibinfo {author} {\bibfnamefont {C.}~\bibnamefont {{Wagner}}}, \bibinfo
  {author} {\bibfnamefont {S.}~\bibnamefont {{Habib}}}, \ and\ \bibinfo
  {author} {\bibfnamefont {D.}~\bibnamefont {{Higdon}}},\ }\href {\doibase
  10.1088/0004-637X/715/1/104} {\bibfield  {journal} {\bibinfo  {journal}
  {\apj}\ }\textbf {\bibinfo {volume} {715}},\ \bibinfo {pages} {104} (\bibinfo
  {year} {2010})},\ \Eprint {http://arxiv.org/abs/0812.1052} {arXiv:0812.1052}
  \BibitemShut {NoStop}%
\bibitem [{\citenamefont {{Heitmann}}\ \emph {et~al.}(2014)\citenamefont
  {{Heitmann}}, \citenamefont {{Lawrence}}, \citenamefont {{Kwan}},
  \citenamefont {{Habib}},\ and\ \citenamefont {{Higdon}}}]{Heitmann:2014}%
  \BibitemOpen
  \bibfield  {author} {\bibinfo {author} {\bibfnamefont {K.}~\bibnamefont
  {{Heitmann}}}, \bibinfo {author} {\bibfnamefont {E.}~\bibnamefont
  {{Lawrence}}}, \bibinfo {author} {\bibfnamefont {J.}~\bibnamefont {{Kwan}}},
  \bibinfo {author} {\bibfnamefont {S.}~\bibnamefont {{Habib}}}, \ and\
  \bibinfo {author} {\bibfnamefont {D.}~\bibnamefont {{Higdon}}},\ }\href
  {\doibase 10.1088/0004-637X/780/1/111} {\bibfield  {journal} {\bibinfo
  {journal} {\apj}\ }\textbf {\bibinfo {volume} {780}},\ \bibinfo {eid} {111}
  (\bibinfo {year} {2014})},\ \Eprint {http://arxiv.org/abs/1304.7849}
  {arXiv:1304.7849} \BibitemShut {NoStop}%
\bibitem [{\citenamefont {{Giannantonio}}\ \emph {et~al.}(2008)\citenamefont
  {{Giannantonio}}, \citenamefont {{Scranton}}, \citenamefont {{Crittenden}},
  \citenamefont {{Nichol}}, \citenamefont {{Boughn}}, \citenamefont {{Myers}},\
  and\ \citenamefont {{Richards}}}]{Giannantonio:2008}%
  \BibitemOpen
  \bibfield  {author} {\bibinfo {author} {\bibfnamefont {T.}~\bibnamefont
  {{Giannantonio}}}, \bibinfo {author} {\bibfnamefont {R.}~\bibnamefont
  {{Scranton}}}, \bibinfo {author} {\bibfnamefont {R.~G.}\ \bibnamefont
  {{Crittenden}}}, \bibinfo {author} {\bibfnamefont {R.~C.}\ \bibnamefont
  {{Nichol}}}, \bibinfo {author} {\bibfnamefont {S.~P.}\ \bibnamefont
  {{Boughn}}}, \bibinfo {author} {\bibfnamefont {A.~D.}\ \bibnamefont
  {{Myers}}}, \ and\ \bibinfo {author} {\bibfnamefont {G.~T.}\ \bibnamefont
  {{Richards}}},\ }\href {\doibase 10.1103/PhysRevD.77.123520} {\bibfield
  {journal} {\bibinfo  {journal} {\prd}\ }\textbf {\bibinfo {volume} {77}},\
  \bibinfo {eid} {123520} (\bibinfo {year} {2008})},\ \Eprint
  {http://arxiv.org/abs/0801.4380} {arXiv:0801.4380} \BibitemShut {NoStop}%
\bibitem [{\citenamefont {{Cabr{\'e}}}\ \emph {et~al.}(2007)\citenamefont
  {{Cabr{\'e}}}, \citenamefont {{Fosalba}}, \citenamefont {{Gazta{\~n}aga}},\
  and\ \citenamefont {{Manera}}}]{Cabre:2007}%
  \BibitemOpen
  \bibfield  {author} {\bibinfo {author} {\bibfnamefont {A.}~\bibnamefont
  {{Cabr{\'e}}}}, \bibinfo {author} {\bibfnamefont {P.}~\bibnamefont
  {{Fosalba}}}, \bibinfo {author} {\bibfnamefont {E.}~\bibnamefont
  {{Gazta{\~n}aga}}}, \ and\ \bibinfo {author} {\bibfnamefont {M.}~\bibnamefont
  {{Manera}}},\ }\href {\doibase 10.1111/j.1365-2966.2007.12280.x} {\bibfield
  {journal} {\bibinfo  {journal} {\mnras}\ }\textbf {\bibinfo {volume} {381}},\
  \bibinfo {pages} {1347} (\bibinfo {year} {2007})},\ \Eprint
  {http://arxiv.org/abs/astro-ph/0701393} {astro-ph/0701393} \BibitemShut
  {NoStop}%
\end{thebibliography}%
                             
\end{document}